\documentclass[a4paper,10pt]{article}
\pdfoutput=1

\usepackage{graphicx}
\usepackage[curve,matrix,arrow,color]{xy}

\usepackage{subcaption}
\usepackage{pstricks}
\usepackage{pstricks-add}
\usepackage{tikz}
\usepackage{verbatim}
\usetikzlibrary{calc}
\usetikzlibrary{decorations.pathreplacing}
\usetikzlibrary{decorations.markings}

\usepackage{mathtools}
\usepackage{upgreek}
\usepackage{amsmath}
\usepackage{amsfonts}
\usepackage[title]{appendix}
\usepackage[margin=2cm]{geometry}

\DeclareMathAlphabet{\oldmathcal}{OMS}{cmsy}{m}{n}

\usepackage{amsmath, amssymb, amsthm}
\usepackage{color,soul}
\usepackage{float}
\usepackage{placeins}
\usepackage{hyperref}
\usetikzlibrary{intersections} 
\usetikzlibrary{er,positioning}
\usetikzlibrary{arrows,shapes}
\usetikzlibrary{decorations.markings}
\usepackage{blkarray}
\usepackage{cite} 

\usepackage{BOONDOX-cal} 

\newcommand{\beq}{\begin{equation}}

\newcommand{\eeq}{\end{equation}}

\newcommand{\q}{\mathfrak{q}}

\newcommand{\aSt}{\mathcal{W}}
\newcommand{\coaSt}{\widetilde{\mathcal{W}}}

\newcommand{\AStTL}[2]{\aSt_{#1,#2}}
\newcommand{\coAStTL}[2]{\coaSt_{#1,#2}}

\newcommand{\bAStTL}[2]{\overline{\mathcal{W}}_{#1,#2}}
\newcommand{\ATL}[1]{\mathsf{T}^{\rm a}_{#1}}

\newcommand{\Verma}[1]{\mathsf{V}_{#1}}
\newcommand{\Vermab}[1] {\overline{\mathsf{V}}_{#1}}
\newcommand{\IrrV}[1]{\mathsf{X}_{#1}}

\newcommand{\FV}[1]{\mathsf{F}_{#1}} 

\newcommand{\VirN}{\hbox{Vir}\otimes\overline{\hbox{Vir}}}
\newcommand{\FF}[1]{\widetilde{\mathsf{V}}_{#1}}
\newcommand{\Vir}{\hbox{Vir}}
\newcommand{\HN}{\mathcal{H}}
\newcommand{\FN}{F}
\newcommand{\PN}{\mathcal{P}}

\newcommand{\m}{\mathfrak{m}}
\newcommand{\Hlatt}{\mathcal{H}}
\newcommand{\Hcont}{H}
\newcommand{\Platt}{\mathcal{P}}
\newcommand{\Pscaled}{{\mathcal p}}

\newcommand{\KSgen}{\mathcal{L}}

\newcommand{\remainder}{\mathcal{R}}
\newcommand{\spindagger}{\dag}
\newcommand{\loopdagger}{\ddag}
\newcommand{\spininner}[2]{\scalebox{1.5}[1]{$\langle$}  #1, #2 \scalebox{1.5}[1]{$\rangle$} }
\newcommand{\loopinner}[2]{( #1, #2 )}
\newcommand{\eXXZ}{f}
\newcommand{\normord}[1]{\boldsymbol{:}\mathrel{#1}\boldsymbol{:}}
\newcommand{\LnMat}[1]{\mathbcal{L}_{#1}}

\newcommand{\LnMatXXZ}[1]{\mathbcal{L}^{\rm XXZ}_{#1}}

\newcommand{\arcsize}{0.2}
\newcommand{\loopU}{\vcenter{\hbox{
	\begin{tikzpicture}
 	\draw[thick, dotted] (-\arcsize -0.05,-\arcsize - 0.05) -- (-\arcsize-0.05,  0.05);
  	\draw[thick, dotted] (3*\arcsize + 0.05,-\arcsize - 0.05) -- (3*\arcsize+ 0.05,  0.05);
  	\draw[thick, dotted] (-\arcsize - 0.05,  0.05) -- (3*\arcsize+ 0.05,  0.05);
  	\draw[thick, dotted] (-\arcsize - 0.05, -\arcsize - 0.05) -- (3*\arcsize+ 0.05, -\arcsize - 0.05);
	\draw[thick] (0,0) arc (-180:0:\arcsize);
	\end{tikzpicture} }}}
\newcommand{\loopJL}{\vcenter{\hbox{
\begin{tikzpicture}
 	\draw[thick, dotted] (-\arcsize -0.05,-\arcsize - 0.05) -- (-\arcsize-0.05,  0.05);
  	\draw[thick, dotted] (3*\arcsize + 0.05,-\arcsize - 0.05) -- (3*\arcsize+ 0.05,  0.05);
  	\draw[thick, dotted] (-\arcsize - 0.05,  0.05) -- (3*\arcsize+ 0.05,  0.05);
  	\draw[thick, dotted] (-\arcsize - 0.05, -\arcsize - 0.05) -- (3*\arcsize+ 0.05, -\arcsize - 0.05);
	\draw[thick] (0,0) arc (0:-90:\arcsize);
	\draw[thick] (\arcsize*2,0) arc (-180:-90:\arcsize);
	\end{tikzpicture} }}  }
\newcommand{\loopII}{\vcenter{\hbox{
\begin{tikzpicture}
 	\draw[thick, dotted] (-\arcsize -0.05,-\arcsize - 0.05) -- (-\arcsize-0.05,  0.05);
  	\draw[thick, dotted] (3*\arcsize + 0.05,-\arcsize - 0.05) -- (3*\arcsize+ 0.05,  0.05);
  	\draw[thick, dotted] (-\arcsize - 0.05,  0.05) -- (3*\arcsize+ 0.05,  0.05);
  	\draw[thick, dotted] (-\arcsize - 0.05, -\arcsize - 0.05) -- (3*\arcsize+ 0.05, -\arcsize - 0.05);
	\draw[thick] (0,0) -- (0,-\arcsize);
	\draw[thick] (\arcsize*2,0) -- (\arcsize*2,-\arcsize) ;
	\end{tikzpicture} }}  }

\usepackage{listings}
\usepackage{stackengine}

\makeatletter
\newcommand*\dashline{\rotatebox[origin=c]{90}{$\dabar@\dabar@\dabar@$}}
\makeatother

\begin{document}

\title{The action of the Virasoro algebra in quantum spin chains. \\ I. The non-rational case}
\date{}

\maketitle

\begin{center}

\vskip 1cm

{\large Linnea Grans-Samuelsson$^{1}$, Jesper Lykke Jacobsen$^{1,2,3,4}$,  and Hubert Saleur$^{1,5}$}

\vspace{1.0cm}
{\sl\small $^1$  Institut de Physique Th\'eorique, Universit\'e Paris Saclay, CEA, CNRS, F-91191 Gif-sur-Yvette, France\\}
{\sl\small $^2$ Laboratoire de Physique de l'\'Ecole Normale Sup\'erieure, ENS, Universit\'e PSL, \\
CNRS, Sorbonne Universit\'e, Universit\'e de Paris, F-75005 Paris, France\\}
{\sl\small $^3$ Sorbonne Universit\'e, \'Ecole Normale Sup\'erieure, CNRS,
Laboratoire de Physique (LPENS), F-75005 Paris, France\\}
{\sl\small $^4$ Institut des Hautes \'Etudes Scientifiques, Universit\'e Paris Saclay, CNRS, \\
Le Bois-Marie, 35 route de Chartres, F-91440 Bures-sur-Yvette, France\\}
{\sl\small $^5$ Department of Physics and Astronomy,
University of Southern California,
Los Angeles, CA 90089, USA\\}

\end{center}

\vskip 2cm


\begin{abstract}
We investigate the action of discretized Virasoro generators, built out of generators of the lattice Temperley-Lieb algebra (``Koo-Saleur generators''\cite{KooSaleur}), in the critical XXZ quantum spin chain. We explore the structure of the continuum-limit Virasoro modules at generic central charge for the XXZ vertex model, paralleling \cite{LoopPaper} for the loop model. We find again indecomposable modules, but this time not logarithmic ones. The limit of the Temperley-Lieb modules $\AStTL{j}{1}$  for $j\neq 0$ contains pairs of ``conjugate states'' with conformal weights $(h_{r,s},h_{r,-s})$ and  $(h_{r,-s},h_{r,s})$ that give rise to dual structures: Verma or co-Verma modules. The limit of $\AStTL{0}{\q^{\pm2}}$ contains diagonal fields $(h_{r,1},h_{r,1})$ and gives rise to either only Verma or only co-Verma modules, depending on the sign of the exponent in $\q^{\pm 2}$. In order to obtain matrix elements of Koo-Saleur generators at large system size $N$ we use Bethe ansatz and Quantum Inverse Scattering methods, computing the form factors for relevant combinations of three neighbouring spin operators. Relations between form factors ensure that the above duality exists already at the lattice level. We also study in which sense Koo-Saleur generators converge to Virasoro generators. We consider convergence in the weak sense, investigating whether the commutator of limits is the same as the limit of the commutator? We find that it coincides only up to the central term. As a side result we compute the ground-state expectation value of two neighbouring Temperley-Lieb generators in the XXZ spin chain.  
\end{abstract}

\tableofcontents


\section{Introduction}

The mechanism responsible for the emergence of the rich  structure of conformal field theories (CFTs)  in the continuum limit of discrete (not necessarily integrable)  lattice models has attracted growing interest in the last few years. There are several  reasons for this. On the one hand, this is part of the more general question of how one can approximate  field theories using discrete systems, with the ultimate goal of carrying out more efficient (quantum) simulations \cite{Vidal,Vidal2,Verstraete}. On the other hand, the possibility of observing, on the lattice, properties ``analogous'' to those of  CFTs opens the door to the introduction of new discrete mathematical tools with a smorgasbord of exciting potential applications \cite{Smirnov,V.Jones}. Finally, the study of non-unitary CFTs---which play, in particular, a crucial role in our description of geometrical statistical models (such as percolation or polymers),  or of transitions between plateaux in the integer quantum Hall effect---is mired in the technical difficulties that emerge when the representation theory of the Virasoro algebra is not semi-simple. In many cases, only a bottom-up approach, where things are studied starting at the level of the lattice model, has made progress possible. An example of this was, at central charge $c=0$, the determination of the logarithmic coupling (also variantly known as indecomposability parameter, beta-invariant, or $b$-number) between the stress-energy tensor and its logarithmic partner in the bulk CFT \cite{GurarieLudwig,DJS,VJS,VGJS}. 

Many  aspects  of CFTs can of course be approached  using lattice-model discretizations. In some cases, results for  well-chosen {\sl finite systems} are already directly relevant to the continuum limit. This is the case, for instance, of modular transformations (studied as early as \cite{PasquierSaleur}; see also \cite{Fendley} for more recent work on this)  and  fusion rules \cite{ReadSaleur07-2,GJS18}.  In other cases, the precise connection with CFT can only be made after the continuum limit is taken. Examples of this include the determination of three-point functions \cite{IkhlefJacobsenSaleur,Vidal3}, and more recently, of  four-point functions \cite{Ribault,JacobsenSaleur,HeGransJacobsenSaleur,HeJacobsenSaleur}. 

Underlying the relationship between lattice discretization and CFT is the central question of the role of conformal transformations (and  thus the Virasoro algebra) in lattice models. In an early work by Koo and Saleur \cite{KooSaleur}, a very simple construction was proposed where the Virasoro algebra $\Vir$ (or the product $\VirN$ of left and right Virasoro algebras) could be  obtained starting with Fourier modes of the local energy and momentum densities in finite spin chains (such as  XXZ), and taking  the limit of large chains while restricting to ``scaling states'', which are states belonging to the continuum limit. The restriction to scaling states is done through a double-limit procedure that we call the ``scaling limit'' and denote by $\mapsto$. (See \cite{KooSaleur} and below for more detail on this). The evidence for the validity of the construction in \cite{KooSaleur} came from exact results in the case of the Ising model combined with rather elementary numerical checks. Quite recently, the construction was put on more serious mathematical footing (and further checked for symplectic fermions) in  \cite{GRS1,GRS2,GRS3}. It was also revisited extensively  for the Ising case using the language of anyons in \cite{Wang}. In \cite{Wang}, a set of conditions for the construction to work more generally---together with more precise definitions of the scaling limit---were also given.

The purpose of this paper is twofold. On the one hand, we wish to revisit the proposal of \cite{KooSaleur} by carrying out much more sophisticated numerics than was possible at the time. This involves in particular the techniques of {\sl lattice form factors}, thanks to which matrix elements of the discrete Virasoro generators can be expressed in closed form using Bethe-ansatz roots. By contrast, in \cite{KooSaleur}, only eigenstates were obtained with the Bethe ansatz, while matrix elements were calculated by brute-force numerics. The quantitative improvement is substantial. While in \cite{KooSaleur} only chains up to length ten or so  were studied, we are easily able in this paper to tackle lengths up to 80. Our results fully confirm the validity of the proposal in \cite{KooSaleur}, and shed some extra light on the nature of the scaling limit, and the convergence of the lattice Virasoro algebra to $\VirN$.

On the other hand, the growing interest in logarithmic CFT has made it crucial to understand in detail the nature of $\VirN$ modules appearing in the continuum limit of specific lattice models when some of the parameters take on values such that the representation theory of $\VirN$ becomes non-semi simple, and the detailed information how irreducible modules are glued in a particular physical model of interest cannot be obtained from general principles. This information is crucial to answer a variety of questions: chief among these is the nature of fields whose conformal weights are in the extended Kac table, $h=h_{r,s}$ with $r,s\in\mathbb{N}^*$. While in unitary CFTs the resulting degenerate behaviour implies the existence of certain differential equations satisfied by the correlators of these fields, such result does not necessarily hold in the non-unitary case, where Virasoro ``norm-squares'' are not positive definite any longer (see more discussion of this below). 
The second purpose of this paper is to 
find out specifically what kind of Virasoro modules occur in the  XXZ chain 
when the Virasoro representations are degenerate---that is, (some) fields belong to the extended Kac table.   We will do this straightforwardly, by exploring the action of the lattice Virasoro generators, and checking directly whether the relevant combinations vanish or not---in technical parlance, whether ``null states'' or ``singular vectors''  are zero indeed. This is of course of utmost importance in practice, as this criterion determines the applicability of the BPZ formalism \cite{BPZ} to the determination of correlation functions, such as the four-point functions currently under investigation \cite{JacobsenSaleur,HeGransJacobsenSaleur,HeJacobsenSaleur}. We shall find some unexpected results, that we hope to complete in a subsequent paper \cite{fullynongenericpaper} by studying the cases when the central charge is rational (e.g., the case $c=0$ with applications to percolation).

A similar investigation in the cognate link-pattern representation of the XXZ chain (relevant for the corresponding loop model) has already appeared in a companion paper \cite{LoopPaper}. We will remark on the important differences between the two representations throughout the present paper.

The paper is organized as follows. Section \ref{sec:disVir} discusses general algebraic features. We consider successively the affine Temperley-Lieb algebra and its representations, the issues of indecomposability, and recall the basic Koo-Saleur conjecture which proposes lattice regularizations for the generators of the Virasoro algebra. In section \ref{sec:contlim} we discuss expected features of the continuum limit, based in part on the Dotsenko-Fateev  or Feigin-Fuchs free-boson construction. We also discuss partly the issue of scalar products---with more remarks on this topic being given in Appendix~\ref{RemScalProd}. In section \ref{sec:Bethe} we recall several aspects of the Bethe-ansatz solution of the XXZ chain, in particular those concerning the form-factor calculations and the definition of scaling states. Relations between the form factors show that duality relations between certain expected continuum modules are already present at finite size. In section 5 we discuss the continuum limit of the Koo-Saleur generators in the {\sl non-degenerate case}, where none of the conjectured conformal weights belong to the extended Kac table. Following this, section \ref{PartlyNonGeneric} discusses their continuum limit in the {\sl degenerate cases} where the conjectured conformal weights take on values in the extended Kac table.  
Section \ref{Anomalies} discusses the nature of the convergence of the Koo-Saleur generators to their Virasoro-algebra continuum limit. Finally, section \ref{Conclusion} contains our concluding remarks.

Several technical aspects are addressed in the appendices. In appendix \ref{RemScalProd} we briefly remind the reader of the physical meaning of the ``conformal scalar product'' for which $L_n^\loopdagger=L_{-n}$ (we reserve the notation $\spindagger$ for another scalar product). In  appendix \ref{FormFactorAppendix} we discuss the form factors for the Koo-Saleur generators. In appendix \ref{NumericsAppendix}  more details about  numerical results  for the action of the Koo-Saleur generators are provided. A proof of \eqref{separate_conj}, an expression involving the ground-state expectation of two neighbouring Temperley-Lieb generators that we initially conjectured based on our numerical results, is provided in appendix \ref{Proof_Appendix}.
In appendix \ref{limits_appendix} we discuss in further detail the continuum limit of commutators of Koo-Saleur generators. Details about a particular commutation relation---which we call the chiral-antichiral commutator---are given in appendix \ref{Chiral-Antichiral}. Finally, the last appendix \ref{UnitaryXXZ} discusses a natural variant of the Koo-Saleur construction in the case of the XXZ chain considered as a system of central charge $c=1$.

\smallskip

\textbf{Main results}: Our main results for relations between form factors and lattice duality are given in equations \eqref{weak} and \eqref{strong}. Our main results for the nature of the modules arising in the continuum limit are given in equations (\ref{MainRes1}) and (\ref{MainRes2}). Our main results for the nature of the convergence of the Koo-Saleur generators are given in \eqref{ndiff0_conj}, \eqref{separate_conj},\eqref{cstar_conjecture} and \eqref{chiral-antichiral_conj}.

\subsection*{Notations and definitions}

We gather here some general notations and definitions that are used throughout the paper:

\begin{itemize}

\item{} $\ATL{N}(\m)$ --- the affine Temperley--Lieb algebra on $N=2L$ sites with parameter $\m$. We shall later parametrize the loop weight as $\m=\q+\q^{-1}$, with $\q=e^{i\gamma}$ and
\begin{equation}
\label{gammaparam}
 \gamma={\pi\over x+1} \,.
\end{equation}

\item{} $\AStTL{j}{e^{i\phi}}$ --- standard module of the affine Temperley-lieb algebra with $2j$ through-lines and pseudomomentum $\phi/2$. We define a corresponding {\em electric charge} as%
\footnote{Note that the twist term in~\cite{PasquierSaleur}, which was denoted there
   $q^{2t}$, reads in these notations as $e^{i\phi}$.  It  corresponds to $e^{2iK}$ in many of our other papers \cite{GRS1,GRS2,GRS3,GRSV}, to $z^2$ in the Graham--Lehrer work~\cite{GL}, and to the parameter $x$ in the work of  Martin--Saleur \cite{MartinSaleur1}.}
\begin{equation} \label{e_phi}
 e_\phi\equiv \frac{\phi}{2\pi} \,.
\end{equation}

\item{} $\Verma{r,s}$ --- Verma module for the conformal weight $h_{r,s}$ when $r,s \notin \mathbb{N}^*$.

\item{} $\Verma{r,s}^{\rm (d)}$ --- the (degenerate) Verma module for the conformal weight $h_{r,s}$ when $r,s\in\mathbb{N}^*$.

\item{} $\FF{r,s}^{\rm (d)}$ --- the (degenerate) co-Verma module for the conformal weight $h_{r,s}$ when $r,s\in\mathbb{N}^*$.

\item{} $\IrrV{r,s}$ --- irreducible Virasoro module for the conformal weight $h_{r,s}$.

\item{} A conformal weight $h_{r,s}$ with $r,s\in\mathbb{N}^*$ will be called degenerate. For such a weight, 
there exists a  descendent state that is also primary: this descendent is often  called a null  or singular vector or state.  We will denote by $A_{r,s}$  the combination of Virasoro generators producing the null state corresponding to the degenerate weight $h_{r,s}$. $A_{r,s}$ is normalized so that the coefficient of $L_{-1}^{rs}$ is equal to unity. Some examples are
\begin{subequations}
\label{A_operators}
\begin{eqnarray}
A_{1,1}&=&L_{-1} \,, \\
A_{1,2}&=&L_{-1}^2-{2(2h_{1,2}+1) \over 3}L_{-2} \,, \\
A_{2,1}&=&L_{-1}^2 -{2(2h_{2,1}+1)\over 3}L_{-2} \,.
\end{eqnarray}
\end{subequations}

\item{} We will in part of this paper use the Dotsenko-Fateev construction \cite{DotsenkoFateev}. For theories with central charge $c=1-{6\over x(x+1)}$, this involves in particular screening charges $\alpha_+=\sqrt{(x+1) /  x}$ and $\alpha_-=-\sqrt{x / (x+1)}$, and the set of vertex operator charges 
$\alpha(e,m)={1\over 2}(e\alpha_+-m\alpha_-)$ and $\bar{\alpha}(e,m)={1\over 2}(e\alpha_++m\alpha_-)$. Instead of ``Coulomb gas'' labels $e,m$ we will use ``Kac'' labels $r,s$, with 
\begin{equation}\label{alpha_rs}
\alpha_{r,s}={1-r\over 2}\alpha_++{1-s\over 2}\alpha_-\,.
\end{equation}
Of course these labels are not independent in a free-field theory with charge at infinity, and one has 
\begin{equation}
\alpha_{-e,m}=\alpha(e,m)+\alpha_0 \,.
\end{equation}
We define the {\em background charge}
\begin{equation} \label{alphazero}
 \alpha_0 = \frac{\alpha_+ + \alpha_-}{2} = {1\over 2\sqrt{x(x+1)}}
\end{equation}
and introduce the notation
\begin{equation}\label{conjugate_charge}
 \alpha^{\rm c} \equiv  2\alpha_0-\alpha
\end{equation}
for the {\em conjugate charge} to a charge $\alpha$. Note that $\alpha_{r,s}^{\rm c}=\alpha_{-r,-s}$. The conformal weight corresponding to a charge $\alpha$ is $ h = \alpha^2-2\alpha\alpha_0 $.

\item{} $\FV{\alpha}$ --- Fock space generated from $\normord{e^{i\alpha\varphi}}$, where $\varphi$ denotes a free bosonic field.

\item{} We will in this paper restrict to generic values of the parameter $\q$ (i.e., $\q$ not a root of unity), and thus to generic values of $x$ (i.e., $x$ irrational). Even in this case, we will encounter situations where some of the modules of interest are not irreducible anymore. We will refer to these situations as ``non-generic'' when applied to modules of the affine Temperley-Lieb algebra, and ``degenerate'' when applied to modules of the Virasoro algebra. In earlier papers, we have referred to such cases as ``partly non-generic'' and ``partly degenerate'', respectively, since having $\q$ a root of unity adds considerably more structure to the modules. We will not do so here, the context clearly excluding $\q$ a root of unity. 

\item{} We shall discuss two scalar products, denoted by $\spininner{-}{-}$ and $\loopinner{-}{-}$, which are defined such that for any two primary states $V_1,V_2$ we have $\spininner{V_1}{L_n V_2} = \spininner{L_n^\spindagger V_1}{V_2}$ and $\loopinner{V_1}{L_n V_2} = \loopinner{L_n^\loopdagger V_1}{V_2}$, where $L_n^\spindagger$ is discussed below and $L_n^\loopdagger=L_{-n}$ is the usual conformal conjugate \cite{BPZ}. The scalar product $\spininner{-}{-}$ is positive definite and will be used for most parts of the paper. When using this scalar product we shall also use the bra-ket notation: $|V\rangle$ denotes a state $V$ (primary or not) and $\langle V|$ its dual, $\langle V_1 | V_2\rangle \equiv \spininner{V_1}{V_2} $ and $\langle V_1| \mathcal{O} |V_2 \rangle \equiv \spininner{V_1}{\mathcal{O}V_2}$ for an operator $\mathcal{O}$ acting on $|V_1\rangle$ (with $V_1,V_2$ being primary or not). 

\item{} We shall call states going over to well defined states in the continuum limit CFT  {\sl scaling states}. The notion of scaling states is made more precise in terms of the Bethe ansatz in Section \ref{Bethe_scaling}.
 We shall call the double limit procedure of restricting to a fixed number of scaling states, and allowing this number to become infinite only after $N\to\infty$, the {\sl scaling limit}, using the notation $\mapsto$. Weak convergence under the condition that everything is restricted to scaling states will be called {\sl scaling-weak convergence}, and is discussed in more detail in Section \ref{scaling_weak}.

\end{itemize}

\section{Discrete Virasoro algebra and the Koo-Saleur formulae}\label{sec:disVir}

While the questions we investigate and the strategy we use are fully general in the context of two-dimensional lattice models having a conformally invariant continuum limit, we focus in this paper specifically on models based on the Temperley-Lieb algebra. As discussed further below, we think of these models as providing some lattice analogue of the Virasoro algebra---or more precisely, since we study systems with closed (i.e., periodic or twisted periodic) boundary conditions, the product of the left and the right Virasoro algebras, $\VirN$---at central charge $c\leq1$. Other types of models could be considered in the same fashion. For instance  models based on the Birman-Wenzl-Murakami algebra  would naturally lead to a lattice analog of the $N=1$ super-Virasoro algebra \cite{Pearce2013}, while models involving higher-rank quantum groups (e.g., $U_\q sl(3)$) would lead to lattice analogs of $W$-algebras \cite{GJprep}.%
\footnote{We note in this respect that a lattice regularization of a $W$-algebra at $c=-2$ was proposed in \cite{GST}.}

\subsection{The Temperley-Lieb algebra in the periodic case}\label{sec:TL-alg-def}

The following two subsections contain material  discussed already in our earlier work on the subject \cite{GRS1,GRS2,GRS3,GRSV}, which we prefer to reproduce here for clarity, completeness and in order to establish notations. 

\subsubsection{The algebra $\ATL{N}(\m)$}

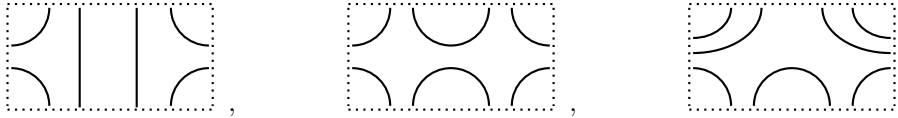
\begin{figure}
\begin{equation*}
 \begin{tikzpicture}
 	\draw[thick, dotted] (-0.05,0.5) arc (0:10:0 and -7.5);
 	\draw[thick, dotted] (-0.05,0.55) -- (2.65,0.55);
 	\draw[thick, dotted] (2.65,0.5) arc (0:10:0 and -7.5);
	\draw[thick, dotted] (-0.05,-0.85) -- (2.65,-0.85);
	\draw[thick] (0,0) arc (-90:0:0.5 and 0.5);
	\draw[thick] (0.9,0.5) arc (0:10:0 and -7.6);
	\draw[thick] (1.65,0.5) arc (0:10:0 and -7.6);
	\draw[thick] (2.6,0) arc (-90:0:-0.5 and 0.5);

	\draw[thick] (0.5,-0.8) arc (0:90:0.5 and 0.5);
	\draw[thick] (2.1,-0.8) arc (0:90:-0.5 and 0.5);
	\end{tikzpicture}\;\;,
	\qquad\qquad
 \begin{tikzpicture}
 	\draw[thick, dotted] (-0.05,0.5) arc (0:10:0 and -7.5);
 	\draw[thick, dotted] (-0.05,0.55) -- (2.65,0.55);
 	\draw[thick, dotted] (2.65,0.5) arc (0:10:0 and -7.5);
	\draw[thick, dotted] (-0.05,-0.85) -- (2.65,-0.85);
	\draw[thick] (0,0) arc (-90:0:0.5 and 0.5);
	\draw[thick] (0.8,0.5) arc (-180:0:0.5 and 0.5);
	\draw[thick] (2.6,0) arc (-90:0:-0.5 and 0.5);

	\draw[thick] (0.5,-0.8) arc (0:90:0.5 and 0.5);
	\draw[thick] (1.8,-0.8) arc (0:180:0.5 and 0.5);
	\draw[thick] (2.1,-0.8) arc (0:90:-0.5 and 0.5);

	\end{tikzpicture}\;\;,
	\qquad\qquad
 \begin{tikzpicture}
 	\draw[thick, dotted] (-0.05,0.5) arc (0:10:0 and -7.5);
 	\draw[thick, dotted] (-0.05,0.55) -- (2.65,0.55);
 	\draw[thick, dotted] (2.65,0.5) arc (0:10:0 and -7.5);
	\draw[thick, dotted] (-0.05,-0.85) -- (2.65,-0.85);
	\draw[thick] (0,0.1) arc (-90:0:0.5 and 0.4);
	\draw[thick] (0,-0.1) arc (-90:0:0.9 and 0.6);
	\draw[thick] (2.6,-0.1) arc (-90:0:-0.9 and 0.6);
	\draw[thick] (2.6,0.1) arc (-90:0:-0.5 and 0.4);
	
	\draw[thick] (0.5,-0.8) arc (0:90:0.5 and 0.5);
	\draw[thick] (1.8,-0.8) arc (0:180:0.5 and 0.5);
	\draw[thick] (2.1,-0.8) arc (0:90:-0.5 and 0.5);
	\end{tikzpicture}
\end{equation*}
\caption{Three examples of affine diagrams for $N=4$, with the left and right sides of the framing rectangle identified. The first diagram represents $e_4$, the second $e_2 e_4$, and expressing the last one is left as an exercise. 
}
\label{fig:aff-diag}
\end{figure}

The algebraic framework for this work is provided by the affine Temperley-Lieb algebra   $\ATL{N}$. A basis for this algebra is provided by particular diagrams,
called \textit{affine diagrams}, drawn on an annulus with $N$ sites on the inner and $N$ on
the outer boundary (we henceforth assume $N$ even), such that 
the sites are pairwise connected by simple curves inside the
annulus that do not cross. Some examples of affine diagrams are shown in Fig.~\ref{fig:aff-diag}; for convenience we have here cut the annulus and transformed
it into a rectangle, which we call \textit{framing}, with the sites labeled from left to right and periodic boundary conditions across.

We define a {\em through-line} as a simple curve connecting
a site on the inner and a site on the outer boundary of the
annulus. Let the number of through-lines be $2j$, and call the $2j$ sites on the inner boundary attached to a through-line {\em free} or
{\em non-contractible}. The inner (resp.\ outer) boundary of the annulus corresponds to the bottom (resp.\ top) side of the framing rectangle.

The multiplication of two affine diagrams, $a$ and $b$, is defined by joining the inner boundary of the annulus containing $a$ to the outer boundary of the annulus containing $b$, and
removing the interior sites. In other words, the product $ab$ is obtained by joining the bottom side of $a$'s framing rectangle to the top side of $b$'s framing
rectangle, and removing the corresponding joined sites. Any closed contractible loop formed in this process
is replaced by its corresponding weight~$\m$.

In abstract terms,
the algebra $\ATL{N}$ is generated by the $e_j$'s together with the identity, subject to the well-known Temperley-Lieb relations \cite{TL71}
\begin{subequations}
\label{TL}
\begin{eqnarray}
e_j^2&=&\m e_j \,, \\
e_je_{j\pm 1}e_j&=&e_j \,, \\
e_je_k&=&e_ke_j\qquad(\mbox{for } j\neq k,~k\pm 1) \,,
\end{eqnarray}
\end{subequations}
where $j=1,\ldots,N$ and the indices are interpreted modulo $N$.
In addition, $\ATL{N}$ contains the elements
 $u$ and $u^{-1}$ generating translations by one site to the right
and to the left, respectively. They obey the following additional defining
relations
\begin{subequations}
\label{TL-u}
\begin{eqnarray}
ue_ju^{-1}&=&e_{j+1} \,, \\
u^2e_{N-1}&=&e_1 \cdots e_{N-1} \,,
\end{eqnarray}
\end{subequations}
and we note that $u^{\pm N}$ is a central element.
The \textit{affine} Temperley--Lieb algebra $\ATL{N}$ is then defined abstractly as the algebra generated by the $e_i$ and $u^{\pm1}$ together with these  relations.

\subsubsection{Standard modules} \label{sec:stdmod}

It is readily checked that for any finite $N$, 
the algebra $\ATL{N}(\m)$ obeying the defining relations \eqref{TL}--\eqref{TL-u} is in fact infinite-dimensional. We wish however to focus on lattice models having a finite number of degrees of freedom per site. Their proper description involves certain finite-dimensional representations of $\ATL{N}$, the so-called {\em standard modules} $\AStTL{j}{e^{i\phi}}$, which depend on two parameters. Diagrammatically, the first parameter defines the number of
through-lines $2j$, with $j=0,1,\ldots, \frac{N}{2}$. In addition to the action of the algebra described in the previous subsection,
we now require that the result of this action be zero in the standard modules whenever the affine diagrams  obtained have a number of
through-lines strictly less than $2j$, i.e., whenever two or more free sites are contracted. Moreover, for any $j>0$ it is possible,
the algebra action can cyclically
permute the free sites. Such cyclic permutations give rise to a
{\em pseudomomentum}, which we parametrize by $\phi$ and define as follows:
Whenever $2j$ through-lines wind counterclockwise around
the annulus $l$ times, we can unwind them at the price of a factor
$e^{ijl\phi}$; and similarly, for clockwise winding, the phase is $e^{-i jl\phi}$ \cite{MartinSaleur,MartinSaleur1}. In other words, there is a phase $e^{\pm i\phi/2}$ attributed to each winding through-line.

To define the representation $\AStTL{j}{e^{i\phi}}$ in more convenient diagrammatic terms, we now make the following remark. As free sites cannot be contracted, the pairwise connections between non-free sites on the inner boundary is unchanged under the algebra action. This part of the diagrammatic information is thus irrelevant and can be omitted. Therefore, it is enough to concentrate on the upper halves of the affine diagrams, obtained by cutting a diagram into two parts across its $2j$ through-lines. Each upper half diagram is then called a {\em link state}. We still call through-lines the cut ``upper half'' through-lines attached to the free sites on the outer boundary (or, equivalently, top side of the framing rectangle). A phase $e^{i\phi/2}$ (resp.\ $e^{-i\phi/2}$) is attributed as before, namely each time one of these through-lines moves through the periodic boundary condition of the framing rectangle in the rightward (resp.\ leftward) direction. It is not difficult to see that the Temperley-Lieb algebra action obtained by stacking the affine diagrams on top of the link states produces exactly the same representations $\AStTL{j}{e^{i\phi}}$ as defined above.
 
To identify the dimensions of these modules $\AStTL{j}{e^{i\phi}}$ over $\ATL{N}(m)$ we simply need to count the link states. The result is
 \begin{equation}\label{eq:dj}
 \hat{d}_{j}=
 \binom{N}{\frac{N}{2}+j}\,
 \end{equation}
for the $j>0$ case, and we shall return to the $j=0$ case below.
Notice that these dimensions are independent of $\phi$ (although representations with
different $e^{i\phi}$ are not isomorphic).
The standard modules $\AStTL{j}{e^{i\phi}}$ are also called
{\em cell} $\ATL{N}(\m)$-modules~\cite{GL}.
 
Let us parametrize $\m=\q+\q^{-1}$. For generic values of $\q$ and $\phi$ the standard modules $\AStTL{j}{e^{i\phi}}$ are irreducible,
but degeneracies appear when the following {\em resonance criterion} is satisfied \cite{MartinSaleur1,GL}:
\footnote{In \cite{GL} a slightly different criterion is given, involving some extra liberty in the form of certain $\pm$ signs. We shall however not need these signs here.}
 \begin{eqnarray}\label{deg-st-mod}
 e^{i\phi}&=&\q^{2j+2k},\qquad
 \mbox{for } k > 0 \mbox{ integer} \,.
 \end{eqnarray}
The representation $\AStTL{j}{\q^{2j+2k}}$ then becomes reducible, and contains a submodule isomorphic to 
 $\AStTL{j+k}{\q^{2j}}$. The quotient of those two is generically irreducible, with
 dimension
 \begin{equation} \label{bard}
  \bar{d}_j := \hat{d}_j-\hat{d}_{j+k} \,, \qquad \mbox{for } j > 0 \,.
 \end{equation}
For $\q$ a root of unity, there are infinitely many solutions to \eqref{deg-st-mod}, leading to a complex pattern of degeneracies whose discussion we defer for now.

As already mentioned, the case $j=0$ is a bit different. There is no pseudomomentum in this case, but representations are still characterized by a parameter other than $j$, specifying now the weight of non-contractible loops. (For obvious topological reasons, non-contractible loops are not possible for $j > 0$.) Upon parametrizing this weight as $z+z^{-1}$, the corresponding standard module of  $\ATL{N}(\m)$ is denoted  $\AStTL{0}{z^2}$. Note that this module is isomorphic to $\AStTL{0}{z^{-2}}$. With the identification $z = e^{i\phi/2}$, the resonance criterion \eqref{deg-st-mod} still applies to the case $j=0$.

It is physically well-motivated to require that $z+z^{-1}=\m$, meaning that contractible and non-contractible loops get the same weight. Imposing this
leads to the module $\AStTL{0}{\q^2}$. Notice that this is reducible
even for generic $\q$, as \eqref{deg-st-mod} is satisfied with $j=0$, $k=1$. Therefore $\AStTL{0}{\q^2}$ contains a submodule isomorphic to
$\AStTL{1}{1}$, and taking the quotient $\AStTL{0}{\q^2}/\AStTL{1}{1}$ leads to a simple module  for generic $\q$ which we denote 
by~$\bAStTL{0}{\q^2}$. This module is isomorphic to $\bAStTL{0}{\q^{-2}}$. Its dimension is
\begin{equation}
 \bar{d}_0=\binom{N}{\frac{N}{2}}-\binom{N}{\frac{N}{2}+1} \,,
\end{equation}
which coincides with the general formula \eqref{bard} for $k=1$.

There is a geometrical significance of the difference between $\AStTL{0}{\q^2}$ and $\bAStTL{0}{\q^2}$. In the latter case, we only register which sites are connected to which in the diagrams, while in the former one also keeps information of how the connectivities wind around the periodic direction of the annulus (this ambiguity does not arise when there are through-lines propagating). The corresponding formal result is the existence of a surjection $\psi$ between different quotients of the $\ATL{N}$ algebra:
\begin{equation}\label{psi-ex}
  \xymatrix@C=8pt@R=1pt@M=-5pt@W=-2pt{
  &&	\mbox{}\quad\xrightarrow{{\mbox{}\quad\psi\quad} }\quad &\\
  & {
 \begin{tikzpicture}
 	\draw[thick, dotted] (-0.05,0.5) arc (0:10:0 and -7.5);
 	\draw[thick, dotted] (-0.05,0.55) -- (2.65,0.55);
 	\draw[thick, dotted] (2.65,0.5) arc (0:10:0 and -7.5);
	\draw[thick, dotted] (-0.05,-0.85) -- (2.65,-0.85);
	\draw[thick] (0,0.1) arc (-90:0:0.5 and 0.4);
	\draw[thick] (0,-0.1) arc (-90:0:0.9 and 0.6);
	\draw[thick] (2.6,-0.1) arc (-90:0:-0.9 and 0.6);
	\draw[thick] (2.6,0.1) arc (-90:0:-0.5 and 0.4);
	\draw[thick] (0.5,-0.8) arc (0:90:0.5 and 0.5);
	\draw[thick] (1.8,-0.8) arc (0:180:0.5 and 0.5);
	\draw[thick] (2.1,-0.8) arc (0:90:-0.5 and 0.5);
	\end{tikzpicture}
\quad}&
	& {	\quad
  \begin{tikzpicture}
 	\draw[thick, dotted] (-0.05,0.5) arc (0:10:0 and -7.5);
 	\draw[thick, dotted] (-0.05,0.55) -- (2.65,0.55);
 	\draw[thick, dotted] (2.65,0.5) arc (0:10:0 and -7.5);
	\draw[thick, dotted] (-0.05,-0.85) -- (2.65,-0.85);
	\draw[thick] (0.5,0.5) arc (-180:0:0.8 and 0.56);
	\draw[thick] (0.8,0.5) arc (-180:0:0.5 and 0.4);
	\draw[thick] (2.1,-0.8) arc (0:180:0.8 and 0.56);
	\draw[thick] (1.8,-0.8) arc (0:180:0.5 and 0.4);
	\end{tikzpicture}}
  }
\end{equation}
The previous definition of link states as the upper halves of the affine diagrams is also meaningful for $j=0$. As before, the representation $\AStTL{0}{\q^2}$ requires keeping
track of whether each pairwise connection between the sites on the outer boundary (or top side of the framing rectangle) goes through the periodic boundary condition, whereas the quotient module $\bAStTL{0}{\q^2}$ omits this information. In either case, it is easy to see that the number of link states coincides with the dimension $\hat{d}_0$ or $\bar{d}_0$, respectively.
\subsection{Physical systems and the Temperley-Lieb Hamiltonian}

Following the original work in \cite{KooSaleur} we now consider systems with Hamiltonians
\begin{equation}\label{H_phi}
\HN = -\frac{\gamma}{\pi \sin \gamma} \sum^{N}_{j=1} (e_j-e_\infty) .
\end{equation}
Here, the prefactor is chosen to ensure relativistic invariance at low energy (see the next section), and we recall that $\gamma\in [0,\pi]$ is defined through $\q=e^{i\gamma}$, so $\m\in [-2,2]$.   $e_\infty$ is a constant energy density 
 added to cancel out extensive contributions to the ground state. Its value (as discussed below) is given by
\begin{equation}\label{e_inf}
e_\infty = \sin \gamma \: I_0,
\end{equation}
with $I_0$ being given by the integral 
\begin{equation} \label{I0-def}
I_0=\int^\infty_{-\infty} \frac{\sinh(\pi-\gamma)t }{\sinh(\pi t)\cosh(\gamma t)}\mathrm{d}t.
\end{equation}
In (\ref{H_phi}), the $e_j$ can be  taken to act in different representations of the $\ATL{N}(\m)$ algebra.

We will consider in this paper the {\em XXZ representation}, in which the $e_j$ act on $\mathbb{C}^{N}$ with 
%
\begin{equation}
e_j = -\sigma_j^{-}\sigma_{j+1}^{+}-\sigma_j^{+}\sigma_{j+1}^{-}
-\frac{\cos\gamma}{2}\sigma_j^{z}\sigma_{j+1}^{z} -\frac{i\sin\gamma}{2}(\sigma_j^{z}-\sigma_{j+1}^{z})+\frac{\cos\gamma}{2} \,,
\label{TLspin}
\end{equation}
where the $\sigma_j$ are the usual Pauli matrices, so the Hamiltonian is the familiar XXZ spin chain 
\begin{equation}\label{Pauliham}
\HN = \frac{\gamma}{2 \pi \sin \gamma}\sum^{N}_{j=1}   \left[ \sigma^x_j \sigma^x_{j+1} + \sigma^y_j \sigma^y_{j+1} + \Delta (\sigma^z_j \sigma^z_{j+1} - 1 ) + 2 e_\infty \right]
\end{equation}
with anisotropy parameter
\begin{equation} \label{anisotropydelta}
 \Delta = \cos \gamma \,.
\end{equation}

In the usual basis where $\left[ \begin{smallmatrix} 1\\ 0 \end{smallmatrix} \right]$ corresponds to spin up in the $z$-direction at a given site, the Temperley-Lieb generator $e_j$ acts on spins $j, j+1$ (with periodic boundary conditions) as 
\begin{equation} e_j = \cdots \otimes\mathbf{1}\otimes
\begin{pmatrix}
0 & 0 & 0 & 0 \\
0 & \q^{-1} & -1 & 0\\
0& -1 & \q & 0\\
0 & 0 & 0 & 0 
\end{pmatrix} \otimes \mathbf{1}\otimes \cdots \,.
\end{equation}

It is also possible to introduce a twist in the spin chain without changing the expression (\ref{H_phi}), 
 by modifying the expression of the Temperley-Lieb generator acting between first and last spin with a twist parametrized by $\phi$.  
 %
In terms of the Pauli matrices, this twist imposes the boundary conditions $\sigma^z_{N+1}=\sigma^z_1$ and $\sigma^{\pm}_{N+1}=e^{\mp i \phi} \sigma^{\pm}_1$.
For technical reasons,  we will later on ``smear out'' the twist by taking $\phi/N$ for \emph{each} Temperley Lieb generator:
\begin{equation}\label{TLmatrix_spin} e_j = \cdots \otimes\mathbf{1}\otimes
\begin{pmatrix}
0 & 0 & 0 & 0 \\
0 & \q^{-1} & -e^{i\phi/N} & 0\\
0 & -e^{-i\phi/N} & \q & 0\\
0 & 0 & 0 & 0 
\end{pmatrix}\otimes \mathbf{1}\otimes \cdots \,.
\end{equation}
This is equivalent, and is done in order to preserve invariance under the usual translation operator, which will be useful in the sections below.  Note that the value of the energy density $e_\infty$ is independent of $\phi$ and remains given by (\ref{e_inf}). In the generic case, the XXZ model with magnetization $S_z=j$ and twist $e^{i\phi}$ provides a representation of the module $\AStTL{j}{e^{i\phi}}$. This is not true in the non-generic case---see below. 


Instead of the XXZ representation, one could also consider the so-called {\em loop representation}, which is simply the representation in terms of affine diagrams introduced in section \ref{sec:TL-alg-def}, or equivalently in terms of the corresponding link states.
This loop representation is useful for describing geometrical problems such as percolation or dense polymers. It is also strictly equivalent to the cluster representation familiar from the study of the $Q$-state Potts model with $Q=\m^2$ \cite{JacobsenSaleur}. Other representations are possible---such as the one involving alternating $3,\bar{3}$ representations of $sl(2|1)$ discussed in \cite{GRSV} to study percolation.

There are many common features of the XXZ and the loop representations. In particular, they have the same ground-state energy and the same ``velocity of sound'' determining the correct multiplicative normalization of the Hamiltonian in (\ref{H_phi}). This reason is that the ground state is found in the same module $\AStTL{j}{e^{i\phi}}$ for both models, or in closely related modules for which the extensive part of the ground-state energy  (and hence the constant $e_\infty$) is identical. However, the XXZ and loop representations generally involve mostly {\sl different modules}. The modules appearing in the XXZ chain depend on the twist angle $\phi$, while for the loop model the modules depend on the rules one wishes to adopt to treat non-contractible loops, or lines winding around the system. For a generic and non-degenerate situation, studying the physics in each irreducible module $\AStTL{j}{e^{i\phi}}$ would suffice to answer all questions about all $\ATL{N}(\m)$ models as well as the related Virasoro modules obtained in the scaling limit. But it turns out, importantly, that degenerate cases are always of relevance to the problems at hand. In such cases, a crucial issue that we will be interested in is how the modules ``break up'' or ``get glued''. This issue is highly model-dependent, and is central to the understanding of logarithmic CFT in particular.


The loop representation is studied in \cite{LoopPaper}, with a main focus on the continuum limit of the various standard modules and the build-up of indecomposable structures. Meanwhile, this paper will focus instead on the XXZ spin chain representation. One goal will be to again establish the continuum limit of standard modules---which will turn out different than in the loop case---and another will be to investigate more closely the nature of the convergence of Koo-Saleur generators towards the Virasoro generators. An important point is that the difference between the loop and XXZ spin chain representations is manifest already at the smallest possible finite size. To see this, we start by a detailed discussion of the module $\AStTL{0}{\q^{\pm 2}}$ at $N=2$ sites.

\subsection{Indecomposability}\label{indecomposability}



Consider the standard module $\AStTL{0}{\q^{\pm 2}}$ for $N=2$, i.e., the loop model for two sites, in the sector with no through-lines and with non-contractible loops given the same weight $\m=\q+\q^{-1}$ as contractible ones. We emphasize that since $\q$ only enters in the combination $\q+\q^{-1}$, the sign of the exponent ($\q^2$ versus $\q^{-2}$) does not matter, motivating the notation $\AStTL{0}{\q^{\pm 2}}$.

In order to illustrate the differences between the XXZ and loop representations, let us first recall from \cite{LoopPaper} how to write the two elements of the Temperley-Lieb algebra in the basis of the two link states $\loopU$ and $\loopJL$:
\begin{equation}
 e_1=\left(\begin{array}{cc}
\q+\q^{-1}&\q+\q^{-1}\\
0&0
\end{array}\right)\,, \qquad e_2=\left(\begin{array}{cc}
0&0\\
\q+\q^{-1}&\q+\q^{-1}
\end{array}\right) \,.
\end{equation}
It is apparent that $e_1(\loopU-\loopJL)=e_2(\loopU-\loopJL)=0$. Meanwhile, the action of $e_1 $ and $e_2$ on the single state $\loopII$ in $\AStTL{1}{1}$ vanishes by definition of the standard module, since the number of through-lines would decrease. By comparison we see that $\AStTL{0}{\q^{\pm 2}}$ admits a submodule, generated by $(\loopU-\loopJL)$, that is isomorphic to $\AStTL{1}{1}$. In pictorial terms we thus have 
\begin{equation}
\label{structstdmod}
\text{$\AStTL{0}{\q^{\pm 2}}$:}
\begin{tikzpicture}[auto, node distance=0.6cm, baseline=(current  bounding  box.center)]
  \node (node1)[align=center] {$\bAStTL{0}{\q^{\pm 2}}$\\$\circ$ } ;
  \node (node2) [below  = of node1,align=center]	{$\bullet$\\$\AStTL{1}{1}$};
  
  \draw[-latex] (node1) edge  (node2); 
\end{tikzpicture},
\end{equation}
where $\bullet$ denotes the submodule and $\circ$ the quotient module. The meaning of the arrow is that within the standard module $\AStTL{0}{\q^{\pm 2}}$ a state in $\AStTL{1}{1}$ can be reached from a state in $\bAStTL{0}{\q^{\pm 2}}$ through the action of the Temperley-Lieb algebra, whereas the opposite is impossible.

We next consider instead the XXZ representation with $S_z=0$ and twisted boundary conditions $e^{i\phi}=\q^{-2}$, here without ``smearing'' of the twist. 
We chose the basis of this sector as $u=|\! \uparrow\downarrow\rangle$ and $v=|\! \downarrow\uparrow\rangle$. We have then 
\begin{eqnarray}\label{stmod}
e_1=\left(\begin{array}{cc}
\q^{-1}&-1\\
-1&\q
\end{array}\right),~
e_2=\left(\begin{array}{cc}
\q &-\q^2\\
-\q^{-2}&\q^{-1}
\end{array}\right) . \end{eqnarray}
We find that $e_1(u+\q^{-1}v)=e_2(u+\q^{-1}v)=0$ while $e_1(u-\q v)=(\q+\q^{-1})(u-\q v)$ and $e_2(u-\q v)=(\q+\q^{-1})(u-\q v)+(\q^3-\q^{-1})(u+\q^{-1}v)$. Considering instead the module $\AStTL{1}{1}$, which is the spin $S_z=1$ sector with no twist and where $e_1=e_2=0$, we see that $(u+\q^{-1}v)$ generates a module isomorphic to $\AStTL{1}{1}$. Meanwhile, $u-\q v$ does not generate a submodule, since $e_2$ acting on this vector yields a component along $u+\q^{-1}v$. However, if we quotient by $u+\q^{-1}v$, we obtain a one-dimensional module where $e_1$ and $e_2$ act as $\q+\q^{-1}$, which is precisely the module $\bAStTL{0}{\q^{\pm2}}$. We thus obtain the same result as for the loop model, i.e., the structure \eqref{structstdmod} of the standard module.

Considering instead $e^{i\phi}=\q^2$, we have  
\begin{eqnarray}
\nonumber\\
~e_1=\left(\begin{array}{cc}
\q^{-1}&-1\\
-1&\q
\end{array}\right),~e_2=\left(\begin{array}{cc}
\q &-\q^{-2}\\
-\q^{2}&\q^{-1}
\end{array}\right) .
\end{eqnarray}
We see that  $e_1(u-\q v)=e_2(u-\q v)=(\q+\q^{-1})(u-\q v)$, while $e_1(u+\q^{-1}v)=0$ and $e_2(u+\q^{-1}v)=(\q-\q^{-3})(u-\q v)$. Hence this time we get a proper $\bAStTL{0}{\q^{\pm 2}}$ module, while we only get $\AStTL{1}{1}$ as a quotient module. The corresponding structure can be represented as 
\begin{equation}\label{costmod}
\text{$\coAStTL{0}{\q^{\pm 2}}$:}
\begin{tikzpicture}[auto, node distance=0.6cm, baseline=(current  bounding  box.center)]
  \node (node1)[align=center] {$\bAStTL{0}{\q^{\pm 2}}$\\$\bullet$ } ;
  \node (node2) [below  = of node1,align=center]	{$\circ$\\$ \AStTL{1}{1}$};
    \draw[latex-] (node1) edge  (node2); 
\end{tikzpicture}.
\end{equation}
Observe that the shapes  in (\ref{stmod}) and $\ref{costmod})$ are related by inverting the (unique in this case) arrows; the module in (\ref{costmod}) is referred to as  ``co-standard'', and we indicate this dual nature by placing a tilde on top of the usual $\AStTL{0}{\q^{\pm 2}}$ notation for the standard module.

To emphasize that in the XXZ chain the standard module $\AStTL{0}{\q^{\pm2}}$ corresponds to the twisted boundary condition $e^{i\phi}=\q^{-2}$, while the co-standard module $\coAStTL{0}{\q^{\pm2}}$ corresponds to the twisted boundary condition $e^{i\phi}=\q^{2}$, we introduce the notations $\AStTL{0}{\q^{-2}} \equiv \AStTL{0}{\q^{\pm2}}$ and $\AStTL{0}{\q^{2}} \equiv \coAStTL{0}{\q^{\pm2}}$. Later on, we shall write diagrams of the type above as
\begin{equation}\label{XXZ_W0q2}
\text{$\AStTL{0}{\q^{- 2}}$:}
\begin{tikzpicture}[auto, node distance=0.5 and 0.3cm, baseline=(current  bounding  box.center)]
  \node (node1) [align=center]{$[0,\q^{-2}]$};
  \node (node2) [below = of node1, align=center]{$[1,1]$};
  \draw[-latex]  (node1) edge  (node2);
\end{tikzpicture},
\hspace*{1cm}
\text{$\AStTL{0}{\q^{ 2}}$:}
\begin{tikzpicture}[auto, node distance=0.5 and 0.3cm, baseline=(current  bounding  box.center)]
  \node (node1) [align=center]{$[0,\q^2]$};
  \node (node2) [below = of node1, align=center]{$[1,1]$};
  \draw[latex-]  (node1) edge  (node2);
\end{tikzpicture},
\end{equation}
where it is implicit that any relevant quotients have been taken.

In summary, from this short exercise we see that while in the generic case the loop and spin representations are isomorphic, this equivalence  breaks down in the  non-generic case, where $\phi$ is such that the resonance criterion \eqref{deg-st-mod} is met. 
Only standard modules are encountered in the loop model \footnote{Of course, the co-standard module would be formally obtained by reversing the arrows, which corresponds formally to propagating ``towards the past'', or acting with the  transpose of the  transfer matrix to build partition and correlation functions. It is not clear what this means physically.} while in the XXZ spin chain both standard and co-standard are encountered. This feature extends to larger $N$, according to a pattern we will discuss below. We will also see what this means in the continuum limit when comparing the XXZ spin chain results to the results in the loop representation seen in \cite{LoopPaper}. We note that in the case where $\q$ is also a root of unity, the distinction between the two representations becomes even more pronounced: in this case the modules in the XXZ chain are no longer isomorphic to standard \emph{or} co-standard modules. This will be further explored in a subsequent paper \cite{fullynongenericpaper}.

\subsection{Discrete Virasoro algebra}\label{Discrete_Vir_Section}




Following \eqref{H_phi} we define the Hamiltonian density as 
${\mathcal h}_j=-\frac{\gamma}{\pi\sin\gamma}e_j$, from which we may construct a lattice momentum density ${\mathcal p}_j = i[{\mathcal h}_j,\mathcal{h}_{j-1}]=$ \phantom{ }
 $ -i \big(\frac{\gamma}{\pi\sin\gamma} \big)^2 [e_{j-1},e_j]$ by using energy conservation \cite{Vidal}. We then define the corresponding momentum operator $\PN$ as 
\begin{equation}\label{P_phi}
\PN = -i \left( \frac{\gamma}{ \pi \sin \gamma} \right)^2  \sum^{N}_{j=1} [e_j, e_{j+1} ].
\end{equation}

From the densities ${\mathcal h}_j$  and ${\mathcal p}_j$ we may build components of a discretized stress tensor as
\begin{subequations}
\begin{eqnarray}
  {\mathcal T}_j=\frac{1}{2}(\mathcal{h}_j+\mathcal{p}_j) \,, \\
  \bar{\mathcal T}_j=\frac{1}{2}(\mathcal{h}_j-\mathcal{p}_j) \,,
 \end{eqnarray}
\end{subequations}
and use those to construct discretized versions of the Virasoro generators in the form of Fourier modes \cite{Vidal}. This construction leads to the Koo-Saleur generators%
\footnote{In the present paper we consistently use calligraphic fonts for the lattice analogs of some key quantities: the Hamiltonian ${\mathcal H}$, the momentum ${\mathcal P}$---with their corresponding densities ${\mathcal h}_j$ and ${\mathcal p}_j$---, the Virasoro generators $\KSgen_n$, $\bar{\KSgen}_n$ and the stress-energy tensor ${\mathcal T}$, $\bar{\mathcal T}$. We denote the corresponding continuum quantities by Roman fonts: $H$, $P$ and $L_n$, $\bar{L}_n$, as well as $T$, $\bar{T}$. One of the paramount questions is of course whether we have the convergence $\KSgen_n, \bar{\KSgen}_n \mapsto L_n, \bar{L}_n$ in the continuum limit $N \to \infty$---and if we do, what precisely is the nature of this convergence.}

\begin{subequations}
\label{generators}
\begin{eqnarray}
\KSgen_n[N] \!\!\! &=& \!\!\! \frac{N}{4\pi} \left[ -\frac{\gamma}{\pi \sin \gamma}  \sum^{N}_{j=1} e^{inj2\pi/N} \left( e_j - e_\infty + \frac{i\gamma}{\pi \sin \gamma} [e_j, e_{j+1} ] \right)\right] + \frac{c}{24} \delta_{n,0} \,, \\
\bar{\KSgen}_n[N] \!\!\! &=& \!\!\! \frac{N}{4\pi} \left[ -\frac{\gamma}{\pi \sin \gamma}  \sum^{N}_{j=1} e^{-inj2\pi/N} \left( e_j - e_\infty - \frac{i\gamma}{\pi \sin \gamma} [e_j, e_{j+1} ] \right)\right] + \frac{c}{24} \delta_{n,0} \,.
\end{eqnarray}
\end{subequations}
first derived by other means in \cite{KooSaleur}.
The crucial additional ingredient in these formulae is the central charge, given by
\begin{equation}
c=1-{6\over x(x+1)} \,, \label{c_value}
\end{equation}
where we remind of the parametrisation \eqref{gammaparam}.
This choice (\ref{c_value}) is known to apply to models with Hamiltonian (\ref{H_phi}), such as the ferromagnetic $Q$-state Potts model with $Q=\m^2$. Note, however, that the identification (\ref{c_value}) is actually a rather subtle question, since it may be affected by boundary conditions. We discuss this aspect in details in the next section, with some further discussion in Appendix \ref{UnitaryXXZ}.

\section{Some features of the continuum limit}\label{sec:contlim}

We recall once more that throughout this paper $\q$ is assumed to take generic values (not a root of unity).
Whenever $\phi$ is such that the resonance criterion \eqref{deg-st-mod} is not met we say that $\phi$ is {\em generic};
and when \eqref{deg-st-mod} is satisfied $\phi$ is referred to as {\em non-generic}.

\subsection{Modules in the continuum}


Choosing $S_z=j$ 
 the XXZ representation for generic $\q$ and $\phi$ provides a faithful representation of the modules $\AStTL{j}{e^{i\phi}}$. The Hamiltonian acting on this module has a  CFT low-energy spectrum, encoding conformal weights $h,\bar{h}$. These weights are known from a variety of techniques like the Bethe-ansatz or  Coulomb-gas mappings, combined with extensive numerical studies \cite{PasquierSaleur,Grimm}. It is convenient to encode their values  by using the trace 
\begin{equation}
\text{Tr}\,e^{-\beta_R \HN}e^{-i\beta_I \PN} \,,
\end{equation}
where $\beta_{R}$ and $\beta_{I}$ are real, and $\beta_R>0$. Introducing the (modular) parameters
\begin{subequations}
\begin{eqnarray}
 q &=& \exp\left[-{2\pi\over N}(\beta_R + i\beta_I)\right] \,, \\
 \bar{q} &=& \exp\left[-{2\pi\over N}(\beta_R - i\beta_I)\right]
\end{eqnarray}
\end{subequations} 
and the Kac-table parametrization of conformal weights
\begin{equation}\label{h_x}
h_{r,s}={[(x+1)r-xs]^2-1\over 4x(x+1)}
\end{equation}
%
we have, in the limit where $N\to\infty$, with $\beta_R,\beta_I\to\infty$ so that  $q$ and $\bar{q}$ remain finite, 
 \begin{equation}\label{trace_eq}
\text{Tr}_{\AStTL{j}{e^{i\phi}}}\,e^{-\beta_R \HN}e^{-i\beta_I \PN}  \xrightarrow{\, N\to\infty\,}\; \FN_{j,e^{i\phi}} \,,
 \end{equation}
where \cite{GRSV}
%
 \begin{equation}\label{F-func}
 \FN_{j,e^{i\phi}} = \frac{q^{-c/24}\bar{q}^{-c/24}}{P(q)P(\bar{q})} \sum_{e \in \mathbb{Z}} q^{h_{(e-e_\phi),-j}}\bar{q}^{\,h_{(e-e_\phi),j}}
\end{equation}
and  
\begin{equation}
\displaystyle P(q) =  \prod_{n=1}^{\infty} (1 - q^n) = q^{-1/24} \eta (q) \,,
\end{equation}
where $\eta(q)$ is the Dedekind eta function. We also recall that $e_\phi = \phi/2\pi$. 


Since $\q$ is generic throughout, both $c$ and its parametrization $x$ from \eqref{c_value} takes generic, irrational values.
The conformal weights may be degenerate or not, depending on the lattice parameters. In the non-degenerate case, which corresponds to generic lattice parameters (the opposite does not always hold) it is natural to expect that the Temperley-Lieb module decomposes accordingly into a direct sum of Verma modules,
%
%
\begin{equation}\label{identcontl}
\AStTL{j}{e^{i\phi}}\mapsto \bigoplus_{e\in\mathbb{Z}} \Verma{e-e_\phi,-j}\otimes\Verma{e-e_\phi,j} \,. 
\end{equation}
%
The symbol $\mapsto$ means that action of the lattice Virasoro generators {\sl restricted to scaling states} on $\AStTL{j}{e^{i\phi}}$ corresponds to the decomposition on the right-hand side when $N\to\infty$. We will try to make this more precise below.
Note that to make notation lighter, we are not indicating explicitly that in $\Verma{}\otimes\Verma{}$ the right tensorand is for the $\overline{\Vir}$ algebra: this should always be obvious from the context.

Recall that a Verma module is a highest-weight representation of the Virasoro algebra
\begin{equation}\label{Virasoro}
[L_m,L_n]=(m-n)L_{m+n}+\frac{c}{12}m(m^2-1)\delta_{n+m,0} \,,
\end{equation}
generated by a highest-weight 
vector $|h\rangle$ satisfying $L_n|h\rangle=0,n>0$,  and for which all the descendants
\begin{equation}
 L_{-n_1}\ldots L_{-n_k}|h\rangle \,, \mbox{ with } 0<n_1\leq n_2\leq \cdots \leq n_k \mbox{ and } k>0
\end{equation}
are considered as independent, subject only to the commutation relations \eqref{Virasoro}. 
In the non-degenerate case where the Verma module is irreducible, it is the only kind of module that can occur, motivating the identification in \eqref{identcontl}. 

Meanwhile, in the degenerate cases
the conformal weights may take degenerate values $h=h_{r,s}$ with $r,s\in\mathbb{N}^*$, in which case a singular vector appears in the Verma module. By definition, a singular vector is a vector that is both a descendent and a highest-weight state. For instance, starting with $|h_{1,1}=0\rangle$ we see, by using the commutation relations \eqref{Virasoro}, that
\begin{equation}
L_1(L_{-1}|h_{1,1}\rangle)=2L_0|h_{1,1}\rangle=0 \,,
\end{equation}
while of course $L_n|h=h_{1,1}\rangle=0$ for $n>1$. Hence $L_{-1}|h=h_{1,1}\rangle$ is a singular vector. The action of the Virasoro algebra on this vector generates a sub-module. For $\q$ generic, this sub-module is irreducible, and thus we have the decomposition
\begin{equation}\label{Vermadecomp}
\text{$\Verma{1,1}^{\rm (d)}$:}
\begin{tikzpicture}[auto, node distance=0.6cm, baseline=(current  bounding  box.center)]
  \node (node1)[align=center] {$\IrrV{1,1}$\\$\circ$ } ;
  \node (node2) [below  = of node1,align=center]	{$\bullet$\\$\Verma{1,-1}$};
  
  \draw[-latex] (node1) edge  (node2); 
\end{tikzpicture},
\hspace*{1.5cm}
\end{equation}
%
where we have introduced the notation $\Verma{}^{\rm (d)}$ to denote the degenerate Verma module, and we also denote by $\IrrV{r,s}$ the irreducible Virasoro module (in this case, technically a ``Kac module''), with generating function of levels 
\begin{equation}
\label{Kac-module}
K_{r,s}=q^{h_{rs}-c/24}~{1-q^{rs}\over P(q)} \,.
\end{equation}
The subtraction of the singular vector at level $r s$ gives rise to a quotient module, and corresponds to the use of an open circle in the diagram \eqref{Vermadecomp}.

We stress that in cases of degenerate conformal weights there is more than one possible module that could appear, and the identification in \eqref{identcontl} may no longer hold. Furthermore the identification is different in the XXZ spin-chain representation of the Temperley-Lieb generators, as compared to the loop-model representation, since these are no longer isomorphic.
In later sections we will discuss which identifications hold for the XXZ spin chain representation. For this purpose let us introduce the notation $\FF{r,s}^{\rm (d)}$ for the dual of the (degenerate) Verma modules, or ``co-Verma'' modules. As an example, the dual of \eqref{Vermadecomp} is
\begin{equation}\label{Vermadecompdual}
\text{$\FF{1,1}^{\rm (d)}$:}
\begin{tikzpicture}[auto, node distance=0.6cm, baseline=(current  bounding  box.center)]
  \node (node1)[align=center] {$\IrrV{1,1}$\\$\bullet$ } ;
  \node (node2) [below  = of node1,align=center]	{$\circ$\\$\Verma{1,-1}$};
  
  \draw[-latex] (node2) edge  (node1); 
\end{tikzpicture}
\hspace*{1.5cm}
\end{equation}



\subsection{Bosonization and expected results}\label{Bosonization}

Many algebraic  aspects of the continuum limit of Temperley-Lieb based models can be understood using bosonization of the underlying XXZ spin chain and its relation with the  free-field (Dotsenko-Fateev) description \cite{DotsenkoFateev} of $c<1$ CFTs.  We start with some basic results here, concerning in particular the identification of the stress-energy tensor and its ``twisted'' version, the role of vertex operators, and the nature of corrections to scaling. These  results will be useful later to formulate conjectures about the continuum limit of modules. 

The free field (FF) representation starts with  a pair of (chiral and anti-chiral) bosonic fields $\varphi,\bar{\varphi}$ with the stress-energy tensors
\begin{subequations} \label{T-FF}
\begin{eqnarray}
T_{\rm FF} &=& -{1\over 4} \normord{(\partial\varphi)^2} \,, \\
\bar{T}_{\rm FF} &=& -{1\over 4} \normord{(\bar{\partial}\bar{\varphi})^2} \,,
\end{eqnarray}
\end{subequations}
where $\normord{-}$ denotes normal order. The Hamiltonian is  
%
\begin{equation}
\Hcont_{\rm FF} =-\int {d\upsigma\over 2\pi} (T_{\rm FF}+\bar{T}_{\rm FF})\label{Hfb} \,,
\end{equation}
and the propagators are
\begin{subequations}
\begin{eqnarray}
\langle \varphi(z)\varphi(z')\rangle &=& -2\ln (z-z') \,, \\
\langle \bar{\varphi}(\bar{z})\bar{\varphi}(\bar{z}')\rangle &=& -2\ln (\bar{z}-\bar{z'}) \,.
\end{eqnarray}
\end{subequations}
Here $z=\upsigma+i\uptau$, where $\upsigma$ is the space coordinate and $\uptau$ the imaginary time coordinate. 

To further analyse this free-field problem we define the vertex operators
%
\begin{equation} \label{FFvertexop}
V_{\alpha(e,m),\bar{\alpha}(e,m)} = \;\; \normord{ \exp\left(i{e\over 2}\alpha_+\Phi-i{m\over 2}\alpha_-\Theta\right) } \,,
\end{equation}
expressed here in terms of the non-chiral components
\begin{subequations}
\begin{eqnarray}
\Phi &\equiv& \varphi+\bar{\varphi} \,, \\
\Theta &\equiv& \varphi-\bar{\varphi} \,.
\end{eqnarray}
\end{subequations}
The integers $e,m\in\mathbb{Z}$ can be interpreted as electric and magnetic charges in the Coulomb gas formalism \cite{DFSZ,CGreview}, and in terms of those
we have
\begin{subequations}
\label{alpha-em}
\begin{eqnarray}
\alpha(e,m) &=& {1\over2}\left(e\alpha_+-m\alpha_-\right) \,, \\
\bar{\alpha}(e,m) &=& {1\over2}\left(e\alpha_++m\alpha_-\right) \,,
\end{eqnarray}
\end{subequations}
where $\alpha_\pm$ are coupling constants related with the compactification radius of the boson. The conformal weights of the vertex operators \eqref{FFvertexop} are then
\begin{subequations}
\begin{eqnarray}
 h_{\rm FF} &=& [\alpha(e,m)]^2 \,, \\
 \bar{h}_{\rm FF} &=& [\bar{\alpha}(e,m)]^2 \,.
\end{eqnarray}
\end{subequations}

We consider specifically  low-energy excitations over the ground state of the antiferromagnetic Hamiltonian (\ref{Pauliham}), which are described by (\ref{Hfb}), with 
\begin{subequations}
\begin{eqnarray}
\alpha_+ &\equiv& \sqrt{x+1\over x} \,, \\
\alpha_- &\equiv& -\sqrt{x\over x+1}
\end{eqnarray}
\end{subequations}
in the parametrization \eqref{gammaparam}. 

%
%
%
Defining
\begin{equation} \label{eXXZgens}
\eXXZ_j = -\sigma_j^{-}\sigma_{j+1}^{+}-\sigma_j^{+}\sigma_{j+1}^{-}
-\frac{\cos\gamma}{2}\sigma_j^{z}\sigma_{j+1}^{z} +\frac{\cos\gamma}{2}
\end{equation}
we note that we can equivalently write the Hamiltonian (\ref{Pauliham}) as 
\begin{equation}
\label{Ham-3.2}
{\mathcal H}=-{\gamma\over\pi\sin\gamma}\sum_{j=1}^N (\eXXZ_j-\eXXZ_\infty) \,,
\end{equation}
where $\eXXZ_\infty=e_\infty$.
Let us now consider the scaling limit of each individual term. We use the basic formulae from the literature  (see, e.g., \cite{Lukyanov})
\begin{subequations}\label{spin_to_boson}
\begin{eqnarray}
\sigma_j^z&=&a{\alpha_+\over 2\pi }{{\rm d} \Phi\over {\rm d} \upsigma}+(-1)^j C_1^z a^{d_{1,0}}\sin{\alpha_+\Phi\over 2}(\upsigma)+\ldots \,, \\
\sigma_j^\pm&=&\exp\left(\pm {i\alpha_-\over 2}\Theta\right)(\upsigma)\left[a^{d_{0,1}} C_0^\pm+a^{d_{1,1}}C_1^\pm(-1)^j \cos{\alpha_+\Phi\over 2}(\upsigma)+\ldots\right] \,,
\end{eqnarray}
\end{subequations}
where $a$ is the cutoff (lattice spacing) and the physical coordinate $\upsigma = j a$.
The $C_1^z,C_0^\pm,C_1^\pm$ are (known) constants that depend only on $x$. The numbers $d_{e,m}$ are the physical dimensions of the operators \eqref{FFvertexop}, namely $d_{e,m} \equiv h_{\rm FF}+\bar{h}_{\rm FF}$. 

Using these formulae, one can write a similar expansion for the elementary Hamiltonians  \cite{Lukyanov}
\begin{equation}
\eXXZ_j-\eXXZ_\infty=a^2 {\sin\gamma\over 2\gamma}(T_{\rm FF}+\bar{T}_{\rm FF})(\upsigma)+C_1 (-1)^j a^{d_{1,0}} \cos {\alpha_+\Phi\over 2}(\upsigma)+\ldots \,. \label{hjbos}
\end{equation}
The important quantity is the dimension
\begin{equation}
d_{1,0}={\alpha_+^2\over 2}={x+1\over 2x} \,.
\end{equation}
In the regime we are interested in, $\gamma\in [0,\pi]$, whence $x\in[0,\infty)$. The leading contribution to (\ref{hjbos}) comes from the first term only when $d_{1,0}>2$, that is $1-{\gamma\over \pi}<{1\over 4}$, or $\gamma\in ]{3\over 4},\pi]$. Equivalently, the anisotropy parameter $\Delta\in [-1,-{\sqrt{2}\over 2}]$ from \eqref{anisotropydelta}, or the conformal parameter $x<{1\over 3}$ from \eqref{gammaparam}. For values inside this interval, we can thus safely write, as $a\to 0$:
\begin{equation}
-{\gamma\over a^2\pi\sin\gamma}(\eXXZ_j-\eXXZ_\infty)\approx -{1\over 2\pi}(T_{\rm FF}+\bar{T}_{\rm FF}) \,.
\end{equation}
Outside this interval---i.e. for $x>{1\over 3}$ (including $x$ integer)---the second term dominates. A very important fact however is that the second term comes with a $(-1)^j$ alternating prefactor, i.e., it only contributes to excitations at lattice momentum near $\pi$. As a result, for all Virasoro generators $L_n$ at finite $n$---and thus at momentum of order $1/N$---the alternating term is effectively scaling with a dimension $d_{1,0}+2$.
For instance, for $L_0$ we have
\begin{equation}
\eXXZ_{j-1}+\eXXZ_{j+1}+2\eXXZ_j=4a^2 {\sin\gamma\over 2\gamma}(T_{\rm FF}+\bar{T}_{\rm FF})(\upsigma)+C_1 (-1)^j a^{d_{1,0}+2} {{\rm d}^2\over {\rm d}\upsigma^2}\cos {\alpha_+\Phi\over 2}(\upsigma)+\ldots \,,
\end{equation}
and we see that all corrections are irrelevant. 

The same analysis can now be carried out  for the Temperley-Lieb generators $e_j$. Comparing \eqref{eXXZgens} and \eqref{TLspin} we see that
\begin{equation}
e_j=\eXXZ_j-i{\sin\gamma\over 2}\left(\sigma_j^z-\sigma_{j+1}^z\right) \,,
\end{equation}
for which we get the continuum limit
\begin{equation}
-{\gamma\over a^2\pi\sin\gamma} (e_j-e_\infty)=-{1\over 2\pi}(T_{\rm FF}+\bar{T}_{\rm FF})(\upsigma) +i{\gamma\over (2\pi)^2} \alpha_+{{\rm d}^2\Phi\over {\rm d} \upsigma^2} \,.
\end{equation}
Since in the models we study, the Temperley-Lieb generators are the fundamental Hamiltonian densities, we must interpret the right-hand side of this equation
as a modified or ``improved'' stress-energy tensor, which is the sum of its free field analog $T_{\rm FF} + \bar{T}_{\rm FF}$ and a new ``deformation'' term.
In terms of the parameter $x$, the ``twist term'' multipliying the  second derivative of $\Phi$ is 
\begin{equation}
{1\over 2\pi}\times {1\over 2(x+1)}\sqrt{x+1\over x}\equiv{1\over 2\pi} \alpha_0 \,,
\end{equation}
where we have introduced the so-called {\em background charge} $\alpha_0$, defined as in \eqref{alphazero},
familiar from the Coulomb-gas analysis \cite{DotsenkoFateev}.
Using that $\partial^2_\upsigma \Phi=\partial^2 \varphi+(\bar{\partial})^2\bar{\varphi}$, we  finally get the expressions for the modified stress-energy tensor
\begin{subequations}
\begin{eqnarray}
T &=& -{1\over 4} \normord{(\partial\varphi)^2 } + \, i\alpha_0\partial^2\varphi \,, \\
\bar{T} &=& -{1\over 4} \normord{ (\bar{\partial}\bar{\varphi})^2 } + \, i\alpha_0\bar{\partial}^2\bar{\varphi} \,.
\end{eqnarray}
\end{subequations}
This is the well known  ``twisted'' stress-energy tensor studied in \cite{FeiginFuchs,FeiginFuchs2,DotsenkoFateev}. 
It is this modified stress-energy tensor---rather than the free-field version $T_{\rm FF}, \bar{T}_{\rm FF}$ of \eqref{T-FF}---that is relevant for a lattice discretization based on the Temperley-Lieb algebra. We shall henceforth consider $T, \bar{T}$ throughout almost all the paper, except in Appendix~\ref{UnitaryXXZ} where we shall give an alternative construction based on the untwisted stress-energy tensor.

With respect to this stress-energy tensor, the vertex operators $V_{\alpha,\bar{\alpha}}$ get the modified conformal weights 
\begin{subequations}\label{twisted_weights}
\begin{eqnarray}
 h &=& \alpha^2-2\alpha\alpha_0 \,, \\
 \bar{h} &=& \bar{\alpha}^2-2\bar{\alpha}\alpha_0 \,,
\end{eqnarray}
\end{subequations}
to be compared to the previous free-field expressions \eqref{alpha-em}.

%
%
The eigenvalues of $\Hlatt$ and $\Platt$  only allow one to determine the conformal weights, not the value of the charges. For a given conformal weight, two values are possible in general, $\alpha$ and $2\alpha_0-\alpha$. Recalling the notation in \eqref{alpha_rs} we now state the result, first shown in \cite{KooSaleur}, which we will justify in detail below:

\vspace*{5pt}
\noindent\fbox{
\hspace*{5pt}\begin{minipage}{\linewidth-20pt}\em
\vspace*{5pt}
In the XXZ spin chain with twisted boundary conditions parametrized by $\phi=2\pi e_\phi$, the scaling states in the sector of magnetization $S_z$ correspond in the scaling limit to primary states $V_{\alpha,\bar{\alpha}}$ with charges on the form
\begin{subequations}\label{alpha}
\begin{eqnarray}
\alpha & = \quad \alpha_{-(e-e_\phi),S_z\phantom{-}}\quad = & \frac{1}{2}(e -  e_\phi )  \alpha_{+}  + \alpha_0 - \frac{1}{2}S_z \alpha_{-}  \\
\bar{\alpha} & = \quad \alpha_{-(e-e_\phi),-S_z} \quad = & \frac{1}{2}  (e - e_\phi ) \alpha_{+}  + \alpha_0 + \frac{1}{2}S_z \alpha_{-} ,
\end{eqnarray}
\end{subequations}
where $e$ is an integer, and their descendants. The conformal weights are given by \eqref{twisted_weights}.

\vspace*{5pt}
\end{minipage}\hspace*{5pt}}
\vspace*{5pt}
~\\
The precise correspondence, including the proper identification of the integers $e$, will be discussed in Section \ref{sec:Bethe}, as well as the exact meaning of the words {\em scaling states} and {\em scaling limit}.

\medskip

We conclude this section by some remarks about corrections to scaling. 
The  leading corrections to $e_j$ look a bit different from (\ref{hjbos}).   We have 
\begin{equation} 
(e_j-e_\infty)=a^2{\sin\gamma\over 2\gamma}(T+\bar{T})(\upsigma)-i\alpha_+ a^2{\sin\gamma\over 4\pi}{{\rm d}^2\Phi\over {\rm d} \upsigma^2}+(-1)^j a^{d_{1,0}}\left(C_1^z\sin{\alpha_+\Phi\over 2}(\upsigma)+2iC_1\cos{\alpha_+\Phi\over 2}(\upsigma)\right)+\ldots \,. \label{combination}
\end{equation}
Meanwhile, it is expected that the leading correction 
should now be given by the operator $\Phi_{2,1}$ with conformal weights $h=\bar{h}=h_{2,1}={x+3\over 4x}$ in the twisted theory. Under twisting, we expect in general the field
\begin{equation}
\normord{\exp(ni\alpha\Phi)} \hbox{ with }~\alpha={\alpha_+\over 2}
\end{equation}
to get the weight
\begin{equation}
h=\alpha^2-2\alpha\alpha_0=h_{1-n,1} \,.
\end{equation}
This means the term $\exp(i\alpha_+\Phi/2)$ should disappear from the combination in (\ref{combination}), leading to a relationship between the two constants 
\begin{equation}
C_1^z=2C_1 \,. \label{desired}
\end{equation}
While \eqref{desired} can be checked to hold in some cases using results in  \cite{Sergei99,Furusaki}, we are not aware of a general proof: more investigation of this question  would be very interesting, but it outside the scope of this paper.

\subsection{The choices of metric}

We have just recovered the well-known result that  the continuum limit of the XXZ spin chain  is made up of sectors of a twisted free-boson theory. The space of states in the continuum limit can be built out of vertex-operator states $V_\alpha = \normord{ e^{i\alpha\varphi} }$ (in this section we shall only consider the chiral part for notational brevity) and derivatives $\partial\varphi, \partial^2\varphi,(\partial\varphi)^2,\ldots$. 

From the free-boson current $J(z)=\frac{1}{4}\partial_z \varphi$, we define $a_n$ as its modes, such that we have the Heisenberg algebra

\begin{equation}\label{Heisenberg}
[a_n,a_m]=n\delta_{n+m,0}.
\end{equation}
We can then equivalently consider the state space to be built from states of the form
\begin{equation}\label{Fock_a_n}
 (a_{-n_1})^{N_1} \cdots (a_{-n_k})^{N_k}  V_\alpha \,.
\end{equation}
The vertex operator $V_\alpha$ is a highest-weight state for the Heisenberg algebra, with 
\begin{subequations}\label{Heis_weight}
\begin{eqnarray}
a_n V_\alpha &=& 0 \,, \quad \forall n> 0 \,, \\
a_0 V_\alpha &=& \sqrt{2}(\alpha-\alpha_0)V_\alpha \,.
\end{eqnarray}
\end{subequations}
In terms of $a_n$ we have for the twisted boson theory
\begin{subequations}\label{Vir_gens}
\begin{eqnarray}
L_n &=& \sum_{k=0}^\infty a_{n-k}a_k -\sqrt{2}\alpha_0 n a_n \,, \quad \mbox{for } n\neq 0 \,, \\
L_0 &=& \sum_{k=1}^\infty a_{-k}a_k + \frac{1}{2}a_0^2 -  \alpha_0^2
\end{eqnarray}
\end{subequations}
for which the 
Virasoro algebra relations \eqref{Virasoro} are readily shown to be satisfied with
\begin{equation}
 c=1-24\alpha_0^2 \,.
\end{equation}

Two possible scalar products can be introduced in the CFT.  The one for which 
$a_n^\dagger = a_{-n}$, denoted in what follows by $\spininner{-}{-}$, is positive definite and corresponds to the  usual positive definite scalar product for the spin chain, where the  conjugation sending a  ket to a bra is anti-linear.  A crucial observation is that for this scalar product  $L^\dagger_n \neq L_{-n}$.  This means that norm squares of descendants cannot be obtained using  Virasoro algebra commutation relations. Also, this scalar product must be used with great care when calculating correlation functions; this point is discussed more in appendix \ref{RemScalProd}. 
Instead of using the Virasoro relations directly we shall use the Heisenberg relations \eqref{Heisenberg}.
 
We shall in the following sections, especially when comparing to the numerical results, refer to the \emph{conjectured} values of various norms. What we refer to is then the value we obtain by considering the states to be given as $V_\alpha$ with $\alpha$ as in \eqref{alpha}, writing any Virasoro generator in terms of $a_n$ using \eqref{Vir_gens}, using the Heisenberg commutation relations \eqref{Heisenberg} to move $a_n$ with $n>0$ to the right and finally applying the highest-weight relation \eqref{Heis_weight}. 

The second scalar product is denoted $\loopinner{-}{-}$ and corresponds to the conjugation $\loopdagger$. Compared to the conjugation $\spindagger$, where we write $L_n$ in terms of $a_n$ as in \eqref{Vir_gens} and use $a_n^\dag = a_{-n}$ to define the conjugate $L_n^\spindagger$, we instead define the conjugate $L_n^\loopdagger$ as simply $L^\loopdagger_n = L_{-n}$. This ``conformal scalar product'' $\loopinner{-}{-}$ 
is known to correspond \cite{DJS,VJS,VGJS}, on the lattice, to the ``loop scalar product'' defined through the Markov trace,
or to a modified scalar product in the XXZ spin chain where $\q$ is treated as a formal, self-conjugate parameter \cite{CJS}.
It is not a positive definite scalar product, and we will not use it much here, as our main goal is to establish whether various quantities are zero or not. 

Of course, the relationship between the two scalar products is a question of great interest: for some recent results about this, see \cite{Gleb}.

\subsection{Feigin-Fuchs modules and conjugate states}\label{FF_modules}


When the lattice parameters are such that the corresponding twisted free boson only involves non-degenerate cases,
the Verma modules are irreducible and coincide with the Fock spaces of the bosonic theory. We now consider what happens in the degenerate case.
As a module over the Heisenberg algebra, the Fock space is irreducible. If we instead wish to consider it as a module over the Virasoro algebra, it will be a Feigin-Fuchs module, which is only irreducible (and then, a Verma module) if $\alpha \neq \alpha_{r,s}$ for any $r,s\in\mathbb{N}^*$, with $\alpha_{r,s}$ defined as in \eqref{alpha_rs}.

To see how Feigin-Fuchs modules differ from Verma modules, it is helpful to introduce the notion of conjugate states: we call states with the same conformal weight $h$ but different charges $\alpha$ conjugates of each other. From \eqref{twisted_weights} we see that a state $V_\alpha$ with charge $\alpha$ has a conjugate state $V_{\alpha^{\rm c}}$  with charge $\alpha^{\rm c} $ defined in \eqref{conjugate_charge}. Note that this conjugation is an involution: $(\alpha^{\rm c})^{\rm c}=\alpha$.

While the conformal weights of a pair of conjugate states are the same, we shall see that their behaviour under the action of the Virasoro algebra is in a sense dual. We illustrate the precise meaning of this statement with the case of the identity state, which has $\alpha=0$, and its conjugate state with $\alpha=2\alpha_0$.
We obtain   $L_{-1} V_\alpha = \partial_z V_\alpha = \sqrt{2} \alpha a_{-1} V_\alpha $, which is zero for $\alpha=0.$
Conversely, the action of $L_1$ on $a_{-1}V_\alpha$ (recall our state space \eqref{Fock_a_n}), yields $L_1 a_{-1} V_{\alpha} = \sqrt{2}(\alpha-2\alpha_0)V_{\alpha}$, which is instead zero for the conjugate charge $\alpha^{\rm c} = 2\alpha_0$.
We thus obtain the following diagrams, where crossed out arrows indicate that the state at the end of the arrow would have zero norm: 
\begin{equation}\label{diagram}
\begin{tikzpicture}[auto, node distance=0.6 and 3cm], baseline=(current  bounding  box.center)]
  \node (node1) [align=center]{$\mathbf{1}$\\ \vspace*{-8pt}};
  \node (node2) [below = of node1, align=center]	{ \vspace*{-8pt} \\$a_{-1}\mathbf{1}$\\ \vspace*{-8pt}  \\ $h=1$};
 %
  \draw[-latex, transform canvas={xshift=-1.5mm}]  (node1) edge  node[left,xshift=-1.5mm]{$L_{-1}$}  (node2); 
 \node(b) [xshift=-1.5mm,red] at ($(node1)!0.4!(node2)$) {$\times$};  
  \draw[latex-, transform canvas={xshift=1.5mm}]  (node1) edge node[right,xshift=1.5mm]  {$L_{1}$} (node2); 
  \node (node3)[right= of node1, align=center] {$V_{2\alpha_0}$\\ \vspace*{-8pt}};
  \node (node4) [below = of node3, align=center]	{\vspace*{-8pt}\\$a_{-1} V_{2\alpha_0}$\\ \vspace*{-8pt}\\ $h=1$};
   \draw[-latex, transform canvas={xshift=-1.5mm}]  (node3) edge node[left,xshift=-1.5mm]  {$L_{-1}$} (node4);  
   \draw[latex-, transform canvas={xshift=1.5mm}]  (node3) edge node[right,xshift=1.5mm]  {$L_{1}$} (node4);  
 \node(c) [xshift=1.5mm,red] at ($(node3)!0.45!(node4)$) {$\times$};     
\end{tikzpicture}.
\end{equation}

We shall later on refrain from writing out such crossed-out arrows at all, using the same type of notation as already seen above for the standard modules. Since we restrict to the case where $\q$ is not a root of unity, there are no other degeneracies in the modules, and we always 
get one of the two following diagrams:
\begin{equation}
\begin{tikzpicture}[auto, node distance=0.4 and 0.3cm, baseline=(current  bounding  box.center)]
  \node (node1) [align=center]{$\bullet$};
  \node (node2) [below = of node1, align=center]{$\circ$};
  \draw[latex-]  (node1) edge  (node2);
\end{tikzpicture}
\hspace*{1cm}
\begin{tikzpicture}[auto, node distance=0.4 and 0.3cm, baseline=(current  bounding  box.center)]
  \node (node1) [align=center]{$\circ$};
  \node (node2) [below = of node1, align=center]{$\bullet$};
  \draw[-latex]  (node1) edge  (node2);
\end{tikzpicture}.
\end{equation}
We note that these can be seen as a co-Verma module and a Verma module. More details will be given  in Section \ref{PartlyNonGeneric}.

\section{Bethe Ansatz picture}\label{sec:Bethe}


In the context of the XXZ spin chain the Bethe ansatz is a  well-adapted tool to carry out the analysis \cite{Alcaraz}.
We present the general picture, and refer the reader to Appendix \ref{FormFactorAppendix} for more detail.

\subsection{Bethe-ansatz and the identification of scaling states}\label{Bethe_scaling}

When  $\q$ not a root of unity, and the resonance criterion \eqref{deg-st-mod} is not satisfied, the XXZ Hamiltonian can be fully diagonalized using a basis of orthonormal Bethe states (see \cite{QQ,QQ1} and references therein). The corresponding Bethe equations are of the form
\begin{equation}\label{bethe_eqs_log}
N\lambda_j=2\pi I_j + \phi - \sum_{k\neq j} \Theta_{\text{XXZ}}(\lambda_j,\lambda_k),
\end{equation}
for $j=1,2,\ldots,\frac{N}{2}-S_z$, and are obtained (after some rewriting) by taking the logarithm of \eqref{BetheEqs} as given in Appendix \ref{FormFactorAppendix}; this defines the scattering kernel $\Theta_{\text{XXZ}}(\lambda_j,\lambda_k)$. 
The Bethe integers $I_j$ corresponding to a given solution shall play an important role in the discussion below. (Note: they are sometimes half-integers, despite their name.) The states have energies
\begin{equation}\label{bethe_energy}
\mathcal{E}(\{I_j\})=-\frac{\gamma}{\pi\sin\gamma}
 \left[2\sum_j ( \cos \gamma + \cos k_j  )   -N e_\infty \right],
\end{equation}
where $k_j$ is related to the Bethe root $\lambda_j$ by $\tan( \gamma/2 )\tan(k_j/2) = \tanh(\lambda_j)$,
%
and momenta 
\begin{equation}\label{bethe_momentum}
\Platt(\{I_j\})=-\frac{2\pi}{N}\sum_{j}I_j + \frac{\phi}{N}\left(\frac{N}{2}-S_z\right).
\end{equation}
For future convenience we also define the rescaled lattice momentum $\Pscaled$ as
\begin{equation}\label{rescaled_momentum}
\Pscaled = \frac{N}{2\pi}\Platt
\end{equation}
such that when $\phi=0$, $\Pscaled$ takes integer values $0,1,\hdots,N-1$. 

The ground state---i.e., the state of lowest energy---depends on the value of $\phi$. It will be convenient in what follows to identify states by their corresponding set of Bethe integers, where the ground state is given by the symmetric, maximally packed set of integers \cite{Alcaraz}. The boundaries of this set of integers are referred to as ``edges'' (by analogy with the Fermi edge in solid state physics).


The third component $S_z$ of the spin is conserved by the Hamiltonian \eqref{Pauliham}, and we can split the problem of diagonalizing ${\mathcal H}$ into subsectors of fixed $S_z$. Within a given subsector, we identify states using the difference between their set of Bethe integers and the one of the lowest-energy state within the same subsector. As $N$ increases, we focus only on {\sl scaling states}, that is,  states for which this difference {\sl measured from the edge} remains fixed and finite. Representing  the set of integers by filled circles, the edge simply refers to the boundary between filled and empty circles in the ground state configuration, as shown here marked by $\dashline$ for $N=10$:
\begin{equation*}
  [-2,-1,0,1,2]  \quad \longleftrightarrow \quad \hdots\circ\circ\circ \hspace*{-2pt}\dashline\hspace*{-2pt}   \bullet\bullet\bullet\bullet\bullet \hspace*{-2pt}\dashline\hspace*{-2pt}  \circ\circ\circ\hdots 
\end{equation*}
For a scaling state there can only be finitely many empty circles between the edges and only finitely many filled circles outside of the edges, and both must occur only at a finite distance from one of the edges.  Examples of scaling states are provided by  the ``electric excitations'' discussed below, where the set of integers from the ground state is shifted by a finite amount $e$.

Of course, the ground states of every finite-$S_z$ sector are scaling states, since their integers coincide with those of the ground state but for $S_z$ of them.\footnote{Strictly speaking, changing the number of integers by an odd number means altering between $I_j$ being integers or half-integers. We expand our definition of scaling states to take this into account.}  A non-scaling state would be, in contrast, a state whose magnetization increases with $N$, for instance the ``ferromagnetic ground state'' will all spins up, $S_z={N\over 2}$. Another example of a non-scaling state is obtained if we make a hole for some finite integer which remains fixed as $N\to\infty$. In this case, the difference from the ground state configuration, measured from the edge, increases linearly with $N$.

We now wish to give a brief motivation for the conjecture \eqref{alpha} given above. To this purpose we first recall the Coulomb gas (CG) picture, where we consider a free field compactified on a circle (see section \ref{Bosonization}). Our fundamental operators are vertex operators (exponentials of the field) and their duals (discontinuities in the field), with conformal weights parametrized by integers $e,m$ called the electric and magnetic charges. The  conformal weights corresponding to these electromagnetic excitations are shown in \eqref{h_x}, with $-e,m$ corresponding to $r,s$ as in \eqref{alpha}. 


In the context of the spin chain we can make a purely magnetic excitation by taking the lowest-energy state within a sector of non-zero total magnetization $S_z$. This means adding to or subtracting from the number of Bethe integers, while still keeping them symmetrical and maximally packed. An electric excitation can then be created within any sector of $S_z$ by shifting all Bethe integers $e$ steps from the symmetric configuration. Examples of such primary states are shown here. We mark the middle of each row of filled circles with a bar, which will intersect the middle circle if their number is odd. This will be helpful for the discussion of descendant states below.
\begin{equation*}
\begin{split}
\hdots\circ\circ\circ\bullet\bullet\bullet \hspace*{-4pt}|\hspace*{-2pt}\ \bullet\bullet\circ\circ\circ\hdots  
\quad  & \leftrightarrow \quad  \text{ground state} \\
\hdots\circ\circ\circ\bullet\bullet \hspace*{-2pt}|\hspace*{-2pt}\bullet\bullet\circ\circ\circ\hdots\hspace*{3.5pt}  \quad  & \leftrightarrow \quad \text{a magnetic excitation ($m=1$)} \\
\hdots\circ\circ\circ\circ\bullet\bullet\bullet\hspace*{-6pt}|\hspace*{-0pt}\bullet\bullet\circ\circ  \hdots \quad  & \leftrightarrow \quad \text{an electric excitation ($e=1$)} \\
\hdots\circ\circ\bullet\bullet\bullet\hspace*{-6pt}|\hspace*{-0pt}\bullet\bullet\circ\circ\circ\circ  \hdots \quad  & \leftrightarrow \quad \text{an electric excitation ($e=-1$)} \\
\hdots\circ\circ\circ\circ\bullet\bullet\hspace*{-0pt}|\hspace*{-2pt}\bullet\bullet\circ\circ\hdots\hspace*{3.5pt}  \quad  & \leftrightarrow \quad \text{an electromagnetic excitation $(e=1, m=1$)} \\
\hdots\circ\circ\bullet\bullet\hspace*{-0pt}|\hspace*{-2pt}\bullet\bullet\circ\circ\circ\circ\hdots\hspace*{3.5pt}  \quad  & \leftrightarrow \quad \text{an electromagnetic excitation $(e=-1, m=1$)} \\
\end{split}
\end{equation*}

Note that the magnetic excitation changes the number of filled circles; this corresponds to alternating between having a set of integers or a set of half-integers.

From \eqref{bethe_momentum} we see that an electric excitation corresponds to a change in momentum. Here we also 
 see that if we introduce twisted boundary conditions, $e_\phi=\phi/2\pi$ enters on the same footing as $e$. This mirrors how in the CG picture a twist can be implemented by inserting electric charges at infinity.\footnote{Note: \emph{numerically}, our momentum only depends on $e$, since we have smeared out the twist. $e_\phi$ instead modifies the Hamiltonian itself.}  With the identification $e\leftrightarrow -( e-e_\phi)$ and $m \leftrightarrow  S_z$ we claim that the scaling states corresponding to electromagnetic excitations in the spin chain can be written in the CG picture as vertex operators
$V_{\alpha,\bar{\alpha}}=\normord{e^{i\alpha\varphi+ i\bar{\alpha}\bar{\varphi}}}$ of charges $\alpha,\bar{\alpha}$ as in \eqref{alpha}. Indeed, we see that with these charges we reproduce the weights \eqref{h_x} for CG electromagnetic excitations.

From any primary state obtained in this fashion, we must make further excitations to reach its descendants. This is done by ``creating holes'' through shifting Bethe integers at the left edge (chiral excitations) or right edge (anti-chiral excitations) of the set. Knowing that the lattice momentum must shift in accordance with the change in conformal spin, we can easily read off which level we reach. If the level has more than one state, we obtain an orthonormal basis for these states.
To better see which excitations are chiral and which are anti-chiral we compare to the bar inserted in the middle of the filled circles. We find the chiral level by first counting, for each filled circle on the left side of the bar, the number of empty circles separating it from the bar, and then adding up these numbers. We find the anti-chiral level in the same manner by considering the right side of the bar. Some examples of descendants:
%
\begin{equation*}
\begin{split}
\hdots\circ\circ\bullet\circ\bullet\bullet \hspace*{-0pt}|\hspace*{-2pt} \bullet\bullet\bullet\circ\circ\circ\hdots\quad &\leftrightarrow \quad \text{chiral level 1}\\
\hdots\circ\circ\circ\bullet\bullet\bullet\hspace*{-0pt}|\hspace*{-2pt} \bullet\bullet\circ\bullet\circ\circ\hdots\quad &\leftrightarrow \quad \text{anti-chiral level 1}\\
\hdots\circ\circ\bullet\circ\bullet\bullet\hspace*{-0pt}|\hspace*{-2pt} \bullet\bullet\circ\bullet\circ\circ\hdots\quad &\leftrightarrow \quad \text{chiral and anti-chiral level 1}\\
    \left. \begin{aligned}
        \hdots\circ\circ\bullet\bullet\circ\bullet\hspace*{-0pt}|\hspace*{-2pt} \bullet\bullet\bullet\circ\circ\circ\hdots\\
        \hdots\circ\bullet\circ\circ\bullet\bullet\hspace*{-0pt}|\hspace*{-2pt} \bullet\bullet\bullet\circ\circ\circ\hdots\\
        \end{aligned}\right\} \hspace*{2pt} &\leftrightarrow \quad \text{chiral level 2}\\
\end{split}
\end{equation*}


In practice, the Bethe integers may come in configurations where some integers coincide. 
We find (via the methods discussed in Appendix \ref{BetheRootAppendix}) the sets of Bethe integers for all relevant scaling states based on the sets listed in \cite{Olga}, in which the possibility of coinciding integers is also briefly discussed. 
As a concrete example, take $N=12,x=\pi,\phi=0$ and consider the primary state  that corresponds to the electromagnetic excitation $[-3,-2,-1,0,1]$. In this case, the level-one chiral excitation does not correspond to $[-4,-2,-1,0,1]$ but rather to
$[-6,-1,0,0,1]$, i.e., the integer $0$ repeats. In such situations the simplified picture above fails to hold. The way to identify the states more generally will be to look at their energy together with the sum of their Bethe integers. A scaling state then corresponds to a state whose sum of Bethe integers is equal to the sum of a ``valid'' configuration, and which is also one of the low-energy states within that sector of lattice momentum. The latter criterion can be quantified by demanding the state to be the $k$th excitation for some $k<k_0$. We only take $k_0\rightarrow\infty$ after $N\rightarrow\infty$, a procedure that is called the double limit in \cite{KooSaleur}. We can see from the picture above the importance of taking $N\rightarrow\infty$ first: to accommodate the shifts in momentum that we get from creating the holes and shifts, we must keep $N$ large enough.


\subsubsection{Overlaps and mixing}\label{mixing_section}

Even when taking the possibility of repeating integers into account, the above picture of scaling states and their conjectured limits is neater than reality. An important example of a more complicated situation, which will be relevant for the numerical results below, is as follows.

Consider two scaling states on the lattice with the same $S_z \neq 0, \phi = 0$ but with opposite electric excitations $e,-e$ (with $e\neq 0$). By conjecture \eqref{alpha} these should correspond in the double limit to two primary states with the conformal weights switched with respect to one another, so that they both have the same energy. 
Following \eqref{bethe_momentum}-\eqref{rescaled_momentum} and taking into account that lattice momentum is defined modulo the system size, the sectors of lattice momentum are only separated by $\Delta \Pscaled  = 2 e S_z$. Making holes to create the descendant states will shift $\Pscaled$ in integer steps, and it is clear that  the momentum sectors of the left descendants of one state will start overlapping with those of the right descendants of the other state at level $e S_z$, no matter how large we take $N$. 

By chiral/anti-chiral symmetry there is in such cases no way to distinguish from which primary a given state descends, and the symmetry may even force us to consider linear combinations of the scaling states as candidates for the descendant states in the limit. We then say that there is \emph{mixing} of the scaling states. To make this issue more clear, let us specialize this example. Let $V_1$ be the primary state corresponding to $S_z=1, e=-1$ and $V_2$ the primary state corresponding to $S_z =1, e = 1$. Let $W_1$ be the level-1 chiral descendant of $V_1$, and let $W_2$ denote the level-1 anti-chiral descendant of $V_2$. The momentum sectors of $W_1$ and $W_2$ are the same. Within this sector, the lowest-energy state (whose energy comes with multiplicity one within this sector) can neither be identified with $W_1$ nor with $W_2$. The reason is that the energies of $W_1,W_2$ in the limit are the same, and ``favouring'' one over the other by assigning it a scaling state with lower energy would be incompatible with the symmetry of the system, which demands that chiral and anti-chiral quantities play equal roles. Instead, we must create $W_1,W_2$ out of linear combinations of more than one scaling state, such that each of them has the same contribution from the lowest-energy state. This phenomenon will be further discussed in Appendix~\ref{spin_untwisted}.

Since we wish to explore the indecomposable structure of the modules, the issue of mixing will be particularly important. The reason is that a primary state with a degenerate conformal weight given by the integers $e$ and $S_z$, related to $r,s$ as in \eqref{alpha}, will have its null state precisely at level $|eS_z|$, in the sector of lattice momentum where mixing can occur.



\subsection{Conjugate states and Bethe roots}\label{ConjugateStates}

In Section \ref{FF_modules} we saw that in the degenerate case there are two possible diagrams for the structure of Virasoro modules, reproduced here for convenience (co-Verma module to the left, Verma module to the right):
\begin{equation}
\begin{tikzpicture}[auto, node distance=0.4 and 0.3cm, baseline=(current  bounding  box.center)]
  \node (node1) [align=center]{$\bullet$};
  \node (node2) [below = of node1, align=center]{$\circ$};
  \draw[latex-]  (node1) edge  (node2);
\end{tikzpicture}
\hspace*{1cm}
\begin{tikzpicture}[auto, node distance=0.4 and 0.3cm, baseline=(current  bounding  box.center)]
  \node (node1) [align=center]{$\circ$};
  \node (node2) [below = of node1, align=center]{$\bullet$};
  \draw[-latex]  (node1) edge  (node2);
\end{tikzpicture}.
\end{equation}
Due to the sign in the conjecture \eqref{alpha}, we see that degenerate chiral weights $h$ (i.e., $h=h_{r,s}$ with $r,s\in\mathbb{N}^*$) are obtained by charges $\alpha$ corresponding to states where $(e-e_\phi)$ and $S_z$ are of \emph{opposite} signs, while degenerate anti-chiral weights $\bar{h}$ are obtained by charges $\bar{\alpha}$ corresponding to states where $(e-e_\phi)$ and $S_z$ are of the same sign.
%
%
With the sign conventions for magnetization and lattice momentum used in this paper, we have found the co-Verma type for 
\begin{subequations}
\begin{eqnarray}
\alpha=\alpha_{r,s} \quad \Leftrightarrow \quad (e-e_\phi)<0 \mbox{ and } S_z>0 \,, &\qquad& \mbox{(chiral)} \\
\bar{\alpha}=\alpha_{r,s} \quad \Leftrightarrow \quad  (e-e_\phi)<0 \mbox{ and } S_z<0 \,, &\qquad& \mbox{(anti-chiral)} \,,
\end{eqnarray}
\end{subequations}
while the Verma type was found for the conjugate charges
\begin{subequations}
\begin{eqnarray}
\alpha=\alpha_{r,s}^{\rm c} \quad \Leftrightarrow \quad  (e-e_\phi)>0 \mbox{ and } S_z<0 \,, &\qquad& \mbox{(chiral)} \\
\bar{\alpha}=\alpha_{r,s}^{\rm c} \quad \Leftrightarrow \quad  (e-e_\phi)>0 \mbox{ and } S_z>0 \,, &\qquad& \mbox{(anti-chiral)} \,,
\end{eqnarray}
\end{subequations}
with $r,s\in\mathbb{N}^*$ as given in \eqref{alpha}. As an example, $(e-e_\phi)=-1,S_z=1 \Rightarrow \alpha = \alpha_{1,1}=0$ while $(e-e_\phi)=1,S_z=-1 \Rightarrow \alpha=\alpha_{1,1}^{\rm c}=2\alpha_0$, which can be compared to \eqref{diagram}. Of course the choice of conventions holds no deeper meaning, and the important takeaway is that \emph{within each sector of $S_z$ we expect to find pairs of conjugate primary states such that one has a chiral null state, the other a anti-chiral one, and the modules are of opposite types \emph{(Verma or co-Verma)}.} 

We now show that such pairs of states, which differ by the sign of $(e-e_\phi)$, correspond to Bethe states that differ only by the sign of their Bethe integers and the sign of their Bethe roots. This can be seen directly from the shape of the Bethe equations \eqref{BetheEqs} in Appendix \ref{FormFactorAppendix}, reproduced here for convenience:
\begin{equation}
\frac{d(\lambda_j)}{a(\lambda_j)}\prod_{k\neq j}\frac{b(\lambda_k,\lambda_j)}{b(\lambda_j,\lambda_k)}=1
\end{equation}
Leaving the definitions of the various terms to the Appendix, we here need only know that $b(-\lambda_k,-\lambda_j) = b(\lambda_j,\lambda_k)$, $a(\lambda)=1$ and that for $\phi=0$, $d(-\lambda) = (d(\lambda))^{-1}$ in the homogeneous limit (our case of interest). When $\phi \neq 0$, the only modification to the Bethe equations is $d(\lambda) \rightarrow e^{i\phi}d(\lambda)$, showing that one must take $e_\phi \rightarrow -e_\phi$ in order for the roots with opposite signs to be a solution.

That the Bethe roots for these pairs of states differ only by their sign will be important in Section \ref{ResultsAboutFF}, where we show that a strong duality 
of the corresponding modules can be seen directly on the lattice using Bethe ansatz techniques. To clarify the meaning of ``strong'', let us for comparison write out a weaker type of duality that is expected based only on the conjecture \eqref{alpha} as discussed above. We shall need a more precise notation for conjugate states than in Section \ref{FF_modules}. 
We write $ V_{\bar{\alpha}^{\rm c},\alpha^{\rm c}} $ for a state whose anti-chiral charge is the conjugate of the chiral charge of $V_{\alpha,\bar{\alpha}}$, and whose chiral charge is the conjugate of the anti-chiral charge of $V_{\alpha,\bar{\alpha}}$.
%
%
In this notation, we have the following result:

\vspace*{5pt}
\noindent\fbox{
\hspace*{5pt}\begin{minipage}{\linewidth-20pt}\em
\vspace*{5pt}
{\bf Weak duality:} 
Whenever $V_{\alpha,\bar{\alpha}}$ has a degenerate chiral charge $\alpha=\alpha_{r,s}$ or $\alpha=\alpha_{r,s}^{\rm c}$ (with $r,s \in \mathbb{N}^*$) we expect that 
\begin{subequations}
\label{weak}
\begin{equation}\label{weak1}
\langle W | A_{r,s} | V_{\alpha,\bar{\alpha}}\rangle = 0 
\quad \Leftrightarrow \quad
\langle V_{\bar{\alpha}^{\rm c},\alpha^{\rm c}} | \bar{A}_{r,s}^\ddag | W^{\rm c}\rangle
= 0 \,,
\end{equation}
where $W, W^{\rm c}$ are the corresponding null states at level $r s$, $A_{r,s}$ the relevant combination of lowering operators as in the examples \eqref{A_operators} and $A_{r,s}^\ddag$ the ``conformal conjugate'' for which $L_n^\ddag =  L_{-n}$. The same type of statements hold when $V_{\alpha,\bar{\alpha}}$ has a degenerate anti-chiral charge $\bar{\alpha}=\alpha_{r,s}$ or $\bar{\alpha}=\alpha_{r,s}^{\rm c}$: 
\begin{equation}\label{weak2}
\langle \overline{W} | \bar{A}_{r,s} | V_{\alpha,\bar{\alpha}} \rangle = 0 
\quad \Leftrightarrow \quad 
\langle V_{\bar{\alpha}^{\rm c},\alpha^{\rm c}} | A_{r,s}^\ddag | \overline{W}^{\rm c} \rangle
= 0 \,.
\end{equation}
\end{subequations}
%
\end{minipage}\hspace*{5pt}}
\vspace*{5pt}
~\\
On the lattice, we expect that the corresponding matrix elements (with the Virasoro generators replaced by the Koo-Saleur generators, and primary/descendant states by their corresponding scaling states) will approach either zero or non-zero values in the limit according to this duality. Comparing to \eqref{weak}, the stronger duality that will be shown below is the statement that two matrix elements \emph{have the same value} already on the lattice.

Before turning to the stronger duality we note that even without the Bethe ansatz we can, in fact, see the weaker duality already on the lattice for one particular case of interest: the modules  $\AStTL{0}{\q^{-2}} $ and $\AStTL{0}{\q^{2}}$ as described in Section \ref{indecomposability}:
\begin{equation}
\text{$\AStTL{0}{\q^{- 2}}$:}
\begin{tikzpicture}[auto, node distance=0.5 and 0.3cm, baseline=(current  bounding  box.center)]
  \node (node1) [align=center]{$[0,\q^{-2}]$};
  \node (node2) [below = of node1, align=center]{$[1,1]$};
  \draw[-latex]  (node1) edge  (node2);
\end{tikzpicture},
\hspace*{1cm}
\text{$\AStTL{0}{\q^{ 2}}$:}
\begin{tikzpicture}[auto, node distance=0.5 and 0.3cm, baseline=(current  bounding  box.center)]
  \node (node1) [align=center]{$[0,\q^2]$};
  \node (node2) [below = of node1, align=center]{$[1,1]$};
  \draw[latex-]  (node1) edge  (node2);
\end{tikzpicture}.
\end{equation}
Within these diagrams, the pair of states corresponding to $S_z=0, |e|=1$, $e_\varphi = \pm \alpha_-/\alpha_+$\footnote{ Note that within \eqref{alpha}, a charge with a twist of $e_\varphi = \pm \alpha_-/\alpha_+$  can be rewritten as a charge with $e_\varphi=0$ where the magnetization $S_z$ is shifted by one, which shows that these charges are on the form $\alpha_{r,s}, \; r,s \in \mathbf{Z}$ leading to degenerate conformal weights. In particular, with $|e|=1$ we obtain the degenerate weight $h_{1,1}$. 
 } 
 can be found within $[0,\q^{\pm 2}]$ while their corresponding level-1 null states can be found within $[1,1]$. The duality under the action of the Koo-Saleur generators on the lattice then follows directly from the duality of the Temperley-Lieb modules, since the Koo-Saleur generators are built out of Temperley-Lieb generators. 

\subsection{Some results about form factors}\label{ResultsAboutFF}


In Appendix \ref{FormFactorAppendix} we give a brief recapitulation of the Quantum Inverse Scattering Method, in which local operators such as $\sigma^z_m$ are expressed in terms of entries of the monodromy matrix $\mathbf{T}= \left( \begin{smallmatrix}
A&B\\
C&D
\end{smallmatrix} \right)$. For a general overview of this method see \cite{Slavnov}. The eigenstates of the XXZ Hamiltonian are given as $|\{\lambda\}\rangle = \prod_{k=1}^nB(\lambda_k)|0\rangle$ for a set of Bethe roots $\{\lambda\}$, their duals as $\langle \{\lambda\}| =\langle 0 | \prod_{k=1}^nC(\lambda_k)$ and we wish to find matrix elements on the form $\langle \{\mu\}|\prod_i\sigma^{a_i}_{m_i}|\{\lambda\}\rangle$, where $a_i \in\{z,+,-\}$ and $m_i\neq m_j$ for $i\neq j$.
 The resulting expressions for these matrix elements in terms of functions of the Bethe roots are called form factors.

If we can find form factors for the Koo-Saleur generators $\KSgen_n$, the work of looking at the action of the Virasoro algebra in the spin chain reduces to evaluating the expressions of these form factors, rather explicitly diagonalizing the finite Hamiltonian and then acting on the resulting eigenstates. For large system sizes $N$, this is a significant advantage: while the size of the Hamiltonian to be diagonalized grows exponentially in $N$, the time needed to evaluate form factors is only polynomial.
A similar form-factor program has already been carried out in \cite{Bondesan} for the case of the $SU(2)$-invariant six-vertex model and its descendants, which in the continuum correspond to $SU(2)_k$ WZW models. However, in this case the program was carried out for the current $J^a(z)$ rather than the Virasoro generators.

Finding form factors for $\KSgen_n$ boils down to finding form factors for $e_i$ and $[e_i,e_{i+1}]$. A priori this will involve form factors for all six permutations of three different neighbouring operators, namely
\begin{subequations}
\label{all6-1}
\begin{eqnarray}
& & \sigma_m^{z}\sigma_{m+1}^{-}\sigma_{m+2}^{+} \,, \\
& & \sigma_m^{z}\sigma_{m+1}^{+}\sigma_{m+2}^{-} \,, \\ 
& & \sigma_{m}^{+}\sigma_{m+1}^{-}\sigma_{m+2}^{z} \,, \\
& & \sigma_m^{-}\sigma_{m+1}^{+}\sigma_{m+2}^{z}
\end{eqnarray}
\end{subequations}
and
\begin{subequations}
\label{all6-2}
\begin{eqnarray}
& & \sigma_m^{-}\sigma_{m+1}^{z}\sigma_{m+2}^{+} \,, \\
& & \sigma_{m}^{+}\sigma_{m+1}^{z}\sigma_{m+2}^{-} \,,
\end{eqnarray}
\end{subequations}
as well as  $\sigma_m^{-}\sigma_{m+1}^{+}, \sigma_m^{+}\sigma_{m+1}^{-}, \sigma_m^{z}\sigma_{m+1}^{z}$ and $\sigma^z$. Luckily we can reduce the number of form factors we need to compute by using various relations that follow from the Bethe ansatz. These relations can also give us some general insight into how the lattice Virasoro generators ${\mathcal L}_n, \bar{\mathcal L}_n$ should act on Bethe states---in particular, we shall soon see how the duality for conjugate states appears already at finite size.

\subsubsection{Properties under conjugation and parity, implications for the modules}\label{conjugate_vs_Bethe}

The site dependence of the form factors does not depend on the choice of operators and can be factorized in the expressions. Following the notation in Appendix \ref{FormFactorAppendix} we let  $F_{\mathcal{O}_1 \cdots \mathcal{O}_j}$ denote the site-independent part of the form factor for $j$ neighboring operators $\mathcal{O}_1 \cdots \mathcal{O}_j$. We wish to find relations between the site-independent part of the form factors of interest.

The first type of relations between the site-independent part of the form factors of interest is due to conjugation. The dual states of on-shell Bethe states are, up to a possible phase\footnote{In the numerics below, the phase is found for a given scaling state at small $N$, by explicitly comparing its dual with its conjugate, and stays the same as $N$ is increased.}, their conjugates,
\begin{equation}\label{conjugation}
\langle \{\lambda \}| = ({\rm phase}) \big( |\{\lambda \}\rangle  \big)^\dagger \,.
\end{equation}
Combining this relation with
\begin{equation}
(\sigma^{+})^\dagger=\sigma^{-},\quad (\sigma^{z})^\dagger=\sigma^{z}
\end{equation}
we obtain 
\begin{subequations}
\begin{eqnarray}
 \langle \{\mu \} | \sigma_m^{+} \sigma_{m+1}^z  \sigma_{m+2}^{-} | \{\lambda \} \rangle 
&=& ({\rm phase}) \Big(\langle \{\lambda \}| \sigma_m^{-} \sigma_{m+1}^z  \sigma_{m+2}^{+} |\{\mu \}\rangle \Big)^* \,, \\
 \langle \{\mu \} | \sigma_m^{z} \sigma_{m+1}^+  \sigma_{m+2}^{-} | \{\lambda \} \rangle 
&=& ({\rm phase}) \Big(\langle \{\lambda \}| \sigma_m^{z} \sigma_{m+1}^-  \sigma_{m+2}^{+} |\{\mu \}\rangle \Big)^* \,,
\end{eqnarray}
\end{subequations}
which relate $F_{\sigma^+\sigma^z\sigma^-}$ to $F_{\sigma^-\sigma^z\sigma^+}$ and $F_{\sigma^z\sigma^+\sigma^-}$ to $F_{\sigma^z\sigma^-\sigma^+}$ through the conjugation of each of the operators. Similarly we can relate $F_{_{\sigma^+\sigma^-}}$ to $F_{\sigma^-\sigma^+}$.

The second type of relations between the site-independent part of the form factors of interest is due to parity.
Following \cite{Doikou} we denote by $\Pi$ the parity operator. It acts on a local operator $X_m$ as
\begin{equation}
\Pi X_m \Pi^{-1}=X_{N+1-m}
\end{equation}
and on the $B$-operators as
\begin{equation}\label{parityB}
\Pi B(\lambda) \Pi^{-1}= (-1)^{N-1}B(-\lambda).
\end{equation}
Thus, parity will act on Bethe states by taking $|\{\lambda\}\rangle$ into $|\{-\lambda\} \rangle$ (up to a possible sign)  and act on $j$ neighboring operators by reversing their order and the sites they act on. We obtain
\begin{subequations}
\begin{eqnarray}
 \langle \{\mu \} | \sigma_m^{-} \sigma_{m+1}^+  \sigma_{m+2}^{z} | \{\lambda \} \rangle 
&=& ({\rm sign}) \langle \{- \mu  \}| \sigma_{N-m-1}^{z} \sigma_{N-m}^+  \sigma_{N-m+1}^{-} |\{ -\lambda\}\rangle \,, \\
 \langle \{\mu \} | \sigma_m^{+} \sigma_{m+1}^-  \sigma_{m+2}^{z} | \{\lambda \} \rangle 
&=& ({\rm sign}) \langle \{- \mu  \}| \sigma_{N-m-1}^{z} \sigma_{N-m}^- \sigma_{N-m +1}^{+} |\{ -\lambda\}\rangle \,,
\end{eqnarray}
\end{subequations}
which relate $F_{\sigma^-\sigma^+\sigma^z}$ to $F_{\sigma^z\sigma^+\sigma^-}$ and $F_{\sigma^+\sigma^-\sigma^z}$ to $F_{\sigma^z\sigma^-\sigma^+}$ through reversing the order of the operators. Altogether, the combined actions of conjugation and parity relate the four form factors \eqref{all6-1} among themselves, and similarly for the remaining two \eqref{all6-2}. Thus, the only form factors involving three operators that we shall need to compute in Appendix \ref{FormFactorAppendix} will be one of each group, here chosen to be $F_{\sigma^z\sigma^-\sigma^+}$ and $F_{\sigma^-\sigma^z\sigma^+}$.






Combining the expression for the Koo-Saleur generators \eqref{generators} with the parity relations for the form factors yields relations for the matrix elements of $\KSgen_n$ in the basis of Bethe states. In particular, we note that the term $[e_j,e_{j+1}]$ in \eqref{generators} will pick up a sign when the order of all operators is reversed, changing $\KSgen_n$ into $\bar{\KSgen}_{-n}$. Thus we have
\begin{equation}\label{parity_duality}
\langle \{\mu\} | \KSgen_n | \{\lambda\} \rangle = \langle \{-\lambda\} | \bar{\KSgen}_{-n} | \{-\mu\} \rangle.
\end{equation}
Meanwhile, the pairs of conjugate states in a given sector of $S_z$ are, as discussed in Section \ref{ConjugateStates}, related through a change of signs for all the Bethe roots, which is precisely the action of parity on states as seen in \eqref{parityB}. Taken together, we find in particular that \eqref{weak1} and \eqref{weak2} can be turned into a much stronger statement:

\vspace*{5pt}
\noindent\fbox{
\hspace*{5pt}\begin{minipage}{\linewidth-20pt}\em
\vspace*{5pt}
{\bf Strong duality:} Let $V_{\alpha,\bar{\alpha}}, V_{\bar{\alpha}^{\rm c},\alpha^{\rm c}}$ and $W, W^{\rm c}$ be a pair of conjugate primary states and their respective null states, as defined in connection to \eqref{weak1}--\eqref{weak2}, and $A_{r,s}$ the corresponding combination of Virasoro generators. We here denote the corresponding scaling states and lattice operators by calligraphic letters, e.g. $\mathcal{V}[N]_{\alpha,\bar{\alpha}}$. At each finite size $N$---large enough to accommodate the states of interest---we have the following equality of the matrix elements:
\begin{subequations}
\label{strong}
\begin{equation}\label{strong1}
\langle\mathcal{ W}[N] | \mathcal{A}_{r,s}[N] | \mathcal{V}_{\alpha,\bar{\alpha}}[N] \rangle = ({\rm phase})
\langle \mathcal{V}_{\bar{\alpha}^{\rm c},\alpha^{\rm c}}[N] | \bar{\mathcal{A}}_{r,s}^\ddag [N]| \mathcal{W}^{\rm c} [N] \rangle \,.
\end{equation}
The same type of duality holds when $V_{\alpha,\bar{\alpha}}$ has a degenerate anti-chiral charge $\bar{\alpha}=\alpha_{r,s}$ or $\bar{\alpha}=\alpha_{r,s}^{\rm c}$:
\begin{equation}\label{strong2}
\langle\overline{\mathcal{W}}[N] | \bar{\mathcal{A}}_{r,s}[N] | \mathcal{V}_{\alpha,\bar{\alpha}}[N] \rangle = ({\rm phase})
\langle \mathcal{V}_{\bar{\alpha}^{\rm c},\alpha^{\rm c}}[N] | \mathcal{A}_{r,s}^\ddag [N]| \overline{\mathcal{W}}^{\rm c} [N] \rangle \,.
\end{equation}
\end{subequations}

\vspace*{5pt}
\end{minipage}\hspace*{5pt}}
\vspace*{5pt}
~\\
Examples of this situation are seen in the numerical results, in Tables \ref{ED_comparisons} and \ref{level1twisted}. 

The considerations above hold for the matrix elements of single Temperley-Lieb generators as well, because of translational invariance (to be discussed in the next section):
%
\begin{equation}
\langle \{\mu\} | e_j | \{\lambda\} \rangle = ({\rm phase}) \langle \{-\lambda\} | e_j | \{-\mu\} \rangle.
\end{equation}
This is in agreement with the result for the modules  $\AStTL{0}{\q^{-2}} $ and $\AStTL{0}{\q^{2}}$ discussed in the end of Section \ref{ConjugateStates}.
%
%
Of the previous examples mentioned, Table \ref{level1twisted} 
has exact results at finite size coming from the Temperley-Lieb structure.

 


\section{Lattice Virasoro in the non-degenerate case}\label{sec:lattVir1}


In this section we give a first example of the numerical results obtained in the XXZ spin-chain representation by the Bethe ansatz. 
We consider matrix elements of the Koo-Saleur generator $\KSgen_{-1}$ in the basis of Bethe states, in the non-degenerate case where neither $\alpha$ nor $\bar{\alpha}$ leads to a degenerate conformal weight $h_{r,s}$. We begin by some general considerations with regard to lattice momentum, which also carry over to the degenerate case.





\subsection{Koo-Saleur generators and lattice momentum}\label{KSandLatticeMomentum}

Thanks to the smearing of the twist in \eqref{TLmatrix_spin} the Bethe states are invariant under the usual translation operator, and we can sort them by their (rescaled) lattice momentum $\Pscaled = 0,1, \hdots , N-1$ defined in \eqref{rescaled_momentum}. We define the matrix $\LnMat{n}[N]$ so that $(\LnMat{n}[N])_{ab}$ is the matrix element $\langle a | \KSgen_n [N] |b\rangle$ of the Koo-Saleur generator $\KSgen_n[N]$ between two Bethe states at system size $N$. $\LnMat{n}[N]$ can then be written as a block-diagonal matrix, with the blocks indexed by values of $\Pscaled$. We now show how, thanks to the relations of the affine Temperley-Lieb algebra, only a few blocks of this matrix can have non-zero elements, namely the ones where the lattice momentum is shifted by precisely $n$ between $|a\rangle$ and $|b\rangle$. 

Recall that by \eqref{TL-u} the generator of lattice translation $u$, fulfils $u e_j u^{-1} = e_{j+1}$. Meanwhile, when acting on a Bethe state $|v\rangle$ belonging to the sector of lattice momentum ${\mathcal p}$, the result is a phase $u|v\rangle=e^{i 2\pi {\mathcal p} /N}|v\rangle$. Together these relations determine the behaviour of ${\mathcal L}_n|v\rangle$ under translation. Let us inspect the first part of the $j$'th term within the expression for $\KSgen_n$ \eqref{generators}: 
\begin{equation}\label{shifting_terms}
u \, e^{inj2\pi/N} e_j |v\rangle 
 = e^{inj2\pi/N}  e_{j+1} e^{i {\mathcal p} 2\pi/N} |v\rangle = e^{i ({\mathcal p} -n)2\pi/N}  e^{in(j+1)2\pi/N}  e_{j+1}|v\rangle .
\end{equation}
Up to a phase this is the corresponding part of the $(j+1)$'th term, which can be turned into the $j$'th by a relabelling of the indices within the sum. Applying the same procedure to the second part of the $j$'th term and summing over $j$ yields in total
\begin{equation}
u\, \KSgen_n |v\rangle = e^{i({\mathcal p} - n)2\pi/N}   \KSgen_n |v\rangle,
\end{equation} 
i.e., we find that ${\mathcal L}_n |v\rangle $ is also a momentum eigenstate, and will therefore be orthogonal to any Bethe state with lattice momentum different than ${\mathcal p}-n$. In other words: already at finite size, the conformal spin must change by the proper integer value when we raise or lower a state using a Koo-Saleur generator.

The relations of the affine Temperley-Lieb algebra can also be used to speed up the numerics by computing only one term of the sum. Consider the blocks within which the matrix elements $(\LnMat{n}[N])_{ab}$ may be non-zero, i.e., where the Bethe states fulfil
$u | b \rangle=e^{i {\mathcal p} 2\pi/N} | b \rangle $ and $u |a \rangle = e^{i ({\mathcal p}-n) 2\pi/N} | a \rangle$, respectively. 
Let us write $\KSgen_n[N]$ as $\sum_j f(e_j,e_{j+1})$. By \eqref{shifting_terms} we can shift $f(e_j,e_{j+1})|b\rangle$ to $f(e_{j+1},e_{j+2})|b\rangle$ by applying $u$, at the price of a phase. Considering the matrix element of a single term and inserting  $\mathbf{1}=u^{-1}u$, we can let $u^{-1}$ act to the left. The phases cancel:
\begin{equation}
\langle a | u^{-1}   u\,  f(e_{j},e_{j+1})  |b \rangle = e^{-i({\mathcal p} - n)2\pi/N}   e^{i({\mathcal p} - n)2\pi/N}   \langle a | f(e_{j+1},e_{j+2})  |b \rangle .
\end{equation} 
We thus see that all terms give the same contributions to the matrix elements within non-zero blocks of $\LnMat{n}[N]$, and we can replace $\sum_j  f(e_j,e_{j+1})$ by $N\, f(e_j,e_{j+1})$ in the numerical evaluations, gaining a factor $N$ in speed.

\subsection{Numerical results for $\KSgen_{-1}$}
\label{sec:KS_Lminus1}

We now show the first non-zero matrix elements of the matrix $\LnMat{-1}[N]$ for increasingly large system size $N$ and compare with the values we would expect from CFT computations. In this example we consider $S_z=1,e=0$ at $\phi=1/10, x=\pi$. (Recall \eqref{gammaparam}: $\q =e^{i\gamma} = e^{i\pi/(x+1)}$, and \eqref{e_phi}: $e_{\phi} = \frac{\phi}{2\pi}$). For non-zero matrix elements, the lattice momentum must shift by $p \rightarrow p -n = p +1$ between the two Bethe states. 

We shall consider the block between states in the momentum sectors $p=0$ (denoted $|u_1\rangle, |u_2\rangle, \hdots$) and $p=1$ (denoted $|v_1\rangle, |v_2\rangle, \hdots$). The set of Bethe integers for the lowest-energy state in the $p=0$ sector, which we denote by $|u_1\rangle$, allows us to identify it with a primary state $V_{\alpha,\bar{\alpha}}$, which by \eqref{alpha}
has a chiral charge $\alpha =   -\frac{1}{40 \pi} \alpha_{+}  + \alpha_0 - \frac{1}{2} \alpha_{-}  = 0.56495148 $. 
The other states we consider will be its descendants. At $N=10$ we find the three lowest states in each sector, sorted by energy, for the following Bethe integers:\footnote{Here we see an example of half-integer ``Bethe integers'', since this is the maximally packed symmetric distribution around zero for four roots.}

~\\
$p=0$: 
\begin{equation}\label{p0states}
\begin{split}
| u_1\rangle \leftrightarrow\left\{-\frac{3}{2},-\frac{1}{2},\frac{1}{2},\frac{3}{2}\right\}& \leftrightarrow\circ\circ\bullet\bullet \hspace*{-2pt}|\hspace*{-2pt} \bullet\bullet\circ\circ \quad\leftrightarrow\quad \text{primary state }V_{\alpha,\bar{\alpha}}\\
| u_2\rangle \leftrightarrow \left\{-\frac{5}{2},-\frac{1}{2},\frac{1}{2},\frac{5}{2}\right\}& \leftrightarrow\circ\bullet\circ\bullet\hspace*{-2pt}|\hspace*{-2pt}\bullet\circ\bullet\circ  \quad\leftrightarrow\quad \text{chiral and anti-chiral level 1}\\
| u_3 \rangle \leftrightarrow\left\{-\frac{5}{2},-\frac{3}{2},\frac{3}{2},\frac{5}{2}\right\}& \leftrightarrow\circ\bullet\bullet\circ\hspace*{-2pt}|\hspace*{-2pt}\circ\bullet\bullet\circ  \quad\leftrightarrow\quad \text{chiral and anti-chiral level 2}\hspace*{19pt}
\end{split}
\end{equation}
~\\
$p=1$:
\begin{equation}\label{p1states}
\begin{split}
| v_1 \rangle \leftrightarrow \left\{-\frac{5}{2},-\frac{1}{2},\frac{1}{2},\frac{3}{2}\right\} \leftrightarrow\circ\bullet\circ\bullet\hspace*{-2pt}|\hspace*{-2pt}\bullet\bullet\circ\circ  \quad &\leftrightarrow \quad\text{chiral level 1}\\
    \left.
     \begin{aligned}
       | v_2\rangle \leftrightarrow \left\{-\frac{5}{2},-\frac{3}{2},\frac{1}{2},\frac{5}{2}\right\}& \leftrightarrow\circ\bullet\bullet\circ\hspace*{-2pt}|\hspace*{-2pt}\bullet\circ\bullet\circ \\
       | v_3\rangle \leftrightarrow\left\{-\frac{7}{2},-\frac{1}{2},\frac{1}{2},\frac{5}{2}\right\}& \leftrightarrow\bullet\circ\circ\bullet\hspace*{-2pt}|\hspace*{-2pt}\bullet\circ\bullet\circ \\
    \end{aligned}
    \right\}\hspace*{1pt}  &\leftrightarrow\quad \text{chiral level 2 and anti-chiral level 1} \hspace*{-10pt}\\
\end{split}
\end{equation}

~\\
These patterns extend to larger $N$ by padding with filled circles in the middle. Taking the example of $|u_2\rangle$ we find $\circ\bullet\circ\bullet\hspace*{-2pt}|\hspace*{-2pt}\bullet\circ\bullet\circ \; \rightarrow \; \circ\bullet\circ\bullet \bullet \hspace*{-4pt}|\bullet\circ\bullet\circ \; \rightarrow \; \circ\bullet\circ\bullet\bullet\hspace*{-0pt}|\hspace*{-2pt}\bullet\bullet\circ\bullet\hspace*{2pt}\circ \; \rightarrow \; \hdots$.

Recall the state space \eqref{Fock_a_n}. Since there is only one state at level 1 we can immediately identify $|v_1 \rangle$ with $a_{-1} V_{\alpha,\bar{\alpha}}$ and $| u_2 \rangle$ with $a_{-1}\bar{a}_{-1}V_{\alpha,\bar{\alpha}}$. Meanwhile at level 2 there are two states, and the states $|v_2\rangle$ and $|v_3\rangle$ will correspond to an orthonormal basis for the two-dimensional vector space of $a_{-1}^2 \bar{a}_{-1} V_{\alpha,\bar{\alpha}}$ and $a_{-2} \bar{a}_{-1} V_{\alpha,\bar{\alpha}}$. Finally $|u_3\rangle$ corresponds to one basis vector in an orthonormal basis for the four-dimensional vector space of states at chiral \emph{and} anti-chiral level 2, the other three basis vectors being found by considering scaling states of higher energy.
~\\

\textbf{Conjectures for the matrix elements:}
\begin{itemize}

\item As in Section \ref{FF_modules} we have $L_{-1} V_{\alpha,\bar{\alpha}}  = \sqrt{2} \alpha a_{-1} V_{\alpha,\bar{\alpha}}  $, leading to the conjectured value $\langle a_{-1}V_{\alpha,\bar{\alpha}}   |L_{-1}|V_{\alpha,\bar{\alpha}} \rangle=\sqrt{2}\alpha=0.79896205$ for the matrix element $(\LnMat{-1}[N])_{v_1,u_1}$ at $N\rightarrow \infty$.  We can also find this value by considering the norm squared $\langle L_{-1} V_{\alpha,\bar{\alpha}} | L_{-1} V_{\alpha,\bar{\alpha}} \rangle = \langle  V_{\alpha,\bar{\alpha}} | L_{-1}^\dagger L_{-1} V_{\alpha,\bar{\alpha}} \rangle$,  where $L_{-1}^\dagger = L_{1} + 2\sqrt{2}\alpha_0 a_1 $ by \eqref{Vir_gens}.   Using \eqref{Heisenberg} and \eqref{Heis_weight} we find $L_{-1}^\dagger L_{-1} V_{\alpha,\bar{\alpha}}  = 2\alpha^2 V_{\alpha,\bar{\alpha}} $.\footnote{In general, the calculations of this type that are seen in this paper have been performed by adapting the Mathematica notebook ``Virasoro'' by M. Headrick, available at  \url{http://people.brandeis.edu/~headrick/Mathematica/index.html}.} 

\item The latter method carries over most easily to the next two matrix elements that we wish to consider: $(\LnMat{-1}[N])_{v_2,u_2}$ and 
$(\LnMat{-1}[N])_{v_3,u_2}$. We find the norm squared $\langle L_{-1}a_{-1}V_{\alpha,\bar{\alpha}}  | L_{-1}a_{-1}V_{\alpha,\bar{\alpha}} \rangle= 2(1+2\alpha^2)$. Since $|v_2\rangle$ and $|v_3\rangle$ provide a basis, we expect to recover the full norm $\sqrt{2(1+2\alpha^2)}  =1.81016041$ by combining the projections on these states as $\sqrt{ (\LnMat{-1}[N])^2_{v_2,u_2} + (\LnMat{-1}[N])_{v_3,u_2}^2 }$ at $N \rightarrow \infty$.

\item All other matrix elements among the states in \eqref{p0states} and \eqref{p1states} are conjectured to be zero in the limit, since the chiral levels do not match.

\end{itemize}

Using the form factors computed in Appendix \ref{FormFactorAppendix} we can obtain the matrix elements $(\LnMat{-1}[N])_{v_iu_j}$ for increasingly large system size $N$. We then perform a polynomial extrapolation in $1/N$ to approximate the value at $N\rightarrow\infty$. This is shown in  table \ref{generic_example} for $i=1,j=1$ and $i=2,j=1$.

\begin{table}[H]
\center
$\begin{array}[t]{ccc}
\begin{tabular}{lll}
$N$      & $\langle v_1 | \KSgen_{-1} | u_1 \rangle$ &  $\langle v_2 | \KSgen_{-1} | u_1 \rangle$ \\
\hline
10 & 0.73934396 & 0.01866525 \\
12 & 0.75628984 & 0.0178073 \\
14 & 0.76689996 & 0.01621625 \\
16 & 0.77397144 & 0.01457925 \\
18 & 0.77891781 & 0.01309015 \\
20 & 0.78251286 & 0.01178876 \\
22 & 0.78520823 & 0.01066614 \\
24 & 0.7872816 & 0.00969992 \\
26 & 0.78891124 & 0.00886622 \\
28 & 0.79021581 & 0.00814359 \\
30 & 0.79127676 & 0.0075138 \\
32 & 0.79215151 & 0.00696179 \\
34 & 0.79288148 & 0.00647521 \\
\vphantom{$\int^0$}\smash[t]{\vdots} & \vphantom{$\int^0$}\smash[t]{\vdots}  & \vphantom{$\int^0$}\smash[t]{\vdots} \\ 
\end{tabular}
&
\begin{tabular}{lll}
\vphantom{$\int^0$}\smash[t]{\vdots} & \vphantom{$\int^0$}\smash[t]{\vdots}  & \vphantom{$\int^0$}\smash[t]{\vdots} \\ 
36 & 0.79349714 & 0.00604396 \\
38 & 0.79402136 & 0.00565977 \\
40 & 0.79447152 & 0.00531584 \\
42 & 0.79486106 & 0.00500654 \\
44 & 0.79520048 & 0.0047272 \\
46 & 0.79549811 & 0.0044739 \\
48 & 0.79576061 & 0.00424336 \\
50 & 0.79599334 & 0.00403281 \\
52 & 0.7962007 & 0.00383989 \\
54 & 0.79638628 & 0.00366257 \\
56 & 0.79655307 & 0.00349914 \\
58 & 0.79670354 & 0.00334811 \\
60 & 0.79683979 & 0.00320817 \\
\vphantom{$\int^0$}\smash[t]{\vdots} & \vphantom{$\int^0$}\smash[t]{\vdots}  & \vphantom{$\int^0$}\smash[t]{\vdots} \\ 
\end{tabular}
&
 \begin{tabular}{lll}
\vphantom{$\int^0$}\smash[t]{\vdots} & \vphantom{$\int^0$}\smash[t]{\vdots}  & \vphantom{$\int^0$}\smash[t]{\vdots} \\ 
62 & 0.79696357 & 0.00307821 \\
64 & 0.79707638 & 0.00295725 \\
66 & 0.79717949 & 0.00284442 \\
68 & 0.79727401 & 0.00273896 \\
70 & 0.79736087 & 0.00264022 \\
72 & 0.79744089 & 0.0025476 \\
74 & 0.79751479 & 0.00246056 \\
76 & 0.79758317 & 0.00237865 \\
78 & 0.79764659 & 0.00230144 \\
80 & 0.79770552 & 0.00222856 \\
$p_{25}$& 0.79896913  &\hspace*{-3.5pt}-$4.846\cdot 10^{-5}$\\
$p_{30}$& 0.79896944  & \hspace*{-3.5pt}-$5.025\cdot 10^{-5}$\\
$p_{35}$& 0.79896858  & \hspace*{-3.5pt}-$4.558\cdot 10^{-5}$\\
 \textbf{conj}& 0.79896205   & 0 \\
 \hline 
\end{tabular}
\\
\end{array}$
\caption{Matrix elements $(\LnMat{-1}[N])_{v_iu_j}$ for $i=1,j=1$ and $i=2,j=1$, where the scaling states $|u_j\rangle$ and $|v_i\rangle$ follow the patterns of Bethe integers shown in \eqref{p0states} and \eqref{p1states}. The numerical values are given for the case of $S_z =  1, e = 0, x=\pi, \phi=1/10$ for system size $N$ up to 80, after which polynomial extrapolations $p_n(1/N)$ of degrees $n= 25, 30, 35$ to all the data points are made in order to approximate the value at $N \rightarrow \infty$.}\label{generic_example}
\end{table}


The other seven matrix elements between the Bethe states listed above are found in the same fashion, and we do not write out the corresponding columns in Table \ref{generic_example}. The results of the extrapolation $p_{35}$ for all nine matrix elements are:
%
%
 %
\begin{equation}\label{spinrep_example}
\LnMat{-1}[N] \xrightarrow[p_{35}]{\, N\to\infty \,} \;
\begin{blockarray}{cccc}
 \Pscaled=0 & \Pscaled=1  & \hdots &   \\
\begin{block}{(ccc)c}
\begin{matrix}&&\\&\text{\Large{0}}&\\&&\end{matrix} &  \begin{matrix}&&\\&\text{\Large{0}}&\\&&\end{matrix} & \begin{matrix}&&\\& \Large{\hdots}&\\&&\end{matrix}  &  \Pscaled=0 \\
\begin{bmatrix*}[c]
   0.79896858      & -3.49\cdot 10^{-6}  &  -5.407\cdot 10^{-5}  & \hdots \\
  -4.558\cdot 10^{-5}   & 1.80729191     &-2.66\cdot 10^{-6}     &\\
  -1.722\cdot 10^{-5}   &  0.1039298     & 2\cdot 10^{-8}   &\\
\vdots &&&\ddots 
\end{bmatrix*} & \begin{matrix}&&\\&\text{\Large{0}}&\\&&\end{matrix} &  &  \Pscaled=1  \\
\begin{matrix}&&\\&\Large{\vdots}&\\&&\end{matrix} &   &    \begin{matrix}&&\\&\Large{\ddots}&\\&&\end{matrix} &  \vdots  \\
\end{block}
\end{blockarray}
\end{equation}
%
We compare the extrapolation of $(\LnMat{-1})_{v_1,u_1}$ to the conjectured value of $0.79896205$, and we compare the total contribution of the extrapolations of $(\LnMat{-1})_{v_2,u_2} $ and $(\LnMat{-1})_{v_3,u_2}$,  $\sqrt{1.80729191^2+0.1039298^2}=1.81027773$, to the conjectured value of $1.81016041$. All other matrix elements are conjectured to be zero. We see that we overall obtain a precision of at least around $10^{-4}$ by considering system sizes up to $N=80$.



We conclude this section with a note on the six matrix elements that are conjectured to be zero, whose values are small but non-zero at finite size. We shall call such matrix elements ``parasitic couplings''. They play an important role when considering products and commutators of Koo-Saleur generators. Consider a matrix element of the product of two Koo-Saleur generators, which can be decomposed into a sum over all Bethe states as $\langle a | {\mathcal L}_n {\mathcal L}_m |b\rangle = \sum_x \langle a | {\mathcal L}_n | x \rangle \langle x | {\mathcal L}_m |b\rangle $. Even if each parasitic coupling disappears in the limit $N \rightarrow \infty$, the \emph{number} of parasitic couplings in the sum will grow rapidly, and may yield a finite contribution. Until this is further explored, one cannot assume that limits of products give the same results as products of limits. As a particular example, this non-interchangeability of limits applies when the products under consideration form a commutator of two generators. 
Indeed, the issue of limits and commutators was raised already in \cite{KooSaleur}, where it was shown that the limit of commutators must sometimes differ from the commutators of limits. We shall return to this discussion in Section \ref{Anomalies}.


\section{Lattice Virasoro in the degenerate case}\label{PartlyNonGeneric}

In this section we turn to one of our main goals of this paper: finding the precise nature of modules occurring in the XXZ spin chain representation in degenerate cases, possibly also with non-generic $\phi$. 
Compared to the loop representation studied in \cite{LoopPaper}, the XXZ spin chain representation allows for both standard and co-standard Temperley-Lieb modules at non-generic $\phi$. The Virasoro modules in the limit may differ from those found in the loop representation both at generic and non-generic $\phi$. 
%
 Note that only the detailed structure of the representations is affected by the non-genericity and degeneracy: eigenvalues of the Hamiltonian and momentum 
---and thus values of the conformal weights---are  perfectly regular at points where  $\phi$ fulfils \eqref{deg-st-mod} or the conformal weights are degenerate (or, in fact, even when $\q$ is a root of unity).

We here consider the modules where $\alpha$ (or $\bar{\alpha}$) is such that $h$ (or $\bar{h}$) is degenerate. In this section we shall take $x=\pi$ as our type-example of $\q$ generic, but we shall also show convergence of the central charge for a range of values $x\notin\mathbb{Q}$.
Cases where $\q$ is a root of unity will be considered in a later paper \cite{fullynongenericpaper}.

We consider two types of situations where degenerate conformal weights appear: $j\neq 0$, $e^{i\phi}=1$ and $j=0$, $e^{i\phi}=\q^{\pm2}$. Note that for the latter, the resonance criterion $e^{i\phi} = \q^{2j+2k}$---see \eqref{deg-st-mod}---is met, but not for the former. 

%



\subsection{Modules $\AStTL{j}{1}$ for $j \neq 0$}

While the modules $\AStTL{j}{1}$ for $j \neq 0$ remain irreducible for generic $\q$, the generating function of levels (see \eqref{F-func}) reads 
\begin{equation}
\FN_{j,1}={q^{-c/24}\bar{q}^{-c/24}\over P(q)P(\bar{q})}\sum_{e\in \mathbb{Z} } q^{h_{e,-j}}\bar{q}^{h_{e,j}}
\end{equation}
and involves degenerate values of the conformal weights. Let us first consider $S_z=j>0$. As discussed in Section \ref{ConjugateStates}, the chiral weight $h$ will be degenerate for $e<0$, and the corresponding module is conjectured to have the co-Verma structure 
\begin{equation}
\text{$\FF{e,-j}^{\rm (d)}$:}
\begin{tikzpicture}[auto, node distance=0.6cm, baseline=(current  bounding  box.center)]
  \node (node1)[align=center] {$\IrrV{e,-j}$\\$\bullet$ } ;
  \node (node2) [below  = of node1,align=center]	{$\circ$\\$\Verma{e,j}$};
  
  \draw[-latex] (node2) edge  (node1); 
\end{tikzpicture}.
\hspace*{1.5cm}
\end{equation}
Meanwhile  the anti-chiral weight $\bar{h}$ will be degenerate for $e>0$, and the corresponding module is conjectured to have the Verma structure 
\begin{equation}
\text{$\Verma{e,j}^{\rm (d)}$:}
\begin{tikzpicture}[auto, node distance=0.6cm, baseline=(current  bounding  box.center)]
  \node (node1)[align=center] {$\IrrV{e,j}$\\$\circ$ } ;
  \node (node2) [below  = of node1,align=center]	{$\bullet$\\$\Verma{e,-j}$};
  
  \draw[-latex] (node1) edge  (node2); 
\end{tikzpicture}.
\hspace*{1.5cm}
\end{equation}
For $S_z=-j<0$ we find the same conjecture up to a switch of the chiral and anti-chiral sectors.


In Appendix \ref{spin_untwisted} we present the numerical results exploring the modules appearing in the scaling limit of $\AStTL{j}{1}$ for $j=1,2$. 
The results are consistent with the conjectured correspondence between the charges in the Coulomb gas and the lattice parameters; see \eqref{alpha}. Based on these results we claim that we have the general result:

%
\vspace*{5pt}
\noindent\fbox{
\hspace*{5pt}\begin{minipage}{\linewidth-20pt}\em
\vspace*{5pt}
{\bf XXZ spin-chain modules with non-zero magnetization:} For $j>0$ we have the scaling limits
\begin{subequations}
\label{MainRes1}
\begin{eqnarray}
S_z=j;~~ \AStTL{j}{1}\mapsto \left( \bigoplus_{e>0} \Verma{e,-j}\otimes \Verma{e,j}^{\rm (d)} \right) \oplus \big(  \Verma{0,-j}\otimes\Verma{0,j}  \big)   \oplus \left( \bigoplus_{e<0} \FF{e,-j}^{\rm (d)}\otimes \Verma{e,j} \right)  \,, \\
S_z=-j;~~ \AStTL{j}{1}\mapsto \left( \bigoplus_{e>0} \Verma{e,j}^{\rm (d)} \otimes \Verma{e,-j} \right)  \oplus \big(  \Verma{0,j}\otimes\Verma{0,-j}  \big)     \oplus   \left( \bigoplus_{e<0} \Verma{e,j}\otimes \FF{e,-j}^{\rm (d)} \right)  
\,.\end{eqnarray}
\end{subequations}
\vspace*{5pt}
\end{minipage}\hspace*{5pt}}
\vspace*{5pt}


The concise notation means that, for $S_z=j>0$, the states with conformal weights $(h_{e,-j},h_{e,j})$ with $e<0$ are annihilated by the combination of chiral Virasoro generators corresponding to the degenerate conformal weight $h_{e,j}$, while for $e>0$ there appears a null state for the antichiral Virasoro algebra at level $ej$. 
%
%
%
Acting with the lowering operators $A_{e,-j},\bar{A}_{e,j}$ on the primary state $V_{\alpha, \bar{\alpha}}(e,S_z)$ (with charges $\alpha(e,S_z), \bar{\alpha}(e,S_z)$ given by \eqref{alpha} for $e_\phi=0$ and $S_z = \pm j$ as specified) we find
\begin{subequations}
\begin{eqnarray}
S_z=j>0 \ : \  \begin{cases}
\bar{A}_{e,j} V_{\alpha, \bar{\alpha}}(e,j) \neq 0 \,, & \mbox{ for } e>0 \quad (\bar{h} = h_{e,j})  \,, \\
A_{e,-j} V_{\alpha, \bar{\alpha}}(e,j) =0 \,, & \mbox{ for } e<0 \quad (h = h_{e,-j} = h_{-e,j}) \,,
\end{cases}
\end{eqnarray}
while for negative $S_z$ we have instead:
\begin{eqnarray}
S_z=-j <0 \ : \ \begin{cases}
A_{e,j} V_{\alpha, \bar{\alpha}}(e,-j) \neq 0 \,, & \mbox{ for } e> 0 \quad (h=h_{e,j}) \,, \\
\bar{A}_{e,-j} V_{\alpha, \bar{\alpha}}(e,-j) =0 \,, & \mbox{ for } e < 0 \quad (\bar{h} = h_{e,-j} = h_{-e,j} ) 
\,.
\end{cases}
\end{eqnarray}
\end{subequations}
The converse holds when acting with the raising operators $A^\ddag_{e,-j},\bar{A}^\ddag_{e,j}$ on the corresponding level $|ej|$ states, as in the example \eqref{diagram} shown in Section \ref{FF_modules} for $|ej|=1$.

%
%
%
%

We observe  that  
$\Hlatt$ given by \eqref{Pauliham} is invariant under $\q\to \q^{-1}$ and $S_z\to -S_z$. It is also invariant under $\q\to \q^{-1}$ and parity $m \to N+1-m$ (where $m$ denotes the lattice coordinate). 
Thus, we expect that the XXZ modules for $S_z$ and $-S_z$ give rise to modules identical up to an exchange of the chiral and antichiral sectors, in agreement with this discussion.  


\subsection{Modules  $\AStTL{0}{\q^{\pm2}}$}\label{Warmup}

We now switch to the modules $\AStTL{0}{\q^{2}}$  and  $\AStTL{0}{\q^{-2}}$ as defined in \eqref{XXZ_W0q2}. There will be several differences compared to the case of $\AStTL{j}{1}$. First, the resonance criterion \eqref{deg-st-mod} is fulfilled. Second, due to having $S_z=0$, the chiral and anti-chiral sectors will have same charges \eqref{alpha} and play the same role. Third, with $S_z=0$ the lattice momentum \eqref{bethe_momentum} will change by $\Delta \Pscaled = \frac{N}{2}$ whenever the Bethe integers are shifted one step. Since the lattice momentum $\Pscaled$ is defined modulo the system size, as are the Bethe integers themselves, we cannot distinguish an electric excitation $e$ from $-e$ and only the absolute value $|e|$ matters. 


For the module  $\AStTL{0}{\q^{2}}$ we have from \eqref{F-func}
\begin{eqnarray}
\FN_{0,\q^{2}}&=&{q^{-c/24}\bar{q}^{-c/24}\over P(q)P(\bar{q})}\sum_{n\in  \mathbb{Z} }q^{h_{n+1/(x+1),0}}\bar{q}^{h_{n+1/(x+1),0}}\nonumber\\
&=& {q^{-c/24}\bar{q}^{-c/24}\over P(q)P(\bar{q})}\sum_{n\in   \mathbb{Z} }q^{h_{n,1}}\bar{q}^{h_{n,1}} \,.
\end{eqnarray}
%
Note that the dual module $\AStTL{0}{\q^{-2}}$ 
leads to the same generating function of levels in the continuum limit. And yet, the nature of the Virasoro modules is profoundly different in both cases, in accordance with the discussion made in Section~\ref{sec:Bethe}. We find the following result: 
%


\vspace*{5pt}
\noindent\fbox{
\hspace*{5pt}\begin{minipage}{\linewidth-20pt}\em
\vspace*{5pt}
{\bf XXZ spin-chain modules with zero magnetization:} We have the scaling limits
\begin{subequations}
\label{MainRes2}
\begin{eqnarray}
\AStTL{0}{\q^2}\mapsto \left( \bigoplus_{n > 0} \FF{n,1}^{\rm (d)}\otimes \FF{n,1}^{\rm (d)}  \right) \oplus 
\big( \Verma{0,1}\otimes \Verma{0,1} \big) \,, \\
\AStTL{0}{\q^{-2}}\mapsto \left( \bigoplus_{n > 0} \Verma{n,1}^{\rm (d)}\otimes \Verma{n,1}^{\rm (d)} \right) \oplus 
\big( \Verma{0,1}\otimes \Verma{0,1} \big) \,.
\end{eqnarray}
\end{subequations}
\vspace*{5pt}
\end{minipage}\hspace*{5pt}}
\vspace*{5pt}
%
%

\noindent
Note that since we are in the $S_z=0$ sector, 
 we expect the problem to be symmetric under the exchange of chiral and antichiral sectors, in agreement with this conjecture.

We discuss numerical checks of this statement below, but it is (partly) a simple consequence of the structure of the Temperley-Lieb algebra modules themselves, since the ${\mathcal L}_n$ are made out of $e_j$. The numerical results in support of \eqref{MainRes2} are presented in Appendix \ref{spin_twisted}.

\begin{figure}
\center
\includegraphics[height=6cm]{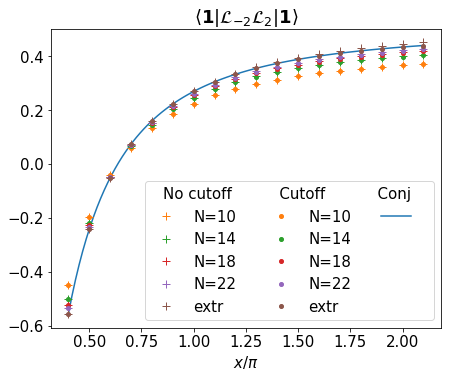} \includegraphics[height=6cm]{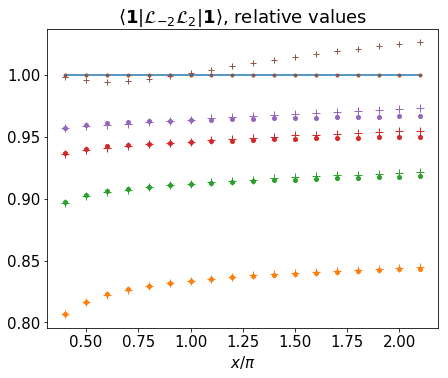}
\caption{Finite-size estimates for the central charge $c$, obtained from \eqref{c_from_L2} for various $N$, plotted against $\frac{x}{\pi}$. The left panel shows the actual estimates for $c$, and the right panel their ratio with respect to the CFT result (``Conj''). We display the results for $N=10,14,\ldots,22$ obtained by two methods, ``No cutoff'' and ``Cutoff'', as explained in the main text. Also shown are extrapolations ``extr'' of the two data sets, obtained from a 7th-order polynomial fit to an extended set of sizes $N=8,10,\ldots,22$.}\label{c_plot}   
\end{figure}

Having shown that the identity state $\mathbf{1}$ belongs to the module $\AStTL{0}{\q^2}$, we can also directly measure the central charge for various values of $x \notin \mathbb{Q}$, through a numerical study of the matrix element
\begin{equation} \label{c_from_L2}
 \langle \mathbf{1} | {\mathcal L}_2 {\mathcal L}_{-2}|\mathbf{1}\rangle = \frac{c}{2} \,.
\end{equation}
While the discussion has so far focussed on matrix elements of a single generator ${\mathcal L}_n$, we encounter here for the first time an example of matrix elements of a product of two generators. To follow up on the issue of ``parasitic couplings''---briefly discussed at the end of Section~\ref{sec:KS_Lminus1}---we employ two different methods to evaluate this matrix element. In the first (denoted ``No cutoff'' in the caption to Figure~\ref{c_plot}), we simply compute the total norm of \eqref{c_from_L2}, whereas in the second (denoted ``Cutoff'') we insert a projector onto the scaling states in-between the product of ${\mathcal L}_2$ and ${\mathcal L}_{-2}$ before computing the norm.
We show in the left panel of Figure \ref{c_plot} how these finite-size estimates converge towards the conjectured CFT value $c=1-24\alpha_0^2$, for $x$ ranging from $4 \pi/10$ to $21 \pi/10$. We also display, in the right panel, the relative convergence, where we have divided through by the conjectured central charge. 
It is seen that although both methods give results close to the conjectured result, only the second one (``Cutoff'') leads to full agreement over the whole range of $x$-values. The reason is that the first method (``No cutoff'') suffers from parasitic couplings to non-scaling states. We defer the full discussion of this crucial phenomenon to Section \ref{Anomalies}. 

\FloatBarrier

\section{Anomalies, and the convergence of the Koo-Saleur generators}\label{Anomalies}

The restriction to scaling states is crucial if one hopes to recover the Virasoro algebra relations \eqref{Virasoro} for the generators ${\mathcal L}_n$, obtained from the Temperley-Lieb algebra by the Koo-Saleur formulae \eqref{generators}. Otherwise---for instance---since the ${\mathcal L}_n$ act on a finite-dimensional Hilbert space for $N$ finite, we would necessarily have that  $\hbox{Tr }[{\mathcal L}_n,{\mathcal L}_{-n}]=0$ (by cyclicity of the trace), preventing the appearance of the central-charge ``anomalous'' term in \eqref{Virasoro}. The procedure to correct this is well known \cite{KacRaina} in free-field theory---e.g., for the Ising model or the free boson---and involves first  restricting to excitations within a certain energy window, then calculating commutators, and finally taking the limit where the energy window goes to infinity.

In other words,  the continuum limit of a commutator is not necessarily the commutator of the continuum limit.  The difference between these two objects arises because of what we will call ``parasitic couplings'', that is, couplings that converge to zero as $N\to\infty$, but are non-zero at finite $N$. In computing, for instance, a matrix element such as \eqref{c_from_L2} we encounter, in principle, an infinity of ``unwanted terms'' where $L_{-2}$ acting on $| \mathbf{1} \rangle$ couples weakly to, in particular, high-energy states, and $L_2$ in turn couples these states back to $\langle \mathbf{1} |$. While we expect each of these unwanted terms to vanish as $N\to \infty$, it is in principle possible that their sum builds up to a finite quantity \cite{Johnson, KacRaina}.

This phenomenon was already observed by Koo and Saleur (KS\footnote{We henceforth refer to equation (x.y) of Koo and Saleur \cite{KooSaleur} in the form KS (x.y).}) \cite{KooSaleur}, whose observations are reproduced and generalized in Figure \ref{c_plot}. We have already discussed the evaluation of the corresponding matrix element \eqref{c_from_L2}, by computing either full norms or projections (denoted ``No cutoff'' and ``Cutoff'' in the figure). Details about precisely how the projected results were obtained will be given below, in section \ref{scaling_weak}.

In KS (3.42) an analytical condition was given under which the central-charge term in \eqref{Virasoro} would be correctly produced for the ${\mathcal L}_n$ generators, despite of the possibility of parasitic couplings. This condition turns out to be satisfied precisely at the values $x=1,2,3$, corresponding respectively to dense polymers, percolation and the Ising model. Close inspection of figure \ref{c_plot} indeed reveals the perfect agreement between full norms and projections 
for $x=1,2,3$ (note that $c=0$ at $x=2$)---however, the agreement is not exact in-between these integers, e.g., at $x=2.5$.

Although the non-generic cases of $x$ integer are not within the scope of the present paper, we here make an exception to briefly comment on the fact that the exact results found in KS can be seen to match our knowledge about the affine TL modules $\bAStTL{0}{\q^{\pm 2}}$ at $x=1,2,3$. For instance, at $x=3$ we observe numerically that there is only one state $V$ at the correct lattice momentum such that both $\langle V|e_i| \mathbf{1} \rangle$ and $\langle \mathbf{1} |e_i|V\rangle$ are non-zero, while at, e.g., $x=4$ there are several. At $x$ generic, the number of such states is observed to be equal to the dimension of the relevant momentum sector of $\bAStTL{0}{\q^{\pm 2}}$, as it should. A more clear-cut example occurs at $x=2$ (i.e., $c=0$), where the identity $| \mathbf{1} \rangle$ is the \emph{only} state in the $[0,\q^2]$ module (in that case, a simple Jones-Temperley-Lieb module of dimension 1), so there clearly is nothing at level 2 to couple to at all. As a consequence, the determination of $c$ by the study of the matrix element \eqref{c_from_L2} is exactly zero for any finite size $N$. Clearly there is much more to be said about these non-generic cases, and this will be the subject of a separate publication \cite{fullynongenericpaper}.


\subsection{``Scaling-weak'' convergence}\label{scaling_weak}

The precise mathematical status of the convergence of the Koo-Saleur generators to the Virasoro generators is clearly a problem beyond the scope of this article. A very conservative statement, which we believe to hold true, is that  1) matrix elements of lattice Virasoro generators converge, when evaluated between scaling states,  to their expected continuum limit and 2)
matrix elements of  {\sl products} of lattice Virasoro generators converge, when evaluated between scaling states, to their expected continuum limit when the products are calculated using only such intermediate states which are scaling states, and using  a double-limit procedure. While the meaning of statement 1) is obvious, the meaning of 2) will be explained in more detail below. 
%
%
We remind the reader that a sequence $f_n$ in a Hilbert space is said to converge weakly to $f$, if the inner products $\langle f_n | g \rangle$ converge to $\langle f | g \rangle$ for all states $g$ in the space. Accordingly we shall refer to the above phenomenon as {\bf scaling-weak convergence}.%
\footnote{This is an example of what is sometimes called ``conditioned weak convergence''.}
It is indeed a weak convergence, as the statement is only for matrix elements, but it is weaker than what is usually called weak convergence because of the restriction to scaling states, in particular in the intermediate states encountered when forming products of operators. In practice, this scaling-weak convergence can be implemented by the  double limit procedure familiar to physicists, as discussed already in \cite{KooSaleur}. We can illustrate it more technically by writing some simple equations. 

As we saw in Figure \ref{c_plot}, computing $\langle \mathbf{1} | L_2L_{-2}|\mathbf{1}\rangle$ via the full norm (``No cutoff'') of the matrix element does not quite give the conjectured result $\frac{c}{2}$.   We can write
\begin{equation}\label{resolve_id}
\langle \mathbf{1} | L_{2}L_{-2} | \mathbf{1} \rangle =\sum_{j=1}^{\mathcal S} \langle \mathbf{1} | L_{2} | v_{(j)} \rangle\langle v_{(j)}|L_{-2} | \mathbf{1} \rangle \,,
\end{equation}
where ${\mathcal S}$ denotes the number of states (in the relevant momentum sector).
We see that even if matrix elements of single Koo-Saleur generators converge towards those of the Virasoro generators, this does not guarantee weak convergence overall. As discussed at the end of Section \ref{sec:KS_Lminus1}, even if each parasitic matrix element $\langle v_{(j)}|L_{-2} | \mathbf{1} \rangle$ in \eqref{resolve_id} converges to zero as $N\rightarrow\infty$, the simultaneous rapid growth of ${\mathcal S}$ can destroy the convergence of the product of generators. To deal with this issue we can consider some fixed cutoff,
\begin{equation}
\sum_{j=1}^{{\mathcal S}_{\rm max}} \langle \mathbf{1} | L_{2} | v_{(j)} \rangle\langle v_{(j)}|L_{-2} | \mathbf{1} \rangle
\end{equation}
(where the intermediate states are supposed to be conveniently ordered),
sending the cutoff ${\mathcal S}_{\rm max} \rightarrow\infty$ \emph{after} taking the scaling limit $N\rightarrow\infty$. The right-hand side of Figure \ref{c_plot} illustrates the most extreme cutoff, where we do not include any parasitic matrix elements at all, that is to say, ``Cutoff'' means that the intermediate states are just the ${\mathcal S}_{\rm max} = 2$ states existing at chiral level 2.

\subsection{A closer look at limits and commutators}\label{limits_section}

A priori, it looks like {\sl any} product of Koo-Saleur generators might be strongly affected by parasitic couplings. We have, however, found serious evidence that only the central charge can come out wrong in calculations. In particular we shall in this section consider commutators, and our belief is that
  the scaling limit of Koo-Saleur commutators {\sl is} the commutator of the scaling limit, {\sl except for the anomalous central charge term}. This is probably expected on general grounds. After all, the Virasoro algebra is just a mode reformulation of the general stress-energy tensor OPE
\begin{equation}
T(z)T(w)={c\over 2(z-w)^4}+{2T(w)\over (z-w)^2}+{\partial T\over z-w}+\hbox{reg.}
\end{equation}
The second term is fixed by the dimension of $T(z)$, which---like for all conserved currents---is not renormalized,
and the third one by consistency under the exchange $z\leftrightarrow w$. Only the first term is anomalous. 
Going back to the original paper by Koo and Saleur \cite{KooSaleur}, there  were some initial checks of how the limit of $[\HN,\PN]$ behaves. In this section we return to this question, exploring this kind of commutator in more detail. Our starting point is KS (3.30), which leads to 
\begin{equation}\label{com_nonzero}
\begin{split}
[\KSgen_{p+n}+\bar{\KSgen}_{-p-n},\KSgen_{-p}-\bar{\KSgen}_{p}] 
= 2 \left({N\over2\pi}\right)^2 \left( \frac{\gamma}{\pi \sin \gamma} \right)^3\Bigg\{ e^{2 i \pi n/N}  \sin \left( \frac{3\pi p + 2\pi n}{N} \right) \sum_{j=1}^{N} e^{2i\pi n j/N}
 [e_j,[e_{j+1},e_{j+2}]] \\
 \quad+ e^{2 i \pi n(1/2)/N}  \sin \left( \frac{\pi p + \pi n}{N} \right) \sum_{j=1}^{N} e^{2i\pi n j/N}
 \sqrt{Q}(e_je_{j+1} + e_{j+1}e_j)  
 -2  \sin \left( \frac{\pi p }{N} \right) \sum_{j=1}^{N} e^{2i\pi n j/N}
 e_{j} 
\Bigg\}.
\end{split}
\end{equation}
If we could exchange freely limits and commutators, the fact that  $\KSgen_n\mapsto L_n$ (resp. $(\bar{\KSgen}_n\mapsto\bar{L}_n$)   would imply that the left-hand side of (\ref{com_nonzero})  converges to 
\begin{subequations}
\label{wishes}
\begin{eqnarray}
 \left[ L_{p+n} + \bar{L}_{-p-n} , L_{-p} - \bar{L}_{p} \right] =  (2p+n) \left( L_n +  \bar{L}_{-n} \right) \,,  \quad \quad &\text{ for } n\neq 0 \,, \label{wishes1} \\
 \left[ L_p + \bar{L}_{-p}, L_{-p} - \bar{L}_{p} \right]=   2p \left( L_0 + \bar{L}_0 - \frac{c}{12} \right) + p^3\frac{c}{6} \,,  \quad \quad &\text{ for } n= 0 \,. \label{wishes2}
\end{eqnarray}
\end{subequations}
It is in fact known that $(\ref{com_nonzero}) \mapsto (\ref{wishes})$ when  $x=1$ \cite{GRS1}, or when  $x=3$ if one restricts to the Ising subspace of the XXZ chain \cite{KooSaleur}. 
It is also known that  $(\ref{com_nonzero}) \not\mapsto (\ref{wishes})$ for other values of $x$  than these and the $x=2$ case (see KS (3.42)): our aim is to investigate in more detail what, exactly, fails.

\subsubsection*{General considerations}

When studying the behaviour of \eqref{com_nonzero} when acting on eigenstates of the lattice translation operator, it is convenient to first recall the discussion in Section \ref{KSandLatticeMomentum}, which yields two facts: 
\begin{itemize}
\item Due to the relative phase $e^{2i\pi nj/N}$, matrix elements of \eqref{com_nonzero} between two states are only non-zero when the lattice momentum $\Pscaled$ of the states differs by precisely $n$.
\item For such non-zero matrix elements, it suffices to evaluate a single summand, say at $j=0$; the matrix element corresponding to the entire sum is then obtained by multiplication with $N$.
\end{itemize}
We shall in the following use the word \emph{term} to refer to the separate contributions for a single, fixed value of $j$. We note that any term that is constant will only contribute when considering matrix elements between a Bethe state and \emph{itself}, since the Bethe states form an orthonormal basis. In particular, this means that whenever $n\neq 0$ we can omit any constant terms, which will allow us to drop normal ordering symbols in the next section. 

We note that by the definition in \eqref{TLmatrix_spin} $e_j$ will act on only two sites in the spin chain, and that the matrix elements are given by $\q$ and $e^{i\phi/N}$---neither of which increases in norm as $N\rightarrow \infty$. As such, the matrix elements of a finite number of Temperley-Lieb generators can at most go towards a constant value as $N \rightarrow \infty$. We shall see that this occurs when matrix elements are between a state and itself---relevant for the $n=0$ case---while they otherwise decrease with $N$. 
%
The numerical results for matrix elements between various states will be shown for the generic example of $x=\pi$. Similar results were found for other generic and non-generic values of $x$. We shall consider matrix elements where the various operators act on either the ground state or one of six low-energy scaling states of various types---electric excitation, magnetic excitation, creating a ``hole'', twisted boundary conditions, and combinations of these. (One of these six excited states is excluded from certain figures, where the matrix elements are exactly zero due to the indecomposability of $\AStTL{0}{\q^{\pm2}}$.)

\begin{figure}[h]
\center
\includegraphics[height=5cm]{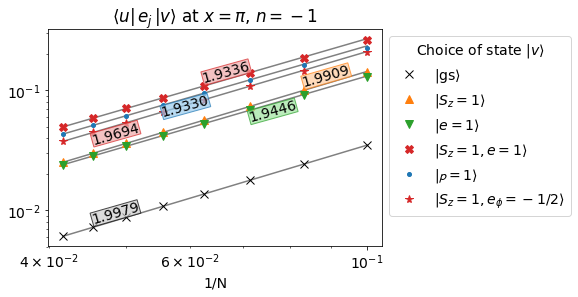}

\includegraphics[height=5cm, trim={0 0 6.7cm 0},clip ]{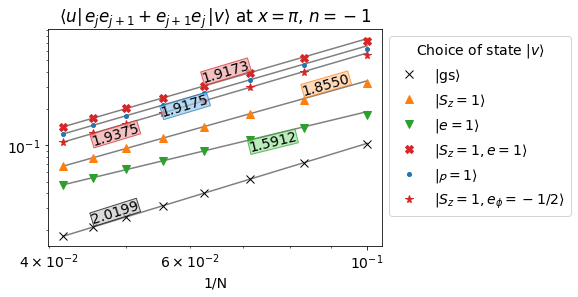}\includegraphics[height=5cm, trim={0 0 6.7cm 0},clip]{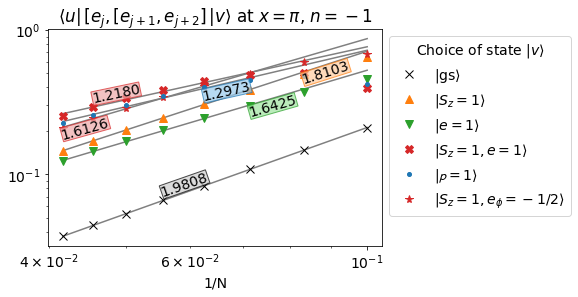}
\caption{Absolute value of matrix elements, plotted with logarithmically scaled axes for sizes $N=10,12,\hdots ,24$. A linear fit is performed, excluding the lowest two sizes. The fit indicates that the matrix elements decay as $\mathcal{O}(1/N^r)$, where the estimated slope $r$ is shown on each curve. The ground state $|\rm{gs}\rangle$ is given for the lattice parameters $S_z=e=e_\phi=\Pscaled=0$. For the other choices of state $|v\rangle$ any non-zero lattice parameter is specified, with the others remaining zero. For each state $|v\rangle$ the state $|u\rangle$ corresponds to the same choice of lattice parameters up to a change in momentum sector, $\Pscaled \to \Pscaled + 1$. The lowest energy state for each choice of lattice parameters is used. The absolute values of the matrix elements are independent of $j$.
}\label{separate_terms}
\end{figure}

\subsubsection{The case of $n \neq 0$}

We now turn back to the comparison between \eqref{com_nonzero} and \eqref{wishes}. The first observation is that the right-hand-side of (\ref{com_nonzero}) contains trigonometric functions of $p/N,n/N$ while the right-hand side of (\ref{wishes}) is linear in $p,n$. While this discrepancy could, by itself, explain why $\eqref{com_nonzero}\not\mapsto (\ref{wishes})$, that is in fact not the case. 
Expanding the sinuses in powers of their arguments (at fixed $p,n$ and with $N$ large),
it turns out that whenever $n\neq 0$ the terms in \eqref{com_nonzero} that are subleading in the expansion fall off fast enough \emph{separately} (this will not be the case at $n=0)$, compared to the leading term $(2p+n)\left\{ 2[e_j,[e_{j+1},e_{j+2}]] + \sqrt{Q}(e_je_{j+1} + e_{j+1}e_j) \right\}$. The leading term by itself goes as $\mathcal{O}(1/N^2)$ since the whole right-hand side of \eqref{com_nonzero} is a term of $\mathcal{O}(1)$. Since the subleading terms come with prefactors $1/N^2, 1/N^4,\ldots$ relative to this leading term, it is enough that $[e_j,[e_{j+1},e_{j+2}]]$, $(e_je_{j+1} + e_{j+1}e_j)$ and $e_j$ on their own have matrix elements that decay with $N$ (i.e., being of order $\mathcal{O}(1/N^r)$ with $r>0$). If they furthermore decay faster than $1/N$ ($r>1$) we may also ignore the subleading terms in the expansion of exponentials in $n/N$, so that we may put all phase factors on the same form. 
Numerical results indicate that this indeed holds: the case of $n=-1$ is shown in Figure \ref{separate_terms}, with slopes around $r=2$, while the case of $n=-2$ is shown in Figure \ref{allthree_2ab} in Appendix \ref{limits_appendix}---here the finite size effects are larger, but the results still indicate $r>1$.
%
%
Keeping terms of leading order only, we have 
\begin{multline}\label{com_leading}
[\KSgen_{p+n}+\bar{\KSgen}_{-p-n},\KSgen_{-p}-\bar{\KSgen}_{p}] 
=   \frac{N }{2\pi} \left( \frac{\gamma}{\pi \sin \gamma} \right)^3\Bigg\{ (3p+2n) 
\sum_{j=1}^{N}  e^{2i\pi n (j+1) /N}
 [e_j,[e_{j+1},e_{j+2}]] 
 \\
+  (p+n) \sqrt{Q}\sum_{j=1}^{2L} e^{2i\pi n (j+1) /N}
(e_je_{j+1} + e_{j+1}e_j)   
  -2  p \sum_{j=1}^{N} e^{2i\pi n (j+1) /N}
 e_{j} 
\Bigg\}+\mathcal{O}\left({1\over N^{r-1}}\right) \,.
\end{multline}
We can now analyze the right-hand side by using a family of operators introduced in \cite{KooSaleur} as alternatives to the basic generators $\KSgen,\bar{\KSgen}$\footnote{These alternatives  were obtained by using the fact that all derivatives of the logarithm of the transfer matrix with respect to the spectral parameter $u$ at $u=0$ produce terms converging, in the weak-scaling sense, to the Hamiltonian and momentum of the associated CFT.}. From the expression of  what is denoted  $\hat{h}^{(3)}$ in KS (2.54) and using the result KS (2.58)\footnote{Generalized to $n\neq 0$ in the same way as in KS (3.33).} we have that 
%
%
%
\begin{equation}\label{h3_normord}
{N\over2\pi}\left({\gamma\over\pi\sin\gamma}\right)^3 \sum_{j=1}^{N}e^{2i\pi n(j+1)/N} \normord{2[e_j,[e_{j+1},e_{j+2}]] + \sqrt{Q}(e_je_{j+1} + e_{j+1}e_j)}~  \mapsto \left( L_n + \bar{L}_{-n}  -\frac{c}{12}\delta_{n,0} \right) \,,
\end{equation}
where normal-order notation here  means that  the average in the ground state has been subtracted.
As discussed at the beginning of this section, this normal ordering does not change the matrix elements when $n \neq 0$, and we shall use the left-hand side of (\ref{h3_normord}) without normal ordering for the rest of this subsection.

Grouping by $p$ and $n$ and dividing   through by $\frac{N}{2\pi}\left( \frac{\gamma}{\pi \sin \gamma} \right)^3$ we have from \eqref{com_leading} the terms
\begin{multline}
2p \Bigg[  \frac{3}{2}\sum_{j=1}^{N}  e^{2i\pi n (j+1) /N}
{ [e_j,[e_{j+1},e_{j+2}]] } 
  + \frac{\sqrt{Q}}{2}  \sum_{j=1}^{N} e^{2i\pi n (j+1) /N}
{e_je_{j+1} + e_{j+1}e_j } 
 - \sum_{j=1}^{N} e^{2i\pi n (j+1)/N} {e_{j}   } 
   \Bigg]\\
  + n \Bigg[
   2 \sum_{j=1}^{N}  e^{2i\pi n (j+1) /N}
{ [e_j,[e_{j+1},e_{j+2}]]} + \sqrt{Q} \sum_{j=1}^{N} e^{2i\pi n (j+1) /N}
{e_je_{j+1} + e_{j+1}e_j} 
  \Bigg] \,, \label{bigeq1}
\end{multline}
from which we subtract 
\begin{multline}
2p \Bigg[ 2\sum_{j=1}^{N}  e^{2i\pi n (j+1) /N}
{ [e_j,[e_{j+1},e_{j+2}]] } 
  +  \sqrt{Q}\sum_{j=1}^{N} e^{2i\pi n (j+1) /N}
{e_je_{j+1} + e_{j+1}e_j} 
   \Bigg]\\
  + n \Bigg[
   2 \sum_{j=1}^{N}  e^{2i\pi n (j+1) /N}
{ [e_j,[e_{j+1},e_{j+2}]] }+ \sqrt{Q}\sum_{j=1}^{N} e^{2i\pi n (j+1) /N}
{e_je_{j+1} + e_{j+1}e_j } \label{bigeq2}
  \Bigg] \,.
\end{multline}
We thus find, thanks to  (\ref{h3_normord}),  that in the scaling limit $
[\KSgen_{p+n}+\bar{\KSgen}_{-p-n},\KSgen_{-p}-\bar{\KSgen}_{p}]$  behaves as $L_n+\bar{L}_{-n}$ (recall, $n\neq 0$ in this section) plus the following expression: 
\begin{multline}\label{rest_sum}
\frac{N}{2\pi}\left({\gamma\over\pi\sin\gamma}\right)^32p\Bigg[ -\frac{1}{2} \sum_{j=1}^{N} e^{2i\pi n (j+1) /N}
 {[e_j,[e_{j+1},e_{j+2}]]} \\
 - \frac{ \sqrt{Q}}{2}  \sum_{j=1}^{N} e^{2i\pi n (j+1) /N}
 { e_je_{j+1} + e_{j+1}e_j}  - \sum_{j=1}^{N} e^{2i\pi n (j+1)/N} {e_{j}} \Bigg] \,.
\end{multline}
In order to have $\eqref{com_nonzero} \mapsto \eqref{wishes}$ at $n\neq 0$, we need  the matrix elements of (\ref{rest_sum}) between any two relevant scaling states ($\Delta \Pscaled = n$) to tend to zero with $N$. 
As discussed in the beginning of this section, it suffices to evaluate a single summand for a fixed value of $j$. For this reason we define the \emph{remainder} $\remainder$ at $n\neq 0$ to be 
%
%
%
%
\begin{equation}\label{rest}
 \remainder = - {[e_j,[e_{j+1},e_{j+2}]] }
 -  \sqrt{Q} ( {e_je_{j+1} + e_{j+1}e_j} ) -  2 {e_{j} }.
\end{equation}
%
In order for the matrix elements of (\ref{rest_sum}) between scaling states to tend to zero with $N$,
we need that matrix elements of $\remainder$ be of $\mathcal{O}(1/N^r)$ for $r$ strictly larger than 2. 
Results for $n=-1$ are shown in Figure \ref{restterm_lvl12}, while results for $n=-2$ are shown in Figure \ref{remainder_2ab} in Appendix \ref{limits_appendix}. These two cases seem to have $r>2$ indeed, which would indicate that the limit of the commutator is indeed the commutator of the limits for the states under consideration. Therefore, we have evidence for the following
materialization of \eqref{wishes1}: 

\vspace*{5pt}
\noindent\fbox{
\hspace*{5pt}\begin{minipage}{\linewidth-20pt}\em
\vspace*{5pt}
{\bf Exchange of commutators and limits:} We have the conjecture
\begin{equation} \label{ndiff0_conj}
[\KSgen_{p+n}+\bar{\KSgen}_{-p-n},\KSgen_{-p}-\bar{\KSgen}_{p}] \mapsto (2p+n) (L_n+\bar{L}_{-n}) \,, \quad \mbox{for } n\neq 0 \,.
\end{equation}
\vspace*{5pt}
\end{minipage}\hspace*{5pt}}
\vspace*{5pt}

\noindent

\begin{figure}[h]
\center
\includegraphics[height=5cm]{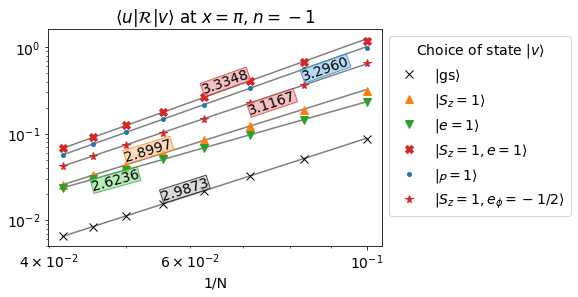}

\caption{Absolute value of matrix elements of $\remainder$ as defined in \eqref{rest}, plotted using the same conventions as in Figure \ref{separate_terms}. %
%
}\label{restterm_lvl12} 
\end{figure}

\FloatBarrier

\subsubsection{The case of $n= 0$}

\begin{figure}[h]
\center
\includegraphics[height=5cm]{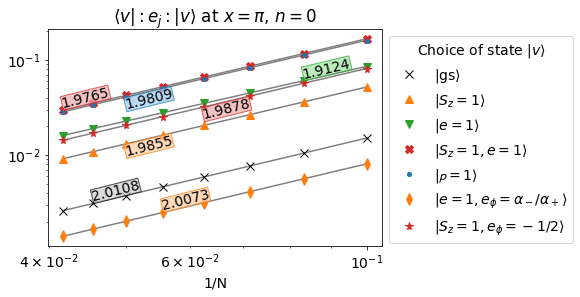}

\includegraphics[height=5cm, trim={0 0 6.8cm 0},clip ]{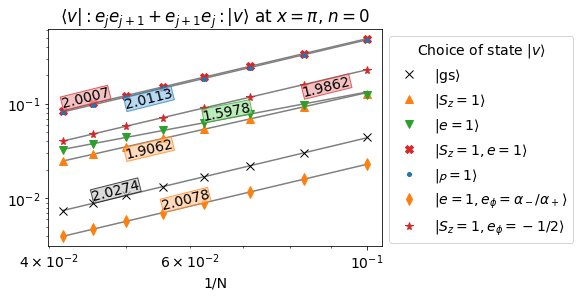}\includegraphics[height=5cm, trim={0 0 6.8cm 0},clip]{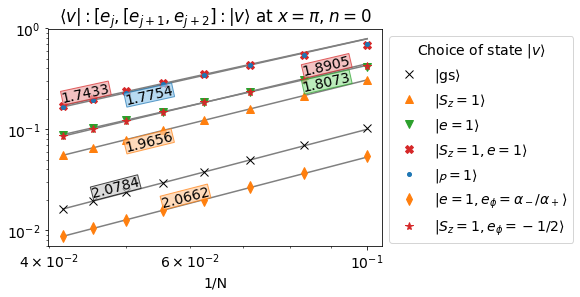}
\caption{
Absolute value of matrix elements, plotted using the same conventions as in Figure \ref{separate_terms} but with the choice of $|u\rangle = |v \rangle$. The operators are here normal ordered: $\normord{\mathcal{O}} \; \equiv O - O_\infty$ with $O_\infty$ being the ground state expectation value as $N \to \infty$. This value is given for $e_j$ in \eqref{e_inf}, and for $e_je_{j+1} + e_{j+1}e_j$ and $[e_j,[e_{j+1},e_{j+2}]]$ in \eqref{separate_conj}. 
%
}\label{allthree_n0} 
\end{figure}

In the case of $n=0$, however, we do \emph{not} obtain the falloff for the individual terms seen in Figure \ref{separate_terms}. Indeed, to stand any chance of recovering the desired central-charge terms in \eqref{wishes2} we must have constant contributions. In Figure \ref{allthree_n0} we see that the matrix elements of the individual terms between the ground state and itself indeed seem to stay constant as $N$ increases, while they decrease as before between the ground state and the first excited state within the same momentum sector (see Figure~\ref{allthree_0b} in Appendix \ref{limits_appendix}).

While we must now consider both linear and cubic terms in the expansion of the sinuses, we can still ignore the quintic and higher terms. We also no longer need to expand any exponentials. This means that most of the derivation of $\remainder$ above is still valid, as long as we apply it to the  \emph{linear} term in $n$. There is only one modification we need to make, compared to \eqref{rest}: When $n=0$ the normal ordering of \eqref{h3_normord} becomes important, and we need to subtract from $2[e_j,[e_{j+1},e_{j+2}]] + \sqrt{Q}(e_je_{j+1} + e_{j+1}e_j)$ its ground-state expectation value $ 4\sin^3 \! \gamma \: I_1$, 
where  the constant $I_1$ follows from the Bethe ansatz \cite{KooSaleur}
\begin{equation} \label{I1-def}
I_1=\int_{-\infty}^\infty t^2 {\sinh(\pi-\gamma)t\over \sinh(\pi t)\cosh(\gamma t)} \, {\rm d}t \,.
\end{equation}
Taking this into account we define the remainder at $n=0$ as
\begin{equation}\label{rest_n0}
 \remainder = - [e_j,[e_{j+1},e_{j+2}]] 
 -   \sqrt{Q} (e_je_{j+1} + e_{j+1}e_j) -  2 e_{j} 
 + 8\sin^3 \! \gamma \: I_1 \,.
\end{equation}

Including the cubic terms in the expansion of the trigonometric functions in \eqref{com_nonzero} yields
\begin{multline}\label{com_n0}
[\KSgen_{p}+\bar{\KSgen}_{-p},\KSgen_{-p}-\bar{\KSgen}_{p}]  
=   \frac{N }{2\pi} \left( \frac{\gamma}{\pi \sin \gamma} \right)^3p~\Bigg\{ 3 \sum_{j=1}^{N} 
 [e_j,[e_{j+1},e_{j+2}]] 
 +  \sum_{j=1}^{N} 
 \Big( \sqrt{Q}(e_je_{j+1} + e_{j+1}e_j) - 2 e_{j} \Big)
\Bigg\}
\\
-  \frac{1}{12}\frac{\pi}{N}  \left( \frac{\gamma}{\pi \sin \gamma} \right)^3p^3~\Bigg\{27 \sum_{j=1}^{N} 
 [e_j,[e_{j+1},e_{j+2}]] 
 +  \sum_{j=1}^{N} 
 \Big( \sqrt{Q}(e_je_{j+1} + e_{j+1}e_j) - 2 e_{j} \Big)
\Bigg\} +\mathcal{O}\left({1\over N^{2}}\right) \,.
\end{multline}
%

\begin{figure}[h]
\center
\includegraphics[height=5cm]{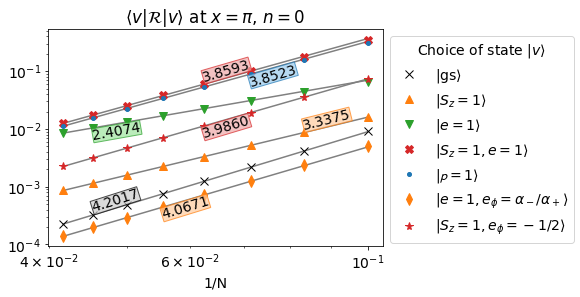}
\caption{Absolute value of matrix elements of $\remainder$ as defined in \eqref{rest}, plotted using the same conventions as in Figure \ref{separate_terms} but with the choice of $|u\rangle = |v \rangle$.}\label{restterm_plot}
\end{figure}

To see whether the term linear in $p$ converges to the desired value we consider matrix elements of \eqref{rest_n0} numerically as shown in Figure \ref{restterm_plot}, which indicates a falloff ${\cal O}(1/N^r)$ for $\langle gs | \remainder | gs \rangle$ with an exponent $r$ around 4.  We show similar figures for the first few excitations above the ground state in Appendix \ref{limits_appendix}. There the value of the slope $r$ in the limit is more unclear due to larger finite-size effects, but the results still indicate $r>2$ for these states. 

We now turn our attention to the term $\propto p^3$ in \eqref{com_n0}, which reads
\begin{equation}\label{pcubeterm}
-  \frac{1}{12}\frac{\pi}{N}  \left( \frac{\gamma}{\pi \sin \gamma} \right)^3p^3~\Bigg\{27 \sum_{j=1}^{N} 
 [e_j,[e_{j+1},e_{j+2}]] 
 +  \sum_{j=1}^{N} 
 \Big( \sqrt{Q}(e_je_{j+1} + e_{j+1}e_j) - 2 e_{j} \Big)
\Bigg\} \,,
\end{equation}
and evaluate its average in the ground state.  From  our study of the $\propto p$ part of \eqref{com_n0}, we know already that  the contribution from the terms  
$\mathcal{E}_p \equiv \langle gs| 3  [e_j,[e_{j+1},e_{j+2}]]+ \sqrt{Q}(e_je_{j+1} + e_{j+1}e_j) - 2 e_{j} |gs\rangle$  vanishes as $N\rightarrow\infty$. This leaves the contribution from $24  [e_j,[e_{j+1},e_{j+2}]]$ in \eqref{pcubeterm} and therefore, in order for $(\ref{com_nonzero}\mapsto (\ref{wishes})$ also for $n=0$, we would need the following miraculous identity to hold
\begin{subequations}
\label{c=cstar}
\begin{equation}\label{three_38}
 \mathcal{E}_{p^3} \equiv [e_j,[e_{j+1},e_{j+2}]]_\infty + \frac{\pi^2 c}{12}\left( \frac{\sin\gamma}{\gamma}\right)^3 \overset{?}{=} 0 \,,
\end{equation}
where we use the notation $[e_j,[e_{j+1},e_{j+2}]]_\infty=\lim_{N\to\infty} \langle gs|[e_j,[e_{j+1},e_{j+2}]]|gs\rangle$. 
Using results from the Bethe ansatz, it  was shown in \cite{KooSaleur}  (section KS 3.3) that \eqref{three_38} is equivalent to
\begin{equation}
 c \overset{?}{=} c^* \,,
\end{equation}
\end{subequations}
where we have defined
\begin{equation}\label{three_42}
c^* =  - \frac{24 \gamma^3I_0}{\pi^2 \sin^2 \! \gamma}
+ \frac{48 \gamma^3 }{\pi^2} I_1.
\end{equation}
It was furthermore shown  in \cite{KooSaleur}  (section  KS 3.3)  that \eqref{c=cstar} holds true indeed when $x=1,2,3$, but not for general $x$. We conjecture that $\mathcal{E}_{p^3}$ will make up for the difference between the left- and right-hand side of criterion \eqref{c=cstar}, that is precisely 
\begin{equation} \label{Ep3conjecture}
 \mathcal{E}_{p^3} = \frac{\pi^2}{12} \left( \frac{\sin \gamma}{\gamma} \right)^3 (c - c^*) \,.
\end{equation}
The numerics indicate that this is true: the extrapolated values of $\mathcal{E}_{p^3} $ are shown for a range of $x$ in Table \ref{difference_c}. 
That is, the limit of \eqref{com_n0} when applied to the ground state is incorrect only for the $p^3$ term, the one that contains the central charge.

\begin{table}[h]
\begin{minipage}{0.25\linewidth}
$x=1$:

\begin{tabular}{ll}
$N$ & $\mathcal{E}_{p^3}$ \\
\hline
6 & -0.09108037 \\
8 & -0.04173017 \\
10 & -0.02441356 \\
12 & -0.01616525 \\
14 & -0.01154442 \\
16 & -0.00867872 \\
18 & -0.00677235 \\
20 & -0.00543712 \\
22 & -0.00446415 \\
24 & -0.00373254 \\
$p_8$ & \phantom{-}0.0000006\\
\bf{conj} &  \phantom{-}0\\
\hline
\end{tabular}
\end{minipage}%
\begin{minipage}{0.25\linewidth}
$x=1.5$:

\begin{tabular}{ll}
$N$ & $\mathcal{E}_{p^3}$ \\
\hline
6 & -0.18750921 \\
8 & -0.08790715 \\
10 & -0.05204436 \\
12 & -0.03471144 \\
14 & -0.0249148 \\
16 & -0.01880414 \\
18 & -0.01472315 \\
20 & -0.0118568 \\
22 & -0.00976384 \\
24 & -0.00818762 \\
$p_8$ & -0.00011095 \\
\bf{conj} &  -0.00011266 \\
\hline
\end{tabular}
\end{minipage}%
\begin{minipage}{0.25\linewidth}
$x=2$:

\begin{tabular}{ll}
$N$ & $\mathcal{E}_{p^3}$ \\
\hline
6 & -0.26295146 \\
8 & -0.12396495 \\
10 & -0.07352658 \\
12 & -0.04905309 \\
14 & -0.03519046 \\
16 & -0.02653245 \\
18 & -0.02074547 \\
20 & -0.0166787 \\
22 & -0.01370811 \\
24 & -0.01147035 \\
$p_8$ & \phantom{-}0.0000024\\
\bf{conj} &  \phantom{-}0\\
\hline
\end{tabular}
\end{minipage}%
\begin{minipage}{0.25\linewidth}
$x=2.5$:

\begin{tabular}{ll}
$N$ & $\mathcal{E}_{p^3}$ \\
\hline
6 & -0.3182588 \\
8 & -0.15047958 \\
10 & -0.08932876 \\
12 & -0.05959601 \\
14 & -0.04273623 \\
16 & -0.03220009 \\
18 & -0.0251554 \\
20 & -0.02020382 \\
22 & -0.01658653 \\
24 & -0.01386144 \\
$p_8$ & \phantom{-}0.00010867\\
\bf{conj} &  \phantom{-}0.00010568\\
\hline
\end{tabular}
\end{minipage}%

~\\
~\\
\begin{minipage}{0.25\linewidth}
$x=3:$

\begin{tabular}{ll}
$N$ & $\mathcal{E}_{p^3}$ \\
\hline
6 & -0.3587194 \\
8 & -0.1700361 \\
10 & -0.10107585 \\
12 & -0.06750028 \\
14 & -0.04844759 \\
16 & -0.03653616 \\
18 & -0.02857005 \\
20 & -0.02297004 \\
22 & -0.01887873 \\
24 & -0.01579641 \\
$p_8$ & \phantom{-}0.00000334\\
\bf{conj} &  \phantom{-}0\\
\hline
\end{tabular}
\end{minipage}%
\begin{minipage}{0.25\linewidth}
$x=4:$

\begin{tabular}{ll}
$N$ & $\mathcal{E}_{p^3}$ \\
\hline
6 & -0.41165366 \\
8 & -0.19603869 \\
10 & -0.11698482 \\
12 & -0.07843206 \\
14 & -0.05653502 \\
16 & -0.04283788 \\
18 & -0.03367441 \\
20 & -0.0272313 \\
22 & -0.02252338 \\
24 & -0.01897619 \\
$p_8$ & -0.00079389 \\
\bf{conj} &  -0.00079737 \\
\hline
\end{tabular}
\end{minipage}%
\begin{minipage}{0.25\linewidth}
$x=6$:

\begin{tabular}{ll}
$N$ & $\mathcal{E}_{p^3}$ \\
\hline
6 & -0.46366442 \\
8 & -0.22244484 \\
10 & -0.13377148 \\
12 & -0.09046753 \\
14 & -0.06585191 \\
16 & -0.05044634 \\
18 & -0.04013642 \\
20 & -0.0328855 \\
22 & -0.02758643 \\
24 & -0.02359336 \\
$p_8$ & -0.00312128 \\
\bf{conj} & -0.00312477\\
\hline
\end{tabular}
\end{minipage}%
\begin{minipage}{0.25\linewidth}
$x=10 \pi:$

\begin{tabular}{ll}
$N$ & $\mathcal{E}_{p^3}$ \\
\hline
6 & -0.52432884 \\
8 & -0.25560287 \\
10 & -0.15658134 \\
12 & -0.10816223 \\
14 & -0.08061811 \\
16 & -0.06337137 \\
18 & -0.05182545 \\
20 & -0.04370333 \\
22 & -0.03776655 \\
24 & -0.03329234 \\
$p_8$ & -0.01034513\\
\bf{conj} & -0.01034891\\
\hline
\end{tabular}
\end{minipage}%

\caption{Comparison of the numerical measures for $\mathcal{E}_{p^3}$, defined in \eqref{three_38}, and its conjectured value which is the right-hand side of \eqref{Ep3conjecture}. The agreement between the extrapolated values and the conjecture is seen to be excellent, with a precision of the order $10^{-6}$ for all values of $x$. 
The same conventions as in Table \ref{generic_example} are used for the extrapolated values $p_8$.
}\label{difference_c}
\end{table}

$\mathcal{E}_{p^3}$ satisfying the conjecture \eqref{Ep3conjecture} is equivalent to the ground-state expectation value of $ [e_j,[e_{j+1},e_{j+2}]] $ being equal to $2\sin\gamma I_0-4\sin^3\gamma I_1$. Combining this with the result KS (3.41)
\begin{equation} \label{KS-3-41}
[e_j,[e_{j+1},e_{j+2}]]_\infty+\frac{\sqrt{Q}}{2}(e_je_{j+1}+e_{j+1}e_j)_\infty=2\sin^3 \! \gamma \: I_1 \,.
\end{equation}
we can improve \eqref{KS-3-41} by conjecturing the values of each of its terms separately:

\vspace*{5pt}
\noindent\fbox{
\hspace*{5pt}\begin{minipage}{\linewidth-20pt}\em
\vspace*{5pt}
{\bf Exchange of commutators and limits:} We have the results
\begin{subequations} \label{separate_conj}
\begin{eqnarray}
 [e_j,[e_{j+1},e_{j+2}]]_\infty&=&2\sin\gamma \: I_0-4\sin^3 \! \gamma \: I_1 \,,  \label{separate_conj_a}\\
 (e_je_{j+1} + e_{j+1}e_j)_\infty&=&{6\sin^3 \! \gamma \: I_1-2\sin\gamma \: I_0\over \cos\gamma} \,, \label{separate_conj_b}
\end{eqnarray}
\end{subequations}
where the integral $I_0$ is defined by \eqref{I0-def} and $I_1$ by \eqref{I1-def}.
\vspace*{5pt}
\end{minipage}\hspace*{5pt}}
\vspace*{5pt}

\noindent
In Appendix \ref{Proof_Appendix} we prove these conjectures (so they are actually theorems), by using known ground state expectation values of spin operators in the XXZ spin chain\cite{Go_Kato}.  In Appendix \ref{gamma_zero_check} we consider the limit $\gamma \to 0$ corresponding to $x \to \infty$, showing that the integrals in \eqref{separate_conj} take the form of polylogarithms in agreement with the known results for ground state expectation values of spin operators in the XXX spin chain\cite{Takahashi}. 
We note that the limit $\gamma\to 0$ is where the theory is the most interacting, and where the anomaly is the largest ($c=1$). Solving the integrals within \eqref{Ep3conjecture} numerically for increasingly large finite values of $x$ indicates that $\mathcal{E}_{p^3}$ converges towards its limit of $\mathcal{E}_{p^3}(x)\big|_{x\to\infty}=-0.011\,114\,954\cdots$ from above, meaning that the magnitude of $\mathcal{E}_{p^3}$ is the largest in this limit. We also see that the effect of parasitic couplings in Figure \ref{c_plot} is the most pronounced at large $x$.  
 


We next turn to the first few excited states. To have the same deviation for the central term we would need the matrix elements $\langle v |  [e_j,[e_{j+1},e_{j+2}]] | v \rangle $, for $|v \rangle$ any scaling state, to go towards the same value as for the ground state---that is, we need $\langle v | \normord{[e_j,[e_{j+1},e_{j+2}]]} | v\rangle $ to go to zero. This matrix element is shown in Figure \ref{allthree_n0}, where the conjectured ground state expectation value in \eqref{separate_conj} is used for the normal ordering. We see that it indeed tends to zero as $N \rightarrow \infty$ for all scaling states under consideration, which indicates that the central term is wrong by a \emph{constant} term, rather than by an operator. This constant deviation corresponds precisely to replacing $c$, as given by \eqref{c_value}, by the slightly different value $c^*$ given in \eqref{three_42} in the cubic term $\propto p^3$. Altogether we have:

\vspace*{5pt}
\noindent\fbox{
\hspace*{5pt}\begin{minipage}{\linewidth-20pt}\em
\vspace*{5pt}
{\bf Exchange of commutators and limits:} We have the conjecture
\begin{equation}\label{cstar_conjecture}
[\KSgen_{p}+\bar{\KSgen}_{-p},\KSgen_{-p}-\bar{\KSgen}_{p}] \mapsto 2p \left(L_0+\bar{L}_{0}-{c\over 12} \right)+ p^3 \frac{c^*}{6}
\end{equation}
\vspace*{5pt}
with $c$ given by \eqref{c_value} and $c^*$ given by \eqref{three_42}.
\end{minipage}\hspace*{5pt}}
\vspace*{5pt}

\begin{figure}[h]
\center

\begin{tikzpicture}

\node (A) at (0.9,0) {\includegraphics[height=6cm]{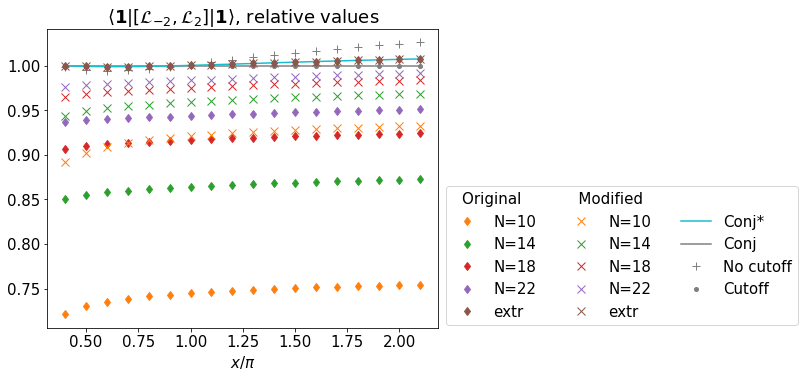}};
\node (B) at (5.2,1.5) {\includegraphics[height=2.5cm]{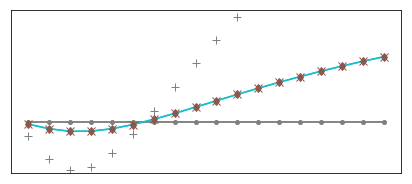}};
\draw[dotted,thick] (1.57,2.26) -- (2.48,2.55);
\draw[dotted,thick] (1.57,1.8) -- (2.48,0.5);
    
\end{tikzpicture}

\caption{Remake of Figure \ref{c_plot}. In this remake the commutator $[\KSgen_{-2},\KSgen_2]$ is used, rather than the product $\KSgen_{-2}\KSgen_2$, and no cutoff is taken. This commutator is denoted ``Original''. We also show the result using a modified version of the Koo-Saleur generators given by \eqref{ZW_generators}, labelled ``Modified''. 
``Conj*'' refers to the result found by applying \eqref{modified_Virasoro}, while ``Conj'' refers to the CFT result as in Figure \ref{c_plot}. 
The same sizes $N$ and method of extrapolation is used as in Figure \ref{c_plot}, and the extrapolated values from Figure \ref{c_plot} are shown for comparison (``No cutoff'' and ``Cutoff''). 
}\label{cstar_plot}
\end{figure}



Having shown by the combination of \eqref{ndiff0_conj} and \eqref{cstar_conjecture} that $[\KSgen_{p+n}  +\bar{\KSgen}_{-p-n},\KSgen_{-p}-\bar{\KSgen}_{p}]$ is conjectured to have the correct limit up to the central term, it is natural to ask if we can eliminate the chiral-antichiral ``cross-terms'' and write the relation for the limit of $[\KSgen_{p+n} ,\KSgen_{-p}]$ on its own. In other words: is the limit of the chiral-antichiral commutator $[\KSgen_{p+n},\bar{\KSgen}_{p}]$ zero? In Appendix \ref{Chiral-Antichiral} we explore this question using the same methodology as in this section, and show that numerical evidence indeed indicates that the chiral-antichiral commutator vanishes in the limit:

\vspace*{5pt}
\noindent\fbox{
\hspace*{5pt}\begin{minipage}{\linewidth-20pt}\em
\vspace*{5pt}
{\bf Exchange of commutators and limits:} We have the conjecture
\begin{equation}\label{chiral-antichiral_conj}
[\KSgen_{p+n},\bar{\KSgen}_{p}] \mapsto 0.
\end{equation}
\vspace*{5pt}
\end{minipage}\hspace*{5pt}}
\vspace*{5pt}
%
%
%
%
%
%

\noindent
Combining the three conjectures \eqref{ndiff0_conj}, \eqref{cstar_conjecture} and \eqref{chiral-antichiral_conj} we thus obtain in the scaling limit a modification of the Virasoro commutation relations \eqref{Virasoro} on the form
\begin{equation} \label{modified_Virasoro}
 [{\mathcal L}_m,{\mathcal L}_n] = (m-n){\mathcal L}_{m+n} + \delta_{m+n,0}\frac{1}{12}(m^3c^*-mc) \,.
\end{equation}
We can remake Figure \ref{c_plot} using this new conjecture. This is done in Figure \ref{cstar_plot}, where it is shown that the effect of the ``parasitic couplings'' is in agreement with \eqref{modified_Virasoro}. We stress that this conjecture applies for the \emph{commutator} $[{\mathcal L}_2,{\mathcal L}_{-2}]$ and not for the product ${\mathcal L}_2 {\mathcal L}_{-2}$ alone, as is seen in the figure. We conclude that the limit of commutators is the same as the commutator of limits up to a modification of the central term.

In Figure \ref{cstar_plot} we also show the modification recently suggested by Shokrian-Zini and Wang in \cite{Wang} (Conjecture 5.5), which amounts to changing the phases within the Koo-Saleur generators \eqref{generators} to
\begin{subequations}
\label{ZW_generators}
\begin{eqnarray}
\KSgen_n[N] \!\!\! &=& \!\!\! \frac{N}{4\pi} \left[ -\frac{\gamma}{\pi \sin \gamma}  \sum^{N}_{j=1} e^{in(j+1/2)2\pi/N} \left( e_j - e_\infty \right) + e^{in(j+1)2\pi/N} \frac{i\gamma}{\pi \sin \gamma} [e_j, e_{j+1} ] \right] + \frac{c}{24} \delta_{n,0} \,, \\
\bar{\KSgen}_n[N] \!\!\! &=& \!\!\! \frac{N}{4\pi} \left[ -\frac{\gamma}{\pi \sin \gamma}  \sum^{N}_{j=1} e^{-in(j+1/2)2\pi/N} \left( e_j-  e_\infty \right) - e^{-in(j+1)2\pi/N} \frac{i\gamma}{\pi \sin \gamma} [e_j, e_{j+1} ] \right] + \frac{c}{24} \delta_{n,0} \,.
\end{eqnarray}
\end{subequations}
We find that while the numerical values show a faster convergence, the change of the central term is the same as with the original Koo-Saleur generators. That the central term must behave in the same way indeed follows directly from the calculations and numerical evidence above, where we in particular find that we may shift the phase by a finite amount without affecting the limit.

Although the relation in \eqref{modified_Virasoro} is no longer the expected relation of the Virasoro algebra at central charge $c$, the Jacobi identity would nevertheless be satisfied:
\begin{align}
& [{\mathcal L}_m,[{\mathcal L}_n,{\mathcal L}_p]] + [{\mathcal L}_n,[{\mathcal L}_p,{\mathcal L}_m]] + [{\mathcal L}_p,[{\mathcal L}_m,{\mathcal L}_n]] \nonumber \\
& \qquad \qquad =(n-p)[{\mathcal L}_m,{\mathcal L}_{n+p}] + (p-m)[{\mathcal L}_n,{\mathcal L}_{p+m}] + (m-n)[{\mathcal L}_p,{\mathcal L}_{m+n}] \nonumber \\
& \qquad \qquad =\delta_{m+n+p,0} \frac{1}{12} \Big\{
 (n-p) (m^3 c^* - mc)
 +(p-m) (n^3 c^* - nc)
 +(m-n) (p^3c^* -pc)
 \Big\} \nonumber \\
 & \qquad \qquad = 0 \,.
\end{align}
Here, the first equality follows from the fact that all ${\mathcal L_{m+n+p}}$ terms cancel; the second equality is the application of \eqref{modified_Virasoro}; and to establish the third equality, note all terms proportional to $c$ inside the parenthesis cancel out for any values of $m,n,p$, while those proportional to $c^*$ cancel because of the constraint $p=-m-n$. We could indeed redefine the generators $\KSgen_0,\bar{\KSgen}_0$ to obtain from \eqref{modified_Virasoro}  the relation of the Virasoro algebra at central charge $c^*$. This would, however, not resolve the underlying difference between limits of commutators and commutators of limits.




\section{Conclusions}\label{Conclusion}

We have first presented in this paper considerable  numerical evidence of the validity of the Koo-Saleur conjecture, namely that $\mathcal{L}_n \mapsto L_n$ as $N \to \infty$, in the scaling-weak sense. Thanks to the systematic use of form factors in the numerical computations, we have been able to increase significantly the size of the systems studied---and thus the accuracy of the checks---as compared with the pioneering paper \cite{KooSaleur}. This has allowed us, in particular, to analyse the structure of the Virasoro modules occurring in the continuum limit of the XXZ spin chain.

In the degenerate case, where the conjectured conformal weights take values in the extended Kac table, one crucial result is that for the XXZ spin chain both Verma and co-Verma modules occur. 
The difference between Verma and co-Verma is related to the existence of two conjugate values of the Coulomb-gas charges giving rise to the same scaling dimension of the corresponding vertex operator. The notion of charge conjugation leads also to the identity of certain matrix elements, exact in finite size, as expressed in the result \eqref{strong} about strong duality.

Our main results for the nature of the modules arising in the continuum limit for the XXZ spin chain representation are given in equations (\ref{MainRes1}) and (\ref{MainRes2}), complementing the results found in \cite{LoopPaper} for the loop representation. 
In our next paper \cite{fullynongenericpaper}, we shall use these techniques to investigate the structure of the Virasoro modules in the case where $\q$ is a root of unity. This will be applied to the understanding of logarithmic CFTs, in particular the determination of  the full ``identity module'' for $c=0$ theories like the $sl(2|1)$ spin chain. 

The main mathematical question raised by our results---the exact nature of the convergence of the Koo-Saleur generators to their continuum limit, and the relation between limits of commutators and commutators of limits---is outside the scope of this work, and certainly deserves further study. As a first step in this direction, we have conjectured that the limit of the commutators of the Koo-Saleur generators is correct {\sl only up to the anomalous central charge term}, a result for which we have given qualitative and numerical evidence, but which we are not able to prove for now. Our results about the exchange of commutators and limits are encompassed in the conjectures \eqref{ndiff0_conj},\eqref{cstar_conjecture} and \eqref{chiral-antichiral_conj} and the result  \eqref{separate_conj}.

Another interesting question deserving more work is the possible relation between the two scalar products we have introduced, and what they have to do with the natural positive definite scalar product in the RSOS case. We also hope to get back to this point in further work.

\subsubsection*{Acknowledgments}
This work was supported by the advanced ERC grant NuQFT. We thank A.~Gainutdinov, E.~Granet, F.~G\"ohmann, Z.~Wang, M.~Shokrian-Zini, P. Fendley and R. Bondesan for many interesting discussions.

~\\
\textbf{Note added.} Motivated by the results of the present paper, we have added a new Appendix C to the latest arXiv version of \cite{LoopPaper}, in which we correct some of our reasoning about the lattice evidence for the structure of the continuum modules. This correction does not change the results in \cite{LoopPaper}, and it also provides further confirmation of the corresponding results about the structure in the current paper.

\appendix
\section{Some remarks on scalar products}\label{RemScalProd}


It is interesting to discuss in more detail why the conformal scalar product (the one corresponding to the conjugation $L_n^\loopdagger=L_{-n}$) is the one relevant for the calculation of correlation functions in the operator formalism, and, in our case, is the one obtained by ``treating $\q$ as a formal variable''. 
Due to the exploratory nature of this Appendix, we shall use here the standard bra-ket notation $\langle - | - \rangle$, although the conformal scalar product
will eventually be denoted by $(-,-)$ in the main text.

We start by discussing a simple example. Consider  a very simple system of two spins $1/2$ coupled via the $U_qsl(2)$ invariant Hamiltonian, which reads (up to an irrelevant scalar factor)
\begin{equation}
\Hlatt=-e=-\left(\begin{array}{cccc}
0&0&0&0\\
0&\q^{-1}&-1&0\\
0&-1&\q&0\\
0&0&0&0
\end{array}\right)
\end{equation}
in the basis $\{++,+-,-+,--\}$. When $\q$ is real, the normalized ground state is given by 
\begin{equation}
|\Omega\rangle={1\over \sqrt{\q+\q^{-1}}}(\q^{-1/2}|+-\rangle-\q^{1/2}|-+\rangle) \,.
\end{equation}
When $\q$ is a complex number of modulus one---the case we are considering here---$\Hlatt$ {\sl is not hermitian}.  The ground state $|\Omega\rangle$ as we have written here is not normalized any more, since its norm square obtained with the usual scalar product is ${2\over \q+\q^{-1}}$. 
 
However, since the  eigenvalues ($0,\q+\q^{-1}$) are real, we see that from
\begin{equation}
\Hlatt^\dagger |E_{\rm L}\rangle=E|E_{\rm L}\rangle
\end{equation}
we get the left-eigenvalue problem
\begin{equation}
\langle E_{\rm L}|\Hlatt=E\langle E_{\rm L}|
\end{equation}
to be compared with the right-eigenvalue problem initially considered
\begin{equation}
\Hlatt|E_{\rm R}\rangle=E|E_{\rm R}\rangle \,.
\end{equation}
In our case we have (where the unprimed states correspond to the singlet, i.e., the ground state, while the primed states correspond to the triplet)
\begin{subequations}
\begin{eqnarray}
|E_{\rm R}\rangle={1\over\sqrt{\q+\q^{-1}}}\left(\q^{-1/2}|+-\rangle-\q^{1/2}|-+\rangle\right) \,, \\
|E'_{\rm R}\rangle={1\over\sqrt{\q+\q^{-1}}}\left(\q^{1/2}|+-\rangle+\q^{-1/2}|-+\rangle\right) \,,
\end{eqnarray}
\end{subequations}
and 
\begin{eqnarray}
|E_{\rm L}\rangle={1\over\sqrt{\q+\q^{-1}}}\left(\q^{1/2}|+-\rangle-\q^{-1/2}|-+\rangle\right)\,,\\
|E'_{\rm L}\rangle={1\over\sqrt{\q+\q^{-1}}}\left(\q^{-1/2}|+-\rangle+\q^{1/2}|-+\rangle\right)\,,
\end{eqnarray}
where we chose normalizations such that left and right eigenstates are orthonormal,
\begin{equation}
\langle E_{\rm L}|E_{\rm R}\rangle = \langle E'_{\rm L}|E'_{\rm R}\rangle = 1 \,, \qquad
\langle E_{\rm L}|E'_{\rm R}\rangle = \langle E'_{\rm L}|E_{\rm R}\rangle = 0 \,,
\end{equation}
and $\langle - | - \rangle$ stands for the ordinary scalar product. We see now that $|\Omega\rangle$ remains properly normalized with these conventions. Moreover, conjugating and switching L for R amounts to treating $\q$ as a formal parameter, by which we mean that it does not undergo any complex conjugation in the process. We thus see that considering scalar products of the form $\langle E_L|E_R\rangle$ (with left and right eigenstates properly distinguished), instead of $\langle E_R|E_R\rangle$ (with no such distinction), amounts to treating ${\mathcal H}$ as self-conjugate, i.e., $\q$ as a formal parameter indeed. 

The second point is to establish what this has to do with physics. We start by demanding that calculations in our quantum (field theory or not) formalism describe well-defined objects in two-dimensional statistical physics. The mapping of the quantum system onto a statistical mechanics one proceeds via an imaginary-time representation, which can be depicted by taking the direction of imaginary time to be upwards. The density operator $\rho$ introduces a horizontal cut in
this time evolution, in the sense that its matrix elements $\rho_{xy}$ are such that $x$ (resp.\ $y$) refers to the degrees of freedom on the lower (resp.\ upper)
lip of this cut. We now argue that one may write the density operator as $\rho=|0_{\rm R}\rangle\langle 0_{\rm L}|$, that is, in terms of the left and right ground states.
Indeed, its matrix elements are of the form
$$
\rho_{xy}\propto\langle +\infty|e^{-\tau H}|y\rangle\langle x|e^{-\tau H}|-\infty\rangle \,,
$$
where $\langle +\infty |$ and $|-\infty\rangle$ denote ``generic states'' (initial conditions that are not orthogonal to the left and right ground states, respectively) introduced at imaginary times $\pm \infty$, that is, at the bottom and top of the time evolution.
Taking $\tau$ large projects onto the lowest-energy eigenstates of ${\mathcal H}$ for the bottom part and ${\mathcal H}^\dagger$ for the top part, so
$$
\rho_{xy}\propto \langle 0_L|y\rangle \langle x|0_R\rangle \,.
$$
With proper normalizationsand it is therefore justified to write $\rho=|0_{\rm R}\rangle\langle 0_{\rm L}|$, as claimed. Correlation functions in the quantum theory are then obtained by tracing $\rho$ with various insertions, and thus correspond to 
\begin{equation}
\langle 0_{\rm L}|\cdots|0_{\rm R}\rangle \,,
\end{equation}
with $\langle - | - \rangle$ denoting the ordinary scalar product.

%
%

\medskip

In CFT, we demand that 
\begin{equation}
\langle\phi|=|\phi\rangle^\dagger \,,
\end{equation}
which leads to (on the surface where $\bar{z}=z^*$, the complex conjugate)
\begin{equation}
[\phi(z,\bar{z})]^\dagger\equiv \bar{z}^{-2h}z^{-2\bar{h}}\phi(1/\bar{z},1/z) \,,
\end{equation}
so that
\begin{eqnarray}
\langle\phi|\phi\rangle &=& \hbox{lim}_{z,\bar{z},w,\bar{w}\to 0}\langle 0|\phi(z,\bar{z})^\dagger\phi(w,\bar{w})|0\rangle\nonumber\\
&=&\hbox{lim}_{\xi,\bar{\xi}\to\infty}\bar{\xi}^{2h}\xi^{2\bar{h}}\langle 0|\phi(\bar{\xi},\xi)\phi(0,0)|0\rangle \nonumber \\
&=&1 \,,
\end{eqnarray}
where in the last step we would generally obtain the residue of the two-point function, if the latter were not normalized. Using this for $T(z)$ leads immediately to $L_n^\dagger=L_{-n}$. In general, we see that this dagger operation is the one that exchanges left and right eigenstates in the non-Hermitian case. 

This conformal scalar product is the continuum limit of the loop scalar product, or the $sl(2|1)$-invariant scalar product as well. It is not, in general, positive definite. 
As already mentioned, it will be denoted $(-,-)$ in the main text.

\section{More on form factors}\label{FormFactorAppendix}


For a general overview of transfer matrix formalism, Bethe ansatz and the Quantum Inverse Scattering Method (QISM), see \cite{Slavnov}. 

\subsection{General framework, notations, conventions}

We shall follow the work in \cite{Terras}. We first fix our notation. The XXZ $R$-matrix is written as
\begin{equation}
R(\lambda,\mu)=\begin{pmatrix}
1&0&0&0\\
0&b(\lambda,\mu)&c(\lambda,\mu)&0\\
0&c(\lambda,\mu)&b(\lambda,\mu)&0\\
0&0&0&1\\
\end{pmatrix} \,,
\end{equation}
with
\begin{subequations}
\begin{eqnarray}
 b(\lambda,\mu) &=& \frac{\sinh(\lambda-\mu)}{\sinh(\lambda-\mu+\eta)} \,, \\
 c(\lambda,\mu) &=& \frac{\sinh(\eta)}{\sinh(\lambda-\mu+\eta)}
\end{eqnarray}
\end{subequations}
and $\eta=i\gamma$. Since $R$ only depends on the difference $\lambda-\mu$, we shall sometimes write it in terms of only one variable, $R(u) \equiv R(\lambda,\mu)$ with $u = \lambda-\mu$. 
%
%
We write the corresponding monodromy matrix as $\mathbf{T}(u)=\left( \begin{smallmatrix}
A(u)&B(u)\\C(u)&D(u)
\end{smallmatrix}\right)$, thereby defining the operators $A,B,C,D$ that will be used below. 
Let $a(\lambda)$ and $ d(\lambda)$ be the eigenvalues of $A(\lambda)$ and $D(\lambda)$ when acting on $|0\rangle$, where $|0\rangle$ denotes the pseudo-vacuum of all spins pointing up. The Bethe equations take the form
\begin{equation}\label{BetheEqs}
\frac{d(\lambda_j)}{a(\lambda_j)}\prod_{k\neq j}\frac{b(\lambda_k,\lambda_j)}{b(\lambda_j,\lambda_k)}=1
\end{equation}
for $j=1,2,\ldots,\frac{N}{2}-S_z$, and an on-shell Bethe state is written as $\prod_j B(\lambda_j)|0\rangle$ with the Bethe roots $\lambda_j$ solving \eqref{BetheEqs}. (If the roots do not solve the Bethe equations, the state is called off-shell.)

%
%
The above holds for periodic boundary conditions, but we also wish to consider boundary conditions parametrized by a twist $\phi$.  Let $\kappa \in GL_2(\mathbb{C})$. Then $\kappa \mathbf{T}$ gives us the monodromy matrix for the case of twisted boundary conditions on the form $\sigma^a_{N+1}=\kappa\sigma^a_1\kappa^{-1}$. For the boundary conditions that we wish to implement (namely $\sigma^z_{N+1}=\sigma^z_1$ and $\sigma^{\pm}_{N+1}=e^{\mp i \phi} \sigma^{\pm}_1$) the proper choice is the following diagonal twist matrix:
\begin{equation}
\kappa=\begin{pmatrix}
1&0\\
0&e^{i\phi }
\end{pmatrix} \,.
\end{equation}
We obtain thus the twisted monodromy matrix
\begin{equation}\label{twistedT}
\kappa\mathbf{T}(u)=
\begin{pmatrix}
1&0\\
0&e^{i\phi }
\end{pmatrix}
\begin{pmatrix}
A(u)&B(u)\\
C(u)&D(u)
\end{pmatrix} =
\begin{pmatrix}
A(u)&B(u)\\
e^{i\phi }C(u)&e^{i\phi }D(u)
\end{pmatrix}.
\end{equation}
Twisting the monondromy matrix $\mathbf{T}(u)\rightarrow\kappa\mathbf{T}(u)$ in this fashion leads to a modified eigenvalue $d(u) \rightarrow e^{i\phi } d(u)$ that will enter into the Bethe equations, since the latter contain the ratios $\frac{d(\lambda_k)}{a(\lambda_k)}$.

With a diagonal twist $\kappa$ as above, any general Bethe considerations for the untwisted case will also be valid for the twisted case.
For instance, while \eqref{twistedT} implies that $B^\dagger(\lambda^*) = C(\lambda)$ for the untwisted case (up to a phase), this same property will still hold after the twist (up to another phase).\footnote{Note that the Bethe roots come in conjugate pairs (or are real, or self-conjugate), so that in total we do indeed get the expression for Bethe states as in eq \eqref{conjugation}.} 
Notice however that this does not carry over to the general case of non-diagonal $\kappa$, where the dual Bethe states (defined from $C$) might no longer be given by the conjugate of the Bethe states (defined from $B$).



\subsection{Quantum Inverse Scattering Method}

In the framework of the quantum inverse scattering method, we express local operators $\mathcal{O}$ in terms of the operators $A,B,C,D$. Let $\mathcal{t}(\frac{\eta}{2}) = A(\frac{\eta}{2}) + D(\frac{\eta}{2})$ denote the transfer matrix, that is, the trace of $\mathbf{T}(u)$ in the homogeneous case $u=\eta/2=i\gamma/2$ (which will sometimes have to be taken as a limit, as we shall see below). 
We recall that the XXZ Hamiltonian \eqref{Pauliham} is recovered from $T^{-1}(u) \frac{\partial T}{\partial u}\Big|_{u=\eta/2}$.
%
Acting on a Bethe state, where $n$ is the number of roots characterizing the state, the transfer matrix has eigenvalues
\begin{equation}
\mathcal{t}\left(\frac{\eta}{2}\right)\prod_{k=1}^n B(\lambda_k)|0\rangle =
%
\left[ 
a\left(\frac{\eta}{2}\right) \prod_{k=1}^n b^{-1}\left(\lambda_k,\frac{\eta}{2}\right)
+ d\left(\frac{\eta}{2}\right)\prod_{k=1}^n b^{-1}\left(\frac{\eta}{2},\lambda_k\right) \right]
\prod_{k=1}^n B(\lambda_k)|0\rangle.
\end{equation}
With the conventions used in \cite{Terras}, $a(\lambda)=1,\,d(\lambda)=\prod_{m=1}^N b(\lambda,\xi_m)$, so that $d(\xi_m)=0$ even before taking the homogeneous limit. We denote by 
$\phi_m\{\lambda\}=\prod_{k=1}^N \left( b^{-1}\left(\lambda_k,\frac{\eta}{2}\right)   \right)^m$ the factor produced by the action of $\mathcal{t}^m \! \left(\frac{\eta}{2}\right)$, which appears in the computation of form factors. We can then express the matrix elements of $j$ neighbouring operators (acting on consecutive sites $m, m+1,\ldots,m+j-1$) in the form
\begin{equation} \label{phiphiF}
 \langle \{\mu\}|\prod_{i=1}^j ({\mathcal{O}_i})_{m+i-1}|\{\lambda\}\rangle = \phi_{m-1}(\{\mu\}) \phi^{-1}_{m+j-1}(\{\lambda\}) F_{{\mathcal{O}}_1,\ldots,{\mathcal{O}}_j}  ,
\end{equation}
where $F_{\mathcal{O}_1,\ldots,\mathcal{O}_j}$ depends on the combination of operators $\mathcal{O}$  whose matrix elements we wish to obtain, while the pre-factors $\phi_m\{\lambda\}$ only depend on how many operators we consider and at which sites. The expression for $F_{\mathcal{O}_1,\ldots,\mathcal{O}_j}$ is site-independent: all dependence on the site $m$ is in the phase coming from the $\phi$ pre-factors in \eqref{phiphiF}. Below we use the shorthand notation $z\leftrightarrow \sigma^z$, $-\leftrightarrow \sigma^-$, $+\leftrightarrow \sigma^+$ to denote the required operators $\mathcal{O}$, and we wish to compute expressions such as $F_{-+}$. 

Because of this site-independence, in numerical applications it is most efficient to compute the relevant $F_{\mathcal{O}_1,\ldots,\mathcal{O}_j}$ only once for each size $N$, and then add up any site-dependent phases (see section \ref{ResultsAboutFF}) in an independent step when using the form factors.%
\footnote{For the specific purpose of computing the Koo-Saleur generators, keeping track of phases actually turns out to be unnecessary, because of the considerations about momentum conservation made in section~\ref{KSandLatticeMomentum}.}
Combined with the parity and conjugation relations mentioned earlier, this significantly reduces the computational load. Furthermore most scalar products $S_n$ (see below) are used in several form factors. This can be taken advantage of as well.



We wish to find $F_{\mathcal{O}_1,\ldots,\mathcal{O}_j}$ for the combinations of Pauli matrices shown in and below \eqref{all6-1}--\eqref{all6-2}. To this end, we shall need to consider the following operators (given here with their shorthand abbreviation): 
\begin{subequations}
\label{operator_correspondence}
\begin{eqnarray}
z &:& \sigma_m^z \leftrightarrow A(\xi_m) - D(\xi_m) \,, \text{ or } 2A(\xi_m)-\mathbf{1} \,, \text{ or }\mathbf{1}-2D(\xi_m) \,, \\
+ &:& \sigma_m^{+} \leftrightarrow C(\xi_m) \,, \\
- &:& \sigma_m^{-} \leftrightarrow B(\xi_m)
\end{eqnarray}
\end{subequations}
in the homogeneous limit $\xi_m \rightarrow \eta/2$. For instance, to get $F_z$ we can compute $F_D=\langle C |D(\xi_m)| B\rangle$, where $\langle C| = \langle 0 | \prod_j C(\mu_j)$ and $|B\rangle=\prod_k B(\lambda_k)$ are Bethe states. We shall always keep $\langle C|$ untouched (and thus still on-shell), commuting the operators above past the string of $B$-operators, so that the result will become expressed in terms of new, off-shell states $|\tilde{B}\rangle$. (See the commutation relations below, where some $B(\lambda_k)$ are swapped into $B\left( \xi_m \right)$ with $\xi_m = \frac{\eta}{2}$.) When computing the final scalar products $S_n=\langle C|\tilde{B}\rangle$, we can then still use relations that hold when one of the states is on-shell.


In order to commute all the operators past the $B$-operators we use the following commutation relations, where $\varphi(\lambda)=\sinh(\lambda)$:
%
%
\begin{subequations}
\begin{eqnarray}
A(\xi_m)\prod_{k=1}^n B(\lambda_k) |0\rangle  &=& \prod_{k=1}^n   
\frac{\varphi(\lambda_k-\xi_m+\eta)}{\varphi(\lambda_k-\xi_m)}
 \prod_{k=1}^n B(\lambda_k) |0\rangle \nonumber \\
& &
-\sum_{a=1}^n \frac{\varphi(\eta)}{\varphi(\lambda_a-\xi_m)} \prod_{k=1,k\neq a}^n \frac{\varphi(\lambda_k-\lambda_a+\eta)}{\varphi(\lambda_k-\lambda_a)} B(\xi_m) \prod_{k=1,k\neq a}^n B(\lambda_k) |0\rangle \,, \\
C(\xi_m)\prod_{k=1}^n B(\lambda_k) |0\rangle &=& \sum_{a=1}^n M_a(\xi_m) \prod_{k=1,k\neq a}^n B(\lambda_k) |0\rangle + \sum_{a\neq b} M_{ab}(\xi_m) B(\xi_m) \prod_{k=1,k\neq a,b}^n B(\lambda_k) |0\rangle \,, \\
D(\xi_m)\prod_{k=1}^n B(\lambda_k) |0\rangle &=& -\sum_{a=1}^n 
\frac{\varphi(\eta)}{\varphi(\xi_m-\lambda_a)} \prod_{k=1,k\neq a}^n \frac{\varphi(\lambda_a-\lambda_k+\eta)}{\varphi(\lambda_a-\lambda_k)} 
 d(\lambda_a) B(\xi_m) \prod_{k=1,k\neq a}^n B(\lambda_k) |0\rangle \,,
\end{eqnarray} 
\end{subequations}
with
\begin{subequations}
\begin{eqnarray}
M_a(\xi_m) &=& \frac{\varphi(\eta)}{\varphi(\lambda_a-\xi_m)}\prod_{k\neq a}\frac{\varphi(\lambda_k-\xi_m+\eta)}{\varphi(\lambda_k-\xi_m)}\frac{\varphi(\lambda_k-\lambda_a+\eta)}{\varphi(\lambda_k-\lambda_a)} \,, \\
M_{ab}(\xi_m)&=& \frac{\varphi(\eta)^2}{\varphi(\lambda_a-\xi_m)\varphi(\lambda_b-\xi_m)}\prod_{k\neq a}\frac{\varphi(\lambda_k-\lambda_a+\eta)}{\varphi(\lambda_k-\lambda_a)}\frac{\varphi(\lambda_k-\lambda_b+\eta)}{\varphi(\lambda_k-\lambda_b)} \,.
\end{eqnarray}
\end{subequations}
We have written out factors of $b^{-1}(\mu,\lambda)$ explicitly to make it clearer later which terms will need to be combined in the homogeneous limit. Note that the $B$-operators commute among themselves. 

Finally we need to divide through with the norms of the Bethe states after obtaining the required form factors, as the states we use are otherwise not normalized. For roots $\{\lambda\}$ solving the Bethe equations, the corresponding state has the norm squared (written using the conventions of \cite{Terras})
\begin{equation}
N_n(\{\lambda\})= \varphi^n(\eta)\prod_{\alpha\neq\beta}\frac{\varphi(\lambda_\alpha-\lambda_\beta+\eta)}{\varphi(\lambda_\alpha-\lambda_\beta)} \det(G(\{\lambda\}) \,,
\end{equation}
with
\begin{equation}
G_{ab}(\{\lambda\})= -\frac{\partial}{\partial\lambda_b} \ln \left(r(\lambda_a) \prod_{k=1,k\neq a}^n  \frac{b(\lambda_a,\lambda_k)}{b(\lambda_k,\lambda_a)}\right).
\end{equation}
Here $r(\lambda)=a(\lambda)/d(\lambda)$.

\subsection{The expressions for the necessary $F_{\mathcal{O}_1,\ldots,\mathcal{O}_j}$ }
In what follows we shall use the notation $\{\lambda\}$ to refer to the set of Bethe roots $\lambda_1,\ldots,\lambda_n$, while $\{\lambda\}_a$ denotes the set with the $a$'th root removed. We use $S_n(\{\mu\},\{\lambda\})$ to refer to scalar products $\langle \{\mu\}| \{\lambda\}\rangle$; these scalar products are given explicitly below.

Note: As when finding the Bethe roots, we must again keep in mind the modified eigenvalue $d(\lambda)\rightarrow e^{i\phi} d(\lambda)$. For the norm squared above any overall phase is ignored in the final normalization. Below, however, the phases $ e^{i\phi}$ do matter. We obtain relative phases between the different terms, and also overall phases that we must keep in mind when we take conjugates as discussed in Section \ref{conjugate_vs_Bethe}.

\subsubsection{$F_z,F_{zz},F_{-+}$}
$F_{-+}$ is taken directly from \cite{Terras}, up to minor changes in notation. For the other two we have chosen to write in terms of $D$, as it has somewhat nicer commutation relations. $F_z,F_{zz}$ are given from $F_D,F_{DD}$ in accordance with \eqref{operator_correspondence}. We have:
\begin{subequations}
\begin{eqnarray}
F_D&=&-\sum_{a=1}^n
\frac{\varphi(\eta)}{\varphi(\xi_m-\lambda_a)} \prod_{k=1,k\neq a}^n \frac{\varphi(\lambda_a-\lambda_k+\eta)}{\varphi(\lambda_a-\lambda_k)} 
{\color{black} e^{i\phi}} d(\lambda_a) S_n(\{\mu\}|\xi_{m},\{\lambda\}_{a}) \,,\\
F_{DD}&=&\sum_{a,b=1,b\neq a}^n 
\frac{\varphi(\eta)^2}{\varphi(\xi_{m+1}-\lambda_a)\varphi(\xi_m-\lambda_b)} \prod_{k=1,k\neq a}^n \frac{\varphi(\lambda_a-\lambda_k+\eta)}{\varphi(\lambda_a-\lambda_k)} 
 {\color{black} e^{i\phi}}d(\lambda_a)
\frac{\varphi(\lambda_b-\xi_{m+1}+\eta)}{\varphi(\lambda_b-\xi_{m+1})} \\
& & \hspace*{4.8cm}
\times 
\prod_{k=1,k\neq a,b}^n \frac{\varphi(\lambda_b-\lambda_k+\eta)}{\varphi(\lambda_b-\lambda_k)} 
 {\color{black} e^{i\phi}}d(\lambda_b) S_n(\{\mu\}|\xi_{m},\xi_{m+1},\{\lambda\}_{a,b}) \,. \nonumber
\end{eqnarray}
\end{subequations}
For $F_{-+}$ we need $F_{BC}$:
\begin{equation}
F_{BC}=\sum_{a=1}^n M_a(\xi_{m+1}) S_n(\{\mu\}|\xi_{m},\{\lambda\}_{a}) + \sum_{b=1,b\neq a}^n M_{ab}(\xi_{m+1}) S_n(\{\mu\}|\xi_{m},\xi_{m+1},\{\lambda\}_{a,b}) \,.
\end{equation}
For all these we can safely let $\xi_m,\xi_{m+1}\rightarrow \eta/2$, as long as we write the scalar products involving multiple $\eta/2$ as shown in section \ref{ScalarProducts}.

\subsubsection{$F_{z-+}$, $F_{-z+}$}
For $F_{z-+}$ we will need $F_{DBC}$:
\begin{equation}
\begin{aligned}
&F_{ {\color{black}{ D }} B  {\color{black}{C}} } =
 -{\color{black}{ \sum_{a=1}^n M_a(\xi_{m+2}) }}   {\color{black}{\sum_{b=1,b\neq a}^{n} \frac{\varphi(\eta)}{\varphi(\xi_m-\lambda_b)} 
\frac{\varphi(\lambda_b-\xi_{m+1}+\eta)}{\varphi(\lambda_b-\xi_{m+1})}
\prod_{k=1,k\neq a,b}^{n} \frac{\varphi(\lambda_b-\lambda_k+\eta)}{\varphi(\lambda_b-\lambda_k)} {\color{black} e^{i\phi}} d(\lambda_b) }} \\
& \hspace*{10cm}
\times
S_n(\{\mu\}|\xi_{m},\xi_{m+1},\{\lambda\}_{a,b})\\
\\
%
&- {\color{black}{ \sum_{a\neq b} M_{ab}(\xi_{m+2})  }} {\color{black}{\sum_{c=1,c\neq a,b}^{n} \frac{\varphi(\eta)}{\varphi(\xi_m-\lambda_c)}
%
\frac{\varphi(\lambda_c-\xi_{m+1}+\eta)}{\varphi(\lambda_c-\xi_{m+1})} 
\frac{\varphi(\lambda_c-\xi_{m+2}+\eta)}{\varphi(\lambda_c-\xi_{m+2})} 
\prod_{k=1,k\neq a,b,c}^{n} \frac{\varphi(\lambda_c-\lambda_k+\eta)}{\varphi(\lambda_c-\lambda_k)} {\color{black} e^{i\phi}} d(\lambda_c) }} \\
& \hspace*{10cm}
\times
S_n(\{\mu\}|\xi_{m},\xi_{m+1},\xi_{m+2},\{\lambda\}_{a,b,c}),\\
\end{aligned}
\end{equation}
where we can again safely let $\xi_m,\xi_{m+1},\xi_{m+2}\rightarrow \eta/2$. 

For $F_{-z+}$ we need both $F_{BDC}$ and $F_{BAC}$: 
\begin{equation}
\begin{aligned}
&F_{ B  {\color{black}{ D }} {\color{black}{C}} } =
 -{\color{black}{ \sum_{a=1}^n M_a(\xi_{m+2}) }}   {\color{black}{\sum_{b=1,b\neq a}^{n} \frac{\varphi(\eta)}{\varphi(\xi_{m+1}-\lambda_b)} 
\prod_{k=1,k\neq a,b}^{n} \frac{\varphi(\lambda_b-\lambda_k+\eta)}{\varphi(\lambda_b-\lambda_k)} {\color{black} e^{i\phi}}d(\lambda_b) }}
S_n(\{\mu\}|\xi_{m},\xi_{m+1},\{\lambda\}_{a,b})\\
\\
%
&- {\color{black}{ \sum_{a\neq b} M_{ab}(\xi_{m+2})  }} {\color{black}{\sum_{c=1,c\neq a,b}^{n}
%
\frac{\varphi(\eta)}{\varphi(\xi_{m+1}-\lambda_c)} 
\frac{\varphi(\lambda_c-\xi_{m+2}+\eta)}{\varphi(\lambda_c-\xi_{m+2})} 
\prod_{k=1,k\neq a,b,c}^{n} \frac{\varphi(\lambda_c-\lambda_k+\eta)}{\varphi(\lambda_c-\lambda_k)}{\color{black} e^{i\phi}} d(\lambda_c) }} \\
& \hspace*{10cm}
\times
S_n(\{\mu\}|\xi_{m},\xi_{m+1},\xi_{m+2},\{\lambda\}_{a,b,c}),\\
\end{aligned}
\end{equation}

\begin{equation}
\begin{aligned}
&F_{ B  {\color{black}{ A }} {\color{black}{C}} } = {\color{black}{ \sum_{a=1}^n M_a(\xi_{m+2}) }}
{\color{black}{ \prod_{k=1,k\neq a}^{n}
\frac{\varphi(\lambda_k-\xi_{m+1}+\eta)}{\varphi(\lambda_k-\xi_{m+1})}
 }} S_n(\{\mu\}|\xi_{m},\{\lambda\}_{a})\\
%
&  -  {\color{black}{ \sum_{a=1}^n M_a(\xi_{m+2}) }} {\color{black}{\sum_{b=1,b\neq a}^{n} \frac{\varphi(\eta)}{\varphi(\lambda_b-\xi_{m+1})} \prod_{k=1,k\neq a,b}^{n} \frac{\varphi(\lambda_k-\lambda_b+\eta)}{\varphi(\lambda_k-\lambda_b)}  }}S_n(\{\mu\}|\xi_{m},\xi_{m+1},\{\lambda\}_{a,b})\\
\\
%
&+ {\color{black}{ \sum_{a\neq b} M_{ab}(\xi_{m+2})  }}
{\color{black}{
\frac{\varphi(\xi_{m+2}-\xi_{m+1}+\eta)}{\varphi(\xi_{m+2}-\xi_{m+1})}
\prod_{k=1,k\neq a,b}^{n} 
\frac{\varphi(\lambda_k-\xi_{m+1}+\eta)}{\varphi(\lambda_k-\xi_{m+1})}
}}
S_n(\{\mu\}|\xi_{m},\xi_{m+2},\{\lambda\}_{a,b})\\
& -{\color{black}{ \sum_{a\neq b} M_{ab}(\xi_{m+2})  }} {\color{black}{ \frac{\varphi(\eta)}{\varphi(\xi_{m+2}-\xi_{m+1})} \prod_{k=1,k\neq a,b}^{n} \frac{\varphi(\lambda_k-\xi_{m+2}+\eta)}{\varphi(\lambda_k-\xi_{m+2})} }} 
S_n(\{\mu\}|\xi_{m},\xi_{m+1},\{\lambda\}_{a,b})\\
& -{\color{black}{ \sum_{a\neq b} M_{ab}(\xi_{m+2})  }} {\color{black}{\sum_{c=1,c\neq a,b}^{n} \frac{\varphi(\eta)}{\varphi(\lambda_c-\xi_{m+1})} \frac{\varphi(\xi_{m+2}-\lambda_c+\eta)}{\varphi(\xi_{m+2}-\lambda_c)} 
\prod_{k=1,k\neq a,b,c}^{n} \frac{\varphi(\lambda_k-\lambda_c+\eta)}{\varphi(\lambda_k-\lambda_c)} }}  \\
& \hspace*{10cm}
\times
S_n(\{\mu\}|\xi_{m},\xi_{m+1},\xi_{m+2},\{\lambda\}_{a,b,c}).\\
\end{aligned}
\end{equation}
In the latter expression, the third and fourth terms are divergent on their own if we take the homogeneous limit. They will, however, combine into a derivative. Let $\xi_{m+1}=\eta/2$, $\xi_{m+2}=\eta/2+x$. We have a common factor $M_{ab}(\xi_{m+2})$ that causes no problem, and for the rest we can write 
\begin{multline}
\frac{1}{\varphi(x)} \left[\text{third} - \text{fourth} \right] \xrightarrow[x\to0]{} \partial_x \Big(\varphi(\eta +x)S_n(\{\mu\}|\eta/2+x,\eta/2,\{\lambda\}_{ab})\Big)\Bigg|_{x=0}\prod\frac{\varphi(\lambda_k+\eta/2)}{\varphi(\lambda_k-\eta/2)} \\
- \Big(\varphi(\eta)S_n(\{\mu\}|\eta/2,\eta/2,\{\lambda\}_{ab})\Big) \partial_x\prod\frac{\varphi(\lambda_k-x+\eta/2)}{\varphi(\lambda_k-x-\eta/2)} \Bigg|_{x=0}.
\end{multline}

\subsubsection{Scalar products and the homogeneous limit}\label{ScalarProducts}

We give below the determinant representation for the scalar products $S_n$. 
Please note than when compared to \cite{Terras} we have taken the factors of $r(\lambda)=a(\lambda)/d(\lambda)$ outside the matrix. This leads to an overall phase $e^{-in\phi}$ in front of the determinant from the modified eigenvalue $d(\lambda)\rightarrow e^{i\phi}d(\lambda)$. 
\begin{equation}
\begin{aligned}
S_n(\{\mu\},\{\lambda\}) \equiv \langle \{\mu\}| \{\lambda\}\rangle = \frac{ {\color{black} e^{-in\phi}} \det(H(\{\mu\},\{\lambda\}))}{\prod_{j>k} \varphi(\mu_k-\mu_j) \prod_{\alpha<\beta}\varphi(\lambda_\beta - \lambda_\alpha) },
\end{aligned}
\end{equation}
with 
\begin{equation}\label{Hab}
\begin{aligned}
H_{ab}(\{\mu\},\{\lambda\})=\frac{\varphi(\eta)}{\varphi(\mu_a-\lambda_b)} \Big(\prod_{k\neq a} \varphi(\mu_k-\lambda_b+\eta) - {\color{black} e^{i\phi}} d(\lambda_b) \prod_{k\neq a} \varphi(\mu_k-\lambda_b-\eta)
\Big).
\end{aligned}
\end{equation}
We shall need the scalar products $S_n(\eta/2,\eta/2,\ldots)$, $S_n(\eta/2,\eta/2,\eta/2,\ldots)$, and the derivative $\partial_x S_n(\{\mu\}|\eta/2+x,\eta/2,\{\lambda\}_{ab})\Big)\Big|_{x=0}$ in the homogeneous limit, which must be taken carefully due to the denominator. 

Subtracting the relevant columns of $H$ from each other before taking the limit, and taking factors of the type $1/\varphi(\xi_m-\xi_{m+1})$ inside, we arrive at the replacement of the first columns by their first and second derivatives. The second term in \eqref{Hab} disappears in these columns, thanks to $d(\xi_m)=0$. Starting with $S_n(\eta/2,\eta/2,\ldots)$, we simply find the first column replaced by its derivative:
\begin{equation}
\begin{aligned}
&H'_{a1}(\{\mu\},\{\lambda\})= -\partial_x \frac{\varphi(\eta)}{\varphi(\mu_a-\eta/2-x)} \prod_{k\neq a} \varphi(\mu_k-x+\eta/2) 
\Bigg|_{0},
\end{aligned}
\end{equation}
and the product in the denominator of $S_n$ is now taken for $\beta>2$. 
%
For $S_n(\eta/2,\eta/2,\eta/2,\ldots)$ we arrive at the following replacements for column 1 and 2:
\begin{subequations}
\begin{eqnarray}
H'_{a2}(\{\mu\},\{\lambda\}) &=& -\partial_x \frac{\varphi(\eta)}{\varphi(\mu_a-\eta/2-x)} \prod_{k\neq a} \varphi(\mu_k-x+\eta/2)  
\Bigg|_{0} \,, \\
H''_{a1}(\{\mu\},\{\lambda\}) &=& \frac{1}{2} \partial^2_x \frac{\varphi(\eta)}{\varphi(\mu_a-\eta/2-x)} \prod_{k\neq a} \varphi(\mu_k-x+\eta/2) 
\Bigg|_{0} \,.
\end{eqnarray}
\end{subequations}
The product in the denominator of $S_n$ is now taken for $\beta>3$. Finally for $\partial_x S_n(\{\mu\}|\eta/2+x,\eta/2,\{\lambda\})$ we can take the derivative inside $\det(H)$ and let it act on the relevant column. Taking the homogeneous limit then leads to further derivatives, so that in total we obtain
\begin{eqnarray}
\partial_x S_n(\{\mu\}|\eta/2+x,\eta/2,\{\lambda\})\Bigg|_0 &=&
-\frac{1}{2} \frac{{\color{black} e^{-in\phi}}\det(\tilde{H}''(\{\mu\}|\eta/2,\eta/2,\{\lambda\}))}{\prod_{j>k} \varphi(\mu_k-\mu_j) \prod_{\alpha<\beta,\beta>2}\varphi(\lambda_\beta - \lambda_\alpha) } \nonumber \\
& & + \sum_{2<\beta} \frac{ \varphi'(\lambda_\beta-\eta/2)}{ \varphi(\lambda_\beta - \eta/2 ) } S_n(\{\mu\}|\eta/2,\eta/2,\{\lambda\}) \,,
\end{eqnarray}
where $\tilde{H}''$ is obtained from $H$  by differentiating the first column twice:
\begin{equation}
\tilde{H}''_{a0}=\partial^2_x \frac{\varphi(\eta)}{\varphi(\mu_a-\eta/2-x)} \prod_{k\neq a} \varphi(\mu_k-x+\eta/2)\Bigg|_{x=0}.
\end{equation}


\section{Numerical results for the degenerate case}\label{NumericsAppendix}

Within this Appendix we provide numerical evidence for the results \eqref{MainRes1} and \eqref{MainRes2} given in Section \ref{PartlyNonGeneric}. Throughout this Appendix we take $x=\pi$, so that $\q$ is not a root of unity. We consider states with degenerate conformal weights $h=h_{r,s}$ and/or $\bar{h}=h_{r,s}$. We shall use the same notation $|u\rangle,|v\rangle,|w\rangle...$ for scaling states and the corresponding states in the continuum.


\subsection{The case of $\AStTL{j}{1}$}\label{spin_untwisted}

The main goal of this section is to provide numerical support for the conjecture \eqref{MainRes1} and the concept of strong duality \eqref{strong}.
We start by showing explicitly in Table \ref{ED_comparisons} the duality for conjugate states for the first few sizes, giving an example of how matrix elements for raising operators will follow from those for the lowering operators. In this example we consider $S_z=1,\,e=\pm1$. We here also see the issue of mixing discussed in section \ref{mixing_section}.

Call the $S_z=1,\,e=-1$ primary state $|u_-\rangle$ and the $S_z=1,\,e=1$ primary state $|u_+\rangle$. The sector of relevant lattice momentum for the chiral level 1 state $|a_{-1}u_-\rangle$ is then the same as that of the anti-chiral level 1 state $|\bar{a}_{-1}u_+\rangle$. Within this sector, the two lowest-energy Bethe states have close but not identical energies. 
We call these two states $|v_1\rangle$ and $|v_2\rangle$. Since nothing in our theory favours chiral over anti-chiral, or vice versa, we must assume that we cannot identify one of $|v_1\rangle,|v_2\rangle$ to $|a_{-1}u_-\rangle$ and the other to $|\bar{a}_{-1}u_+\rangle$. Instead, to keep their energy on the lattice identical, we must consider them as linear combinations, containing equal parts of the two states. We identify $|\bar{a}_{-1} u_+\rangle =  \frac{1}{\sqrt{2}}(|v_1\rangle + |v_2\rangle)$ and $|a_{-1} u_-\rangle = \frac{1}{\sqrt{2}} (|v_1\rangle - |v_2\rangle)$ (with the phases of the eigenvectors $|v_1\rangle,|v_2\rangle$ fixed to give the same phase for the relevant matrix elements.) We then obtain the results in Table  \ref{ED_comparisons_unmixed}, which are in line with our conjectures.


\begin{table}[H]
\center
\begin{minipage}{0.5\linewidth}
\center
\begin{tabular}{llll}
 \vspace*{5pt}     & $\langle v_1 | {\mathcal L}_{-1} | u_- \rangle$ & $\langle v_2 | {\mathcal L}_{-1} | u_- \rangle$ & $\langle \bar{a}_{-1}u_- |\bar{\mathcal L}_{-1}| u_- \rangle$    \\
$N$     &$\langle u_+ |\bar{\mathcal L}_{1}| v_1 \rangle$  & $\langle u_+ |\bar{\mathcal L}_{1}| v_2 \rangle$  & $\langle u_+| {\mathcal L}_{1} | a_{-1}u_+ \rangle$    \\
\hline
8 & 0.0526552 & 0.08033931 & 0.71811371 \\
10 & 0.04744019 & 0.07094947 & 0.87777283 \\
12 & 0.04357369 & 0.06336333 & 0.9762023 \\
14 & 0.04048884 & 0.0573236 & 1.03974472 \\
16 & 0.03793402 & 0.05245015 & 1.08270653 \\
18 & 0.03576776 & 0.04844714 & 1.11294418 \\
20 & 0.03389957 & 0.04510245 & 1.13495942 \\
22 & 0.03226707 & 0.04226491 & 1.15145137 \\
$p_{7}$ & 0.00580449 & 0.00549332 & 1.23163297 \\
 \bf{conj }& 0   & 0    & 1.23170369 \\
\hline
\end{tabular}
\end{minipage}%
\begin{minipage}{0.5\linewidth}
\center
\begin{tabular}{llll}
 \vspace*{5pt}     & $\langle v_1 | \bar{\mathcal L}_{-1} | u_+ \rangle$ & $\langle v_2 | \bar{\mathcal L}_{-1} | u_+ \rangle$ & $\langle {a}_{-1}u_+ |{\mathcal L}_{-1}| u_+ \rangle$    \\
$N$     &$\langle u_- |{\mathcal L}_{1}| v_1 \rangle$  & $\langle u_- |{\mathcal L}_{1}| v_2 \rangle$  & $\langle u_-| \bar{\mathcal L}_{1} | \bar{a}_{-1}u_- \rangle$    \\
\hline
8 & 0.19977481 & 0.30480884 & 0.99943154 \\
10 & 0.20854526 & 0.31189113 & 1.19506516 \\
12 & 0.21572193 & 0.31369525 & 1.31480472 \\
14 & 0.22148634 & 0.31357761 & 1.39180979 \\
16 & 0.22616506 & 0.31271118 & 1.44376243 \\
18 & 0.23002327 & 0.3115647 & 1.48028264 \\
20 & 0.23325554 & 0.31034011 & 1.50685293 \\
22 & 0.23600264 & 0.30912729 & 1.52674922 \\
$p_{7}$ & 0.27168088 & 0.2829496 & 1.62371173 \\
\bf{conj}  & 0.27723073 & 0.27723073 & 1.62376715 \\
\hline
\end{tabular}
\end{minipage}%

\caption{Matrix elements of $\KSgen_{\pm 1}$ and $\bar{\KSgen}_{\pm 1}$ in the sector of $S_z=1$, at $x=\pi$. We call $|u_-\rangle$ the primary state at $e=-1$ and $|u_+\rangle$ the primary state at $e=1$ (both in the $N\to\infty$ limit), with charges given by \eqref{alpha}. $|v_1\rangle$ and $|v_2\rangle$ are linear combinations of $|a_{-1} u_-\rangle $ and $|\bar{a}_{-1} u_+\rangle$ containing equal parts of both states, as discussed in Section \ref{mixing_section}. We here denote the corresponding scaling states on the lattice by the same labels as the states in the limits. 
The matrix elements are computed at increasing lattice size $N$, after which polynomial extrapolations $p_n(1/N)$ of degree $n= 7$ to all the data points is made in order to approximate the value at $N \rightarrow \infty$. The CFT value (``conj'') $\langle a_{-1}u|L_{-1}u\rangle$ for a primary state $u$ with charge $\alpha$ is $\sqrt{2}\alpha$, as described in Section \ref{sec:KS_Lminus1}. Due to $|v_1\rangle,|v_2\rangle$ being mixed states, the nonzero matrix elements involving these states are conjectured to come with a factor of $\frac{1}{\sqrt{2}}$: while the full value is 0.39206346 ($\alpha=2\alpha_0$) we here instead conjecture a value of 0.27723073 each. 
%
%
}
\label{ED_comparisons}
\end{table}

\begin{table}[H]
\center
\begin{minipage}{0.5\linewidth}
\center
\begin{tabular}{lll}
$N$    & $\langle \bar{a}_{-1} u_+  | {\mathcal L}_{-1} | u_- \rangle$ & $\langle a_{-1} u_- | {\mathcal L}_{-1} | u_- \rangle$   \\
\hline
8 & 0.09388641 & 0.02030561 \\
10 & 0.08364162 & 0.01698464 \\
12 & 0.07557809 & 0.01419611 \\
14 & 0.06914237 & 0.01202808 \\
16 & 0.0638982 & 0.01034544 \\
18 & 0.05954055 & 0.00902118 \\
20 & 0.05585724 & 0.00796118 \\
22 & 0.05269816 & 0.00709863 \\
$p_7$ & 0.00798875 & \hspace*{-3.5pt}-0.00022102 \\
\bf{conj} & 0 & 0 \\
\hline
\end{tabular}
\end{minipage}%
\begin{minipage}{0.5\linewidth}
\center
\begin{tabular}{lll}
$N$   & $\langle \bar{a}_{-1} u_+ | \bar{\mathcal L}_{-1} | u_+ \rangle$ & $\langle a_{-1} u_-  | \bar{\mathcal L}_{-1} | u_+ \rangle$  \\
\hline
8 & 0.35679453 & 0.07427027 \\
10 & 0.3680041 & 0.07307657 \\
12 & 0.37435448 & 0.0692776 \\
14 & 0.37834734 & 0.06511836 \\
16 & 0.38104305 & 0.06119735 \\
18 & 0.38296053 & 0.0576585 \\
20 & 0.38438017 & 0.05450702 \\
22 & 0.38546507 & 0.05170693 \\
$p_7$ & 0.39218298 & 0.00796819 \\
\bf{conj} & 0.39206346 & 0 \\
\hline
\end{tabular}
\end{minipage}%

\vspace*{10pt}

\begin{minipage}{0.5\linewidth}
\center
\begin{tabular}{lll}
$N$     &$\langle u_+ |\bar{\mathcal L}_{1}| \bar{a}_{-1} u_+  \rangle$ & $\langle u_+ |\bar{\mathcal L}_{1}| a_{-1} u_- \rangle$  \\
\hline
8 & 0.01957562 & 0.09404132 \\
10 & 0.01662357 & 0.08371413 \\
12 & 0.01399339 & 0.07561589 \\
14 & 0.01190397 & 0.06916384 \\
16 & 0.01026446 & 0.06391126 \\
18 & 0.00896568 & 0.05954893 \\
20 & 0.00792163 & 0.05586287 \\
22 & 0.00706954 & 0.05270207 \\
$p_7$ & \hspace*{-3.5pt}-0.00022003 & 0.00798876 \\
\bf{conj} & 0 & 0 \\
\hline
\end{tabular}
\end{minipage}%
\begin{minipage}{0.5\linewidth}
\center
\begin{tabular}{lll}
$N$     &$\langle u_- |{\mathcal L}_{1}|\bar{a}_{-1} u_+ \rangle$  & $\langle u_- |{\mathcal L}_{1}| a_{-1} u_-  \rangle$   \\
\hline
8 & 0.07703985 & 0.35620678 \\
10 & 0.07466382 & 0.36768535 \\
12 & 0.07028121 & 0.37416736 \\
14 & 0.06579726 & 0.37822987 \\
16 & 0.06168015 & 0.38096519 \\
18 & 0.0580154 & 0.38290662 \\
20 & 0.05477911 & 0.38434149 \\
22 & 0.05191965 & 0.38543648 \\
$p_7$ & 0.00796872 & 0.39218313 \\
\bf{conj} & 0  & 0.39206346 \\
\hline
\end{tabular}
\end{minipage}%

\caption{Matrix elements of $\KSgen_{\pm 1}$ and $\bar{\KSgen}_{\pm 1}$ in the sector of $S_z=1$, at $x=\pi$, with the same conventions as in Table \ref{ED_comparisons}. Having separated $|\bar{a}_{-1} u_+\rangle$ and $| a_{-1} u_- \rangle$ from each other we here obtain the full conjectured norm. 
We note that strong duality does not apply to this table, since we no longer deal with single Bethe states but rather linear combinations thereof.
}\label{ED_comparisons_unmixed}
\end{table}

\FloatBarrier

With this result in mind we then restrict our attention to only considering lowering operators in tables \ref{level1}--\ref{level2opposite}. In these four tables we look at the null states at level 1 and 2 for $\AStTL{1}{1}$ and $\AStTL{2}{1}$; the use of form factors here enables us to access higher sizes.
For comparison we show in each table both the action of $A_{r,s}$ and $\bar{A}_{r,s}$, even though only one of the conformal weights of the primary state is degenerate. 
 
We also restrict our attention to $|v_1\rangle$ in the case of mixing at level 1, having already seen in tables \ref{ED_comparisons}--\ref{ED_comparisons_unmixed} that we recover the full norm when taking both $|v_1\rangle$ and $|v_2\rangle$ into account. The reason for this restriction is technical: when using form factors instead of exact diagonalization, $|v_2\rangle$ corresponds to a singular Bethe state and would require regularization that in turn would perturb the numerical results. 
We note that when excluding singular states in this way in the cases of overlap, the results at both level 1 and 2 do not match the conjecture as well as in the cases of no overlap. Meanwhile we see in Table \ref{ED_comparisons_unmixed} that the agreement when including all singular states and taking the proper linear combinations is comparable to the cases of no overlap. 


\subsubsection{Considerations at level 2}\label{level2_considerations}
At level 2 we typically expect two orthogonal states, which we call $|w_1\rangle$ and $|w_2\rangle$. To get the total projection onto level 2 we consider
\begin{equation}\label{level2norm}
| \chi |_2 \equiv \sqrt{ |\langle w_1 | \chi \rangle|^2 + |\langle w_2 | \chi \rangle|^2}
\end{equation}
and same for the anti-chiral quantities. In the case of degenerate conformal weights we furthermore again have the issue of overlap, 
so that we once again must take into account twice as many states. However, the fourth state is a singular Bethe state, so we do not include it in our form-factor computations to avoid regularization. We label the norms with and without this fourth state as
\begin{subequations}\label{level2norm2}
\begin{eqnarray}
| \chi |_2^{*} & \equiv & \sqrt{ |\langle w_1 | \chi \rangle|^2 + |\langle w_2 | \chi \rangle|^2+ |\langle w_3 | \chi \rangle|^2+ |\langle w_4 | \chi \rangle|^2} \,, \\  \label{level2norm2a}
| \chi |_2^{**} &\equiv & \sqrt{ |\langle w_1 | \chi \rangle|^2 + |\langle w_2 | \chi \rangle|^2+ |\langle w_3 | \chi \rangle|^2} \,.   \label{level2norm2b}
\end{eqnarray}
\end{subequations}
We show the result for $| \chi |_2^{*}$ obtained from exact diagonalization in Table \ref{level2_all4} at $S_z=2,e=1$, before restricting to $| \chi |_2^{**}$ when going to higher sizes with form factors. 

When using form factors, we need to know what conjecture to consider when we exclude the fourth, singular Bethe state. It turns out that the second and third Bethe states, which are found in a degenerate eigenspace of the Hamiltonian, have an almost zero contribution to the norm (or order $\mathcal{O}(10^{-3})$, respectively $\mathcal{O}(10^{-5})$, when extrapolating the form-factor results). Meanwhile the first and fourth states are at different energies, and so by the same argument as for level 1 (i.e., that we cannot favour chiral over anti-chiral, and so their energy must be the same) we expect that the two Bethe states contribute equally. Excluding the fourth state will thus roughly remove half the norm squared at $S_z=2,e=1$, up to the $\mathcal{O}(10^{-3})$ contribution from the second and third states. Meanwhile, when we consider the chiral side at $S_z=2,e=-1$ the conjecture simply remains zero.

\begin{table}[H]
\center
\begin{tabular}{lll}
$N$ & $| \chi |_2^{*}$ & Full norm \\
\hline
10 & 0.45377412 &  0.45904775   \\
12 & 0.47363006 &  0.47883132   \\
14 & 0.48751301 &  0.49234698   \\
16 & 0.4976002 &  0.50267043   \\
18 & 0.50516327 &  0.51126303   \\
20 & 0.51098601 &  0.51895827   \\
22 & 0.51557136 &  0.52628211   \\
$p_5$ & 0.54746072  & 0.84670007 \\
$p_6$ & 0.54780548  & 1.02083064 \\
\bf{conj} & 0.54675127 \\
%
\hline
\end{tabular}
\caption{Norm at level 2 in the case of overlap. Shown is projection onto the four relevant states. In the right column is the full norm for comparison. The presence of parasitic couplings in the latter case leads to a different result, which moreover cannot be reliably extrapolated. The conventions used for the extrapolations $p_5$ and $p_6$ are the same as in Table \ref{generic_example}.}\label{level2_all4}
\end{table}

\subsubsection{Form factor results for $\AStTL{j}{1}$}\label{FF_results_Wj1}


For all form factor results we take the most extreme cutoff in any product of Koo-Saleur generators. (See further the discussion in Section \ref{scaling_weak}.)

\begin{table}[H]
\center
\begin{minipage}{0.5\textwidth} 
\center
\begin{tabular}{lll}
$N$      & $\langle v_1| {\mathcal L}_{-1} | u_- \rangle$  & $\langle \bar{a}_{-1}u_- |\bar{\mathcal L}_{-1}| u_- \rangle$    \\
\hline
 10   & 0.04744019 & 0.87777283 \\
 12   & 0.04357369 & 0.9762023  \\
 14   & 0.04048884 & 1.03974472 \\
 16   & 0.03793402 & 1.08270653 \\
 18   & 0.03576776 & 1.11294418 \\
 20   & 0.03389957 & 1.13495942 \\
 22   & 0.03226707 & 1.15145137 \\
 24   & 0.0308251  & 1.16410715 \\
 26   & 0.02953989 & 1.17402095 \\
 28   & 0.02838551 & 1.18192562 \\
 30   & 0.02734164 & 1.18832605 \\
 32   & 0.02639211 & 1.19357881 \\
 34   & 0.02552386 & 1.19794122 \\
 36   & 0.02472619 & 1.20160266 \\
 38   & 0.02399028 & 1.20470491 \\
 40   & 0.02330873 & 1.20735576 \\
 42   & 0.02267536 & 1.20963831 \\
 44   & 0.02208487 & 1.21161746 \\
 46   & 0.02153276 & 1.21334445 \\
 48   & 0.02101516 & 1.21486018 \\
\vdots & \vdots & \vdots  \\ 
\end{tabular}
\end{minipage}
\begin{minipage}{0.5\textwidth} 
\center
 \begin{tabular}{lll}
  $N$    & $\langle v_1 | {\mathcal L}_{-1} | u_- \rangle$  & $\langle \bar{a}_{-1}u_- |\bar{\mathcal L}_{-1}| u_- \rangle$    \\
\hline
\vdots & \vdots & \vdots  \\ 
 50   & 0.02052871 & 1.21619761 \\
 52   & 0.02007049 & 1.21738353 \\
 54   & 0.01963794 & 1.21843989 \\
 56   & 0.01922881 & 1.21938481 \\
 58   & 0.01884112 & 1.22023337 \\
 60   & 0.0184731  & 1.22099819 \\
 62   & 0.0181232  & 1.22168988 \\
 64   & 0.01779002 & 1.22231744 \\
 66   & 0.01747229 & 1.22288852 \\
 68   & 0.0171689  & 1.22340967 \\
 70   & 0.01687882 & 1.22388653 \\
 72   & 0.01660114 & 1.22432395 \\
 74   & 0.01633502 & 1.22472614 \\
 76   & 0.0160797  & 1.22509676 \\
 78   & 0.01583449 & 1.22543904 \\
 80   & 0.01559876 & 1.22575576 \\
 $p_{25}$ & 0.00214322 & 1.23169818 \\
 $p_{30}$ & 0.00218049 & 1.23169693 \\
 $p_{35}$ & 0.00208063 & 1.23170322 \\
 \bf{conj }& 0       & 1.23170369 \\
\end{tabular}
\end{minipage} \hfill
\caption{Matrix elements of $\KSgen_{\pm 1}$ and $\bar{\KSgen}_{\pm 1}$ in the sector of $S_z=1$, at $x=\pi$, with the same conventions as in Table \ref{ED_comparisons}. In this table $e=-1$ is considered again. Polynomial extrapolations 
 $p_n(1/N)$ of degrees $n= 25, 30, 35$ to all the data points are made in order to approximate the value at $N \rightarrow \infty$. 
%
}\label{level1}
\end{table}

\begin{table}[H]
\center
\begin{minipage}{0.5\textwidth} 
\center
\begin{tabular}{lll}
$N$      & $\langle a_{-1}u_+ | {\mathcal L}_{-1} | u_+ \rangle$  & $\langle v_1 |\bar{\mathcal L}_{-1}| u_+ \rangle$    \\
\hline
10   & 1.19506516 & 0.20854526 \\
 12   & 1.31480472 & 0.21572193 \\
 14   & 1.39180979 & 0.22148634 \\
 16   & 1.44376243 & 0.22616506 \\
 18   & 1.48028264 & 0.23002327 \\
 20   & 1.50685293 & 0.23325554 \\
 22   & 1.52674922 & 0.23600264 \\
 24   & 1.5420146  & 0.23836731 \\
 26   & 1.55397203 & 0.24042566 \\
 28   & 1.56350664 & 0.24223501 \\
 30   & 1.57122775 & 0.24383928 \\
 32   & 1.57756543 & 0.2452726  \\
 34   & 1.58282995 & 0.2465619  \\
 36   & 1.58724958 & 0.24772868 \\
 38   & 1.59099515 & 0.24879034 \\
 40   & 1.59419657 & 0.24976108 \\
 42   & 1.59695398 & 0.25065264 \\
 44   & 1.59934555 & 0.25147477 \\
 46   & 1.60143303 & 0.25223568 \\
 48   & 1.60326571 & 0.25294229 \\
\vdots & \vdots & \vdots  \\ 
\end{tabular}
\end{minipage}
\begin{minipage}{0.5\textwidth} 
\center
 \begin{tabular}{lll}
 $N$    & $\langle a_{-1}u_+ | {\mathcal L}_{-1} | u_+ \rangle$  & $\langle v_1 |\bar{\mathcal L}_{-1}| u_+ \rangle$    \\
\hline
\vdots & \vdots & \vdots  \\ 
 50   & 1.60488329 & 0.25360052 \\
 52   & 1.60631806 & 0.25421543 \\
 54   & 1.60759648 & 0.25479138 \\
 56   & 1.60874039 & 0.25533217 \\
 58   & 1.60976797 & 0.25584109 \\
 60   & 1.61069443 & 0.25632105 \\
 62   & 1.61153256 & 0.25677457 \\
 64   & 1.61229323 & 0.25720392 \\
 66   & 1.61298565 & 0.25761109 \\
 68   & 1.61361774 & 0.25799785 \\
 70   & 1.61419627 & 0.25836579 \\
 72   & 1.61472713 & 0.25871633 \\
 74   & 1.61521538 & 0.25905076 \\
 76   & 1.61566545 & 0.25937022 \\
 78   & 1.61608122 & 0.25967576 \\
 80   & 1.61646607 & 0.25996833 \\
 $p_{25}$ & 1.6237699  & 0.27512229 \\
 $p_{30}$ & 1.62376682 & 0.27508581 \\
 $p_{35}$ & 1.62378719 & 0.27518577 \\
 \bf{conj }& 1.62376715  & 0.27723073\\
\end{tabular}
\end{minipage} \hfill
\caption{
Matrix elements of $\KSgen_{\pm 1}$ and $\bar{\KSgen}_{\pm 1}$ in the sector of $S_z=1$, at $x=\pi$, with the same conventions as in Table \ref{ED_comparisons}. In this table $e=1$ is considered again. Polynomial extrapolations 
 $p_n(1/N)$ of degrees $n= 25, 30, 35$ to all the data points are made in order to approximate the value at $N \rightarrow \infty$. As discussed in Section \ref{spin_untwisted}, the state $|v_2\rangle$ is not included here. The resulting additional factor of $\frac{1}{\sqrt{2}}$ has been taken into account in the conjecture.
%
}\label{level1_opposite}
\end{table}

\begin{table}[H]
\center
\begin{minipage}{0.5\textwidth} 
\center
\begin{tabular}{lll}
$N$      & $| \chi |_2^{**}$  & $|\bar{\chi} |_2$    \\
\hline
 10   & 0.24726801 & 1.20983864 \\
 12   & 0.20066837 & 1.56533417 \\
 14   & 0.16972038 & 1.9115293  \\
 16   & 0.14727663 & 2.23720271 \\
 18   & 0.13029595 & 2.52679885 \\
 20   & 0.11707361 & 2.77534399 \\
 22   & 0.10653922 & 2.98500127 \\
 24   & 0.09798136 & 3.16074314 \\
 26   & 0.09090996 & 3.30802769 \\
 28   & 0.08497837 & 3.43184718 \\
 30   & 0.07993595 & 3.53644327 \\
 32   & 0.07559799 & 3.62530048 \\
 34   & 0.07182607 & 3.70123804 \\
 36   & 0.06851485 & 3.76652378 \\
 38   & 0.06558299 & 3.82298088 \\
 40   & 0.06296693 & 3.8720785  \\
 42   & 0.06061633 & 3.91500532 \\
 44   & 0.05849094 & 3.95272782 \\
 46   & 0.0565582  & 3.98603608 \\
 48   & 0.05479155 & 4.01557964 \\
\vdots & \vdots & \vdots  \\ 
\end{tabular}
\end{minipage}
\begin{minipage}{0.5\textwidth} 
\center
 \begin{tabular}{lll}
  $N$    & $| \chi |_2^{**}$  & $|\bar{\chi} |_2$    \\
\hline
\vdots & \vdots & \vdots  \\ 
 50   & 0.0531691  & 4.04189553 \\
 52   & 0.05167268 & 4.06543021 \\
 54   & 0.05028709 & 4.08655693 \\
 56   & 0.0489995  & 4.10558933 \\
 58   & 0.04779905 & 4.12279235 \\
 60   & 0.04667642 & 4.13839083 \\
 62   & 0.04562364 & 4.15257652 \\
 64   & 0.0446338  & 4.16551368 \\
 66   & 0.04370091 & 4.17734357 \\
 68   & 0.04281975 & 4.18818823 \\
 70   & 0.04198571 & 4.19815347 \\
 72   & 0.04119475 & 4.20733135 \\
 74   & 0.04044331 & 4.21580231 \\
 76   & 0.03972819 & 4.22363682 \\
 78   & 0.03904658 & 4.23089684 \\
 80   & 0.03839593 & 4.237637   \\
 $p_{25}$ & 0.00508552 & 4.37283102 \\
 $p_{30}$ & 0.0051748  & 4.37283425 \\
 $p_{35}$ & 0.00493341 & 4.37283415 \\
 \bf{conj }& 0     & 4.37266058 \\ 
\end{tabular}
\end{minipage} \hfill
\caption{
Values of $|\chi|_2,|\chi|_2^{**}$ as defined in \eqref{level2norm},\eqref{level2norm2b}, in the sector of $S_z=2$ at $x=\pi$. The state $\chi$ in \eqref{level2norm2} is here taken to be the result of acting upon the primary state obtained for $S_z=2, e=-1$ with the lowering operator $\mathcal{A}_{1,2}$ defined from \eqref{A_operators}.
The same conventions as Table \ref{ED_comparisons} are used for the extrapolation, and the CFT values (``conj'') are computed using the general method described in Section \ref{sec:KS_Lminus1}.
Note that on the chiral side we do not project on the fourth relevant state, as discussed in Section \ref{level2_considerations}.
}\label{level2}
\end{table}

\begin{table}[H]
\center
\begin{minipage}{0.5\textwidth} 
\center
\begin{tabular}{lll}
$N$      & $| \chi |_2$  & $|\bar{\chi} |_2^{**}$    \\
\hline
 10   & 2.31045896 & 0.26408829 \\
 12   & 2.87163371 & 0.27340469 \\
 14   & 3.37941892 & 0.28280413 \\
 16   & 3.83197544 & 0.29119674 \\
 18   & 4.22184772 & 0.29847485 \\
 20   & 4.5502387  & 0.30475163 \\
 22   & 4.82397901 & 0.31017865 \\
 24   & 5.0515897  & 0.31489717 \\
 26   & 5.24122863 & 0.31902686 \\
 28   & 5.39994176 & 0.3226658  \\
 30   & 5.53353594 & 0.32589353 \\
 32   & 5.64669381 & 0.32877441 \\
 34   & 5.74315732 & 0.3313607  \\
 36   & 5.82591007 & 0.33369509 \\
 38   & 5.89733494 & 0.33581262 \\
 40   & 5.95934224 & 0.33774227 \\
 42   & 6.01347121 & 0.33950818 \\
 44   & 6.06096911 & 0.34113058 \\
 46   & 6.10285264 & 0.34262654 \\
 48   & 6.13995559 & 0.34401053 \\
\vdots & \vdots & \vdots  \\ 
\end{tabular}
\end{minipage}
\begin{minipage}{0.5\textwidth} 
\center
 \begin{tabular}{lll}
  $N$    & $| \chi |_2$  & $|\bar{\chi} |_2^{**}$    \\
\hline
\vdots & \vdots & \vdots  \\ 
 50   & 6.17296586 & 0.34529493 \\
 52   & 6.20245427 & 0.34649035 \\
 54   & 6.2288972  & 0.34760596 \\
 56   & 6.25269438 & 0.34864969 \\
 58   & 6.27418296 & 0.34962847 \\
 60   & 6.29364876 & 0.35054836 \\
 62   & 6.31133522 & 0.35141469 \\
 64   & 6.32745061 & 0.35223215 \\
 66   & 6.34217393 & 0.3530049  \\
 68   & 6.35565962 & 0.35373665 \\
 70   & 6.36804149 & 0.35443068 \\
 72   & 6.37943587 & 0.35508996 \\
 74   & 6.38994429 & 0.35571713 \\
 76   & 6.39965564 & 0.35631456 \\
 78   & 6.40864803 & 0.35688442 \\
 80   & 6.41699028 & 0.35742864 \\
 $p_{25}$ & 6.58058318 & 0.38348308 \\
 $p_{30}$ & 6.580578   & 0.38343063 \\
 $p_{35}$ & 6.58061875 & 0.38357299 \\
 \bf{conj}& 6.5805709 & 0.387$+\mathcal{O}(10^{-3})$ \\ 
\end{tabular}
\end{minipage} \hfill
\caption{
Values of $|\chi|_2,|\chi|_2^{**}$ as defined in \eqref{level2norm},\eqref{level2norm2b}, in the sector of $S_z=2$ at $x=\pi$. The state $\chi$ in \eqref{level2norm2} is here taken to be the result of acting upon the primary state obtained for $S_z=2, e=1$ with the lowering operator $\mathcal{A}_{1,2}$ defined from \eqref{A_operators}.
The same conventions as Table \ref{ED_comparisons} are used for the extrapolation, and the CFT values (``conj'') are computed using the general method described in Section \ref{sec:KS_Lminus1}.  
Note that on the anti-chiral side we do not project on the fourth relevant state. The effect of this on our conjecture is explained in Section \ref{level2_considerations}.
%
}\label{level2opposite}
\end{table}

\FloatBarrier

\subsection{The case of $\AStTL{0}{\q^{\pm2}}$}\label{spin_twisted}

We here wish to provide, in particular, support for the conjecture \eqref{MainRes2}.
The results are shown in Table \ref{level1twisted}. 
We here show how the indecomposability appears already at finite size in this particular case. We also again see the duality of conjugate states. While our focus here is on small system sizes and the indecomposability of the Temperley-Lieb modules, we still show the extrapolation as well---we see that already for these sizes it is quite close to the conjectured value.

\begin{table}[H]
\center
\begin{minipage}{.5\textwidth}
\center

\begin{tabular}{lll}
 \vspace*{5pt}    & $\langle a_{-1}u_{\q^2} | {\mathcal L}_{-1} | u_{\q^2} \rangle$  & $\langle \bar{a}_{-1}u_{\q^2} |\bar{\mathcal L}_{-1}| u_{\q^2} \rangle$    \\
   $N$  & $\langle u_{\q^{-2}} |\bar{\mathcal L}_{1}| \bar{a}_{-1}u_{\q^{-2}} \rangle$     & $\langle u_{\q^{-2}}| {\mathcal L}_{1} | a_{-1}u_{\q^{-2}} \rangle$   \\
\hline
 \\
  \\
   \\
   \\
8-22 &$ \mathcal{O}(10^{-15})$&$ \mathcal{O}(10^{-15})$\\
\\
 \\
  \\
  \\
 \bf{conj}  & 0  & 0    \\
\hline
\end{tabular}
\end{minipage}%
\begin{minipage}{.5\textwidth}
\center

\begin{tabular}{lll}
  \vspace*{5pt}     & $\langle a_{-1}u_{\q^{-2}} | {\mathcal L}_{-1} | u_{\q^{-2}} \rangle$  & $\langle \bar{a}_{-1}u_{\q^{-2}} |\bar{\mathcal L}_{-1}| u_{\q^{-2}} \rangle$   
   \\
     $N$   & $\langle u_{\q^2} |\bar{\mathcal L}_{1}| \bar{a}_{-1}u_{\q^2} \rangle$    & $\langle u_{\q^2}| {\mathcal L}_{1} | a_{-1}u_{\q^2} \rangle$   \\
\hline
8 & 0.37384355 & 0.37384355 \\
10 & 0.38058992 & 0.38058992 \\
12 & 0.3842338 & 0.3842338 \\
14 & 0.38641194 & 0.38641194 \\
16 & 0.38781199 & 0.38781199 \\
18 & 0.38876228 & 0.38876228 \\
20 & 0.38943519 & 0.38943519 \\
22 & 0.3899281 & 0.3899281 \\
$p_7$ & 0.39204719 & 0.39204719 \\
  \bf{conj}  & 0.39206346  & 0.39206346 \\
\hline
\end{tabular}
\end{minipage}
\caption{Matrix elements  of $\KSgen_{\pm 1}$ and $\bar{\KSgen}_{\pm 1}$ in the sector of $S_z=0$, at $x=\pi$, $|e| =  1$ and twisted boundary conditions, with the same conventions as in Table \ref{ED_comparisons}. We call $|u_{\q^2}\rangle$ the primary state at twisted boundary conditions $e_\phi=\alpha_{-}/\alpha_{+}$, corresponding to the module $\AStTL{0}{\q^{2}}$, and $|u_{\q^{-2}}\rangle$ the primary state at twisted boundary conditions $e_\phi=-\alpha_{-}/\alpha_{+}$, corresponding to the module $\AStTL{0}{\q^{-2}}$.
}\label{level1twisted}
\end{table}

\FloatBarrier

\newpage
\subsection{Relevant Bethe roots}\label{BetheRootAppendix}



In this section we list the Bethe roots used in the numerical results at $x=\pi,N=10$ and with the twist $\phi$ as specified in each case. The roots for other values of $x$ and the twist $\phi$ can be reached numerically by gradually modifying $x,\phi$, using the previous roots as the starting guess of the numerical solver in each step. The roots at larger system sizes $N$ are found in the same fashion. These roots can be inserted into the form factors listed in Appendix \ref{FormFactorAppendix} in order to reproduce our numerical results. To find these sets of roots and integers we have used the values listed in \cite{Olga} at $\Delta=0.7$ as starting guesses. We remind the reader that as discussed in \cite{Olga} it is expected that some Bethe integers may coincide.

We first show Bethe roots corresponding to the non-degenerate example in Section~\ref{sec:lattVir1}, in which $\phi=1/10$, $S_z=1$. 
The roots corresponding to the states in \eqref{p0states} are given by
\begin{equation}
\begin{split}
| u_1\rangle \leftrightarrow &\;\{ -0.26458064 , -0.07736986 , 0.0733926 , 0.25830959 \} \,, \\
| u_2\rangle \leftrightarrow &\;\{ -0.6743697 , -0.07413158 , 0.07080173 , 0.65194697 \} \,, \\
| u_3\rangle \leftrightarrow&\;\{ -0.66175052 , -0.24704638 , 0.24235079 , 0.64024589 \} \,, \\
\end{split}
\end{equation}
while the roots corresponding to  the states in \eqref{p1states} are given by
\begin{equation}
\begin{split}
| v_1\rangle \leftrightarrow&\;\{ -0.65697428 , -0.06400932 , 0.08388252 , 0.26957264 \} \,, \\
| v_2\rangle \leftrightarrow&\;\{ -0.6485393 , -0.23998851 , 0.07699763 , 0.66517052 \} \,, \\
| v_3\rangle \leftrightarrow&\;\{ -0.29843157+i \pi/2 , -0.04047671 , 0.0969293 , 0.69411263 \} \,.\\
\end{split}
\end{equation}
These are also used for Table \ref{hi} in Appendix \ref{UnitaryXXZ}.

Meanwhile for the degenerate examples the roots used to produce the results in Appendix \ref{FF_results_Wj1} are given as follows.

At $S_z=1, e=-1$, the Bethe roots necessary to reproduce Table \ref{level1} are
\begin{equation}
\begin{split}
| u_- \rangle \leftrightarrow& \;  \{ -0.59190986 , -0.2125926 , 0.10678414 , -0.04262483 \}
 \;\text{ with integers}\; \left\{ -\frac{5}{2}, -\frac{3}{2},\frac{1}{2},-\frac{1}{2} \right\} \,, \\
| \bar{a}_{-1} u_-\rangle \leftrightarrow&\; \{ -0.60569326 , -0.22013419 , 0.28934619 , -0.0495778 \}
 \;\text{ with integers}\; \left\{ -\frac{5}{2}, -\frac{3}{2},\frac{3}{2},-\frac{1}{2} \right\} \,, \\
| v_1\rangle \leftrightarrow& \; \{ -0.1439582 , 0.1439582 , i\pi/2 , 0 \}  \;\text{ with integers}\; \left\{ -\frac{1}{2}, \frac{1}{2},\frac{9}{2},-\frac{1}{2} \right\} \,. \\
\end{split}
\end{equation}
The states $| u_+\rangle,  | {a}_{-1} u_+ \rangle $ relevant for Table \ref{level1_opposite} are found by taking the opposite signs of the Bethe roots above. (State $|v_1\rangle$ remains the same under this action---recall that the root $i\pi/2$ is self-conjugate.)

At $S_z=2, e=-1$  let us denote the primary state by $|u_{-}^{(2)}\rangle$. In the context of forming the norms \eqref{level2norm} and \eqref{level2norm2b} we now distinguish descendants on the chiral resp.\ anti-chiral sides by superscripts: $| w_1^{l} \rangle, | w_2^{l} \rangle, | w_3^{l} \rangle$ resp.\ $| w_1^{r} \rangle, | w_2^{r} \rangle$. The Bethe roots necessary to reproduce Table \ref{level2} are
\begin{equation}
\begin{split}
| u_-^{(2)} \rangle \leftrightarrow&\; \{ 0.0199709 , -0.12746759 , -0.33759576 \} \;\text{ with integers}\; \{0,-1,-2\} \\
|a_{-1} u_-^{(2)} \rangle \leftrightarrow&\; \{ 0.03251715 , -0.10949347 , -0.9283113 \} \;\text{ with integers}\;  \{0,-1,-3\}\\
|\bar{a}_{-1} u_-^{(2)} \rangle \leftrightarrow&\; \{ 0.16535175 , -0.13383418 , -0.34635301 \} \;\text{ with integers}\; \{1,-1,-2\} \\
|  w_1^{l} \rangle \leftrightarrow&\; \{ -0.06631605 , 0.06631605 , i\pi/2 \} \;\text{ with integers}\; \{0,0,-5\} \,, \\
|  w_2^{l} \rangle \leftrightarrow&\; \{ -0.03902337 , 0.28513457 , 0.88928688 \}
 \;\text{ with integers}\; \{0,2,3\} \,, \\
|  w_3^{l} \rangle \leftrightarrow&\; \{ 0.03902337 , -0.28513457 , -0.88928688 \} \;\text{ with integers}\; \{0,-2,-3\} \,, \\
|  w_1^{r} \rangle \leftrightarrow&\; \{ 0.15878868 , 0.00783529 , -0.35707322 \} \;\text{ with integers}\; \{1,0,-2\} \,, \\
|  w_2^{r} \rangle \leftrightarrow&\; \{ 0.37558835 , -0.14134827 , -0.35581938 \} \;\text{ with integers}\;  \{2,-1,-2\} \,.\\
\end{split}
\end{equation}

Once more the states relevant for Table \ref{level2opposite} are found by taking the opposite signs of these Bethe roots. (The set of states $| w_1^{l} \rangle,|  w_2^{l} \rangle ,|  w_3^{l} \rangle $ is invariant under this action---the first has real roots symmetric around zero and a self-conjugate imaginary root $i\pi/2$, while the other two have roots with opposite signs to one another.)



\FloatBarrier



\newpage
\section{Proof of \eqref{separate_conj}}\label{Proof_Appendix}

We wish to prove \eqref{separate_conj_b}, from which \eqref{separate_conj_a} then also follows thanks to the known result \eqref{KS-3-41}. 
Within this Appendix, we let $\langle \mathcal{O}\rangle$ refer to the ground-state expectation value of an operator $\mathcal{O}$ in the limit $N\to\infty$. Using the parity of the ground state we can restate \eqref{separate_conj_b} as
\begin{equation}
 \langle e_je_{j+1} \rangle = {3\sin^3 \! \gamma \: I_1-\sin\gamma \: I_0\over \cos\gamma}.
\end{equation}

In \cite{Go_Kato} the following ground-state expectation values are given for spin operators $S^a=\sigma^a/2,\; a=x,y,z$:
\begin{subequations}\label{spin_expectation}
\begin{eqnarray}
\langle  S_j^x S_{j+2}^x \rangle &=& 
-\frac{1}{2\pi\sin 2\gamma} I^{(0)} - \frac{3\cos 2\gamma \tan \gamma }{4\pi^3}I^{(2)}+\frac{\cos 2\gamma}{4\pi^2} I^{(1)} + \frac{\sin^2\!\gamma}{4\pi^4}I^{(3)},
\\
\langle  S_j^z S_{j+2}^z \rangle &=& 
\frac{1}{4} + \frac{\cot 2\gamma}{\pi} I^{(0)} + \frac{3 \tan \gamma}{2\pi^3} I^{(2)} 
- \frac{1}{2\pi^2}I^{(1)} - \frac{\sin^2\!\gamma}{2\pi^4} I^{(3)},\\ 
\langle  S_j^x S_{j+1}^x \rangle &=& 
-\frac{1}{4\pi \sin\gamma} I^{(0)} + \frac{\cos\gamma}{4\pi^2} I^{(1)},\\
\langle  S_j^z S_{j+1}^z \rangle &=&
\frac{1}{4}+\frac{\cot\gamma}{2\pi} I^{(0)} - \frac{1}{2\pi^2}I^{(1)},
\end{eqnarray}
\end{subequations}
where we have introduced the short-hand notations
\begin{subequations}
\begin{eqnarray}
I^{(0)} &=& \int_{-\infty}^\infty \frac{\sinh (1-\nu)t}{\cosh \nu t \, \sinh t} \, \mathrm{d} t, \\
I^{(1)} &=& \int_{-\infty}^\infty t \frac{\cosh t }{\cosh^2\!\nu t \, \sinh t}  \, \mathrm{d} t,
\\
I^{(2)} &=& \int_{-\infty}^\infty t^2 \frac{\sinh (1-\nu)t}{\cosh \nu t \, \sinh t} \, \mathrm{d} t ,\\
I^{(3)} &=& \int_{-\infty}^\infty t^3 \frac{\cosh t }{\cosh^2\!\nu t \, \sinh t}  \, \mathrm{d} t,
\end{eqnarray}
\end{subequations}
with $\gamma=\pi\nu$. 
By comparison with \eqref{I0-def} and \eqref{I1-def} we see that $I_0 = I^{(0)}/\pi$ and $I_1 = I^{(2)}/\pi^3$.

Using the expression \eqref{TLspin} of the Temperley-Lieb generators in terms of Pauli matrices we now rewrite $\langle e_je_{j+1} \rangle$ in terms of spin operators. We use  $\sigma^a\sigma^b = \delta_{ab}\mathbf{1}+i\epsilon_{abc}\sigma^c$ to simplify the products, and by symmetry we may discard any resulting term that involves an odd number of any given spin operator. We obtain
\begin{equation}
\begin{split}
4 \langle e_j  e_{j+1} \rangle &= \Big\langle \Big(
\sigma^x \sigma_{j+1}^x + \sigma_j^y \sigma_{j+1}^y + \cos \gamma \left( \sigma_j^z \sigma_{j+1}^z - \mathbf{1} \right) + i \sin\gamma \left( \sigma_j^z - \sigma_{j+1}^z \right)  
\Big)\\
& \hspace*{3cm}
\Big(
\sigma_{j+1}^x \sigma_{j+2}^x + \sigma_{j+1}^y \sigma_{j+2}^y + \cos \gamma \left( \sigma_{j+1}^z \sigma_{j+2}^z -  \mathbf{1} \right) + i \sin\gamma \left( \sigma_{j+1}^z - \sigma_{j+2}^z \right) 
\Big) \Big\rangle \\
&= 
%
%
\langle \sigma_j^x \sigma_{j+2}^x \rangle +\langle \sigma_j^y \sigma_{j+2}^y \rangle + \cos^2\!\gamma \left( \langle \sigma_j^z \sigma_{j+2}^z \rangle +1 \right)\\
&\quad\;
- \cos\gamma  \left( \langle \sigma_j^x \sigma_{j+1}^x \rangle 
+\langle \sigma_{j+1}^x \sigma_{j+2}^x \rangle \right) 
- \cos\gamma  \left( \langle \sigma_j^y \sigma_{j+1}^y \rangle 
+\langle \sigma_{j+1}^y \sigma_{j+2}^y \rangle \right) 
\\
&\quad\;
-\cos^2\!\gamma \left(  \langle \sigma_j^z \sigma_{j+1}^z \rangle 
+\langle \sigma_{j+1}^z \sigma_{j+2}^z \rangle   \right)
+ \sin^2\!\gamma \left( \langle \sigma_j^z \sigma_{j+2}^z \rangle +1  \right)
- \sin^2\!\gamma \left(  \langle \sigma_j^z \sigma_{j+1}^z \rangle 
+\langle \sigma_{j+1}^z \sigma_{j+2}^z \rangle   \right).
\end{split}
\end{equation}
Using translation invariance of the ground state and $U(1)$ symmetry we rewrite this as
\begin{equation}\label{TLprod}
\langle e_j e_{j+1} \rangle  =2 \langle S_j^x S_{j+2}^x\rangle + \langle S_j^z S_{j+2}^z\rangle  + \frac{1}{4}    -4 \cos\gamma  \langle S_j^x S_{j+1}^x \rangle- 2  \langle S_j^z S_{j+1}^z\rangle.
\end{equation}
Inserting the results from \eqref{spin_expectation} we see immediately that the terms with $I^{(3)}$ cancel, as well as the terms involving no integrals at all, leaving
\begin{equation}
\begin{split}
\langle e_j e_{j+1} \rangle  &= 
\left( - 2\frac{1}{2\pi \sin 2\gamma}   
+ \frac{\cot 2\gamma}{\pi}
+4\cos\gamma \frac{1}{4\pi\sin\gamma}
-2 \frac{\cot \gamma}{2\pi}
 \right) I^{(0)}   \\
&\quad\;+ \left( 
2 \frac{\cos 2\gamma}{4\pi^2}
-\frac{1}{2\pi^2} - 4\cos \gamma \frac{\cos\gamma}{4\pi^2} +2 \frac{1}{2\pi^2}
  \right) I^{(1)} \\
&\quad\;+ \left( 
-2\frac{3\cos 2\gamma \tan \gamma}{4\pi^3}
+ \frac{3 \tan \gamma}{2 \pi^3}
  \right) I^{(2)}.
 \\
\end{split}
\end{equation}
By trigonometric identities we then see that the terms involving $I^{(1)}$ cancel as well, and that we finally obtain 
\begin{equation}
\langle e_j e_{j+1} \rangle  = \frac{\sin\gamma}{\pi\cos\gamma}I^{(0)}  + \frac{3 \sin^3\!\gamma}{\pi^3 \cos\gamma} I^{(2)} 
=
 {3\sin^3 \! \gamma \: I_1-\sin\gamma \: I_0\over \cos\gamma},
\end{equation}
proving \eqref{separate_conj}.

\subsection{The limit $\gamma \to 0$}\label{gamma_zero_check}
In the limit $x\to \infty,\; \gamma\to 0$ we expect that the integrals in \eqref{separate_conj} can be expressed in terms of the polylogarithms $\rm{Li}_{2n+1}(-1)$, following the result for the XXX spin chain \cite{Takahashi}
\begin{subequations}\label{spin_correlators}
\begin{eqnarray}
\langle S^z_j S^z_{j+1} \rangle &=& \frac{1}{12} - \frac{1}{3} \zeta_a(1) \,,\\
\langle S^z_j S^z_{j+2} \rangle &=& \frac{1}{12} - \frac{4}{3} \zeta_a(1) + \zeta_a(3) \, ,
\end{eqnarray}
\end{subequations}
here written in terms of the alternating zeta function
\begin{equation}
\zeta_a(s) = \sum_{n>0} \frac{(-1)^{n-1}}{n^s} = -\rm{Li}_s(-1) \,.
\end{equation}
More precisely, as $\gamma \to 0$ \eqref{TLprod} simplifies to
\begin{equation}
\langle e_je_{j+1} \rangle = 3 \langle S_j^z S_{j+2}^z\rangle  + \frac{1}{4} - 6 \langle S_j^z S_{j+1}^z\rangle \,,
\end{equation}
such that by inserting \eqref{spin_correlators} we expect that \eqref{separate_conj_b} can be written as
\begin{equation}\label{spin_correl_result}
\langle e_je_{j+1} + e_{j+1}e_j \rangle = 
\frac{1}{2}\big(1-2 + 8\zeta_a(1) + 1- 16\zeta_a(1) + 12 \zeta_a(3)\big)
= 6 \zeta_a(3) - 4 \zeta_a(1).
\end{equation}
Comparing the rescaled integrals $\sin^{2n-1}\gamma \; I_n$ in the limit of $x\to\infty$ to the integral representation
\begin{equation}
-{\rm Li}_s(-1) = \frac{1}{\Gamma(s)} \int_0^\infty \frac{t^{s-1}}{e^t+1} \, {\rm d}t \,
\end{equation}
we find
\begin{multline}
\sin^{2n-1}\gamma \; I_n   
= 2 \frac{\sin^{n-1}(\gamma) }{\gamma^{n-1}} \int_0^{\infty} t^{2n}  \frac{\sinh(xt)}{\sinh((x+1)t)\cosh(t)} \, {\rm d}t 
= 4 \frac{\sin^{2n-1}(\gamma) }{\gamma^{2n-1}} \int_0^{\infty} \frac{t^{2n}}{e^{2t} +1} \left( \frac{1-e^{-2xt}}{1-e^{-2(x+1)t}} \, {\rm d}t \right)\\
\;   \xrightarrow[\; x\to\infty\;]{} \;  \frac{4}{2^{2n +1 }} \int_0^\infty \frac{t^{2n}}{e^{t} +1} \, {\rm d}t
 = -  \frac{4}{2^{2n +1 }}  \Gamma(2n+1) \rm{Li}_{2n+1}(-1) \,.
\end{multline}
In particular $\sin\gamma \; I_0 \to  -2 \rm{Li}_1(-1) = 2 \zeta_a(1) $ and $\sin^3 \! \gamma \; I_1 \to  - \rm{Li}_3(-1) = \zeta_a(3) $, such that
\begin{equation}
 \langle e_j e_{j+1} + e_{j+1}e_j \rangle = \frac{ 6 \sin^3 \! \gamma \; I_1 -2 \sin \gamma \; I_0  }{\cos\gamma}  \;\xrightarrow[\; \gamma \to 0\;]{} \;
  6\zeta_a(3) - 4\zeta_a(1) \,
\end{equation}
indeed.

\FloatBarrier

\section{Further numerical results for Section \ref{limits_section}}
\label{limits_appendix}

\begin{figure}[h!]
\hspace*{0.4cm}\includegraphics[height=5cm, trim={0 0 6.7cm 0},clip ]{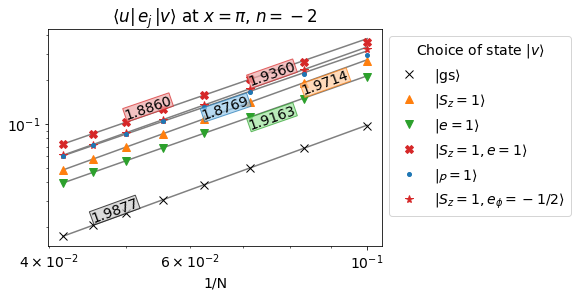}
\includegraphics[height=5cm, trim={0 0 6.8cm 0},clip ]{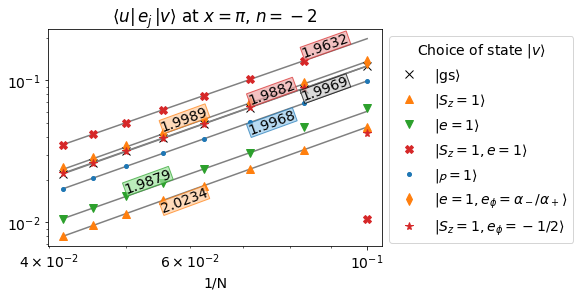}

\hspace*{0.4cm}\includegraphics[height=5cm, trim={0 0 6.7cm 0},clip ]{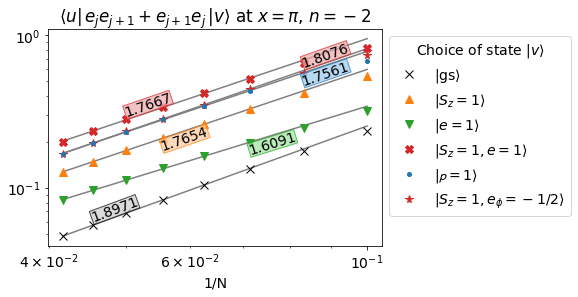}
\includegraphics[height=5cm, trim={0 0 6.8cm 0},clip ]{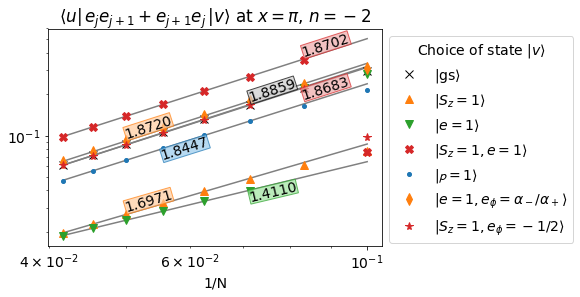}

\includegraphics[height=5cm, trim={0 0 6.7cm 0},clip ]{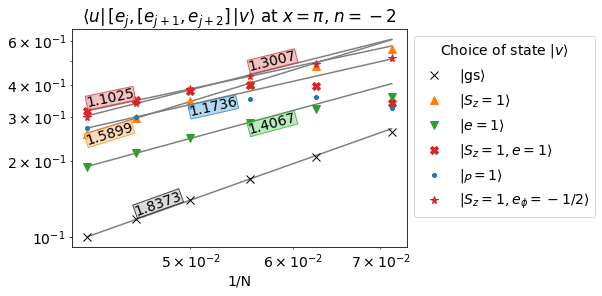}
\includegraphics[height=5cm, trim={0 0 0cm 0},clip ]{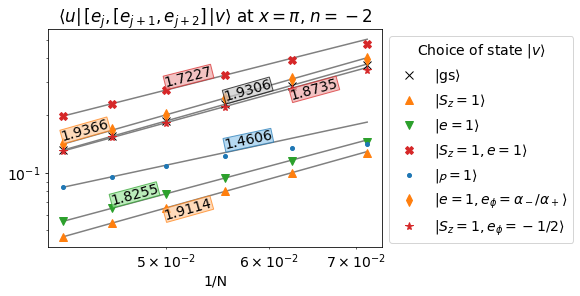}
\caption{Absolute value of matrix elements, plotted using the same conventions as in Figure \ref{separate_terms} up to the choice of $|u\rangle$: For each state $|v\rangle$ the state $|u\rangle$ corresponds to the same choice of lattice parameters up to a change in momentum sector, $\Pscaled \to \Pscaled + 2$. To the left, the lowest energy state for each choice of lattice parameters is used for $|u\rangle$. To the right, the first excited energy state is used for  $|u\rangle$ instead.
%
In case of choices of $|v\rangle$ showing particularly strong finite size effects ($|S_z=1,e=1\rangle$ and $|\mathcal{p}=1\rangle$ in the lower left plot, $|\mathcal{p}=1\rangle$ in the lower right plot) the linear fit is performed using only the two leftmost points.
}\label{allthree_2ab} 
\end{figure}

\begin{figure}[h!]
\center
\includegraphics[height=5cm, trim={0 0 6.7cm 0},clip ]{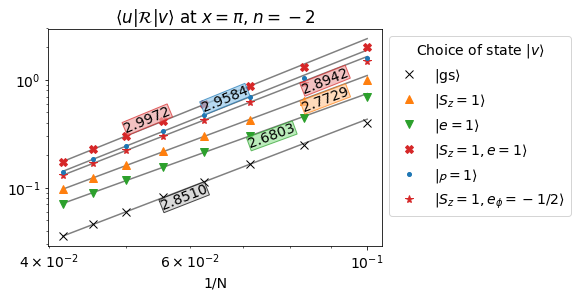} \includegraphics[height=5cm]{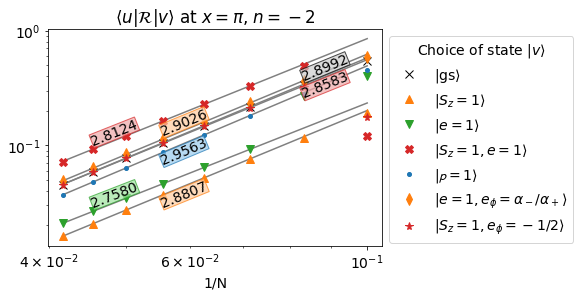}
\caption{Absolute value of matrix elements of $\remainder$ as defined in \eqref{rest}, plotted using the same conventions as in Figure \ref{separate_terms} up to the choice of $|u\rangle$: For each state $|v\rangle$ the state $|u\rangle$ corresponds to the same choice of lattice parameters up to a change in momentum sector, $\Pscaled \to \Pscaled + 2$. To the left, the lowest energy state for each choice of lattice parameters is used for $|u\rangle$. To the right, the first excited energy state is used for  $|u\rangle$ instead. 
}\label{remainder_2ab} 
\end{figure}

\begin{figure}[h!]
\center

\includegraphics[height=5cm, trim={0 0 6.7cm 0},clip ]{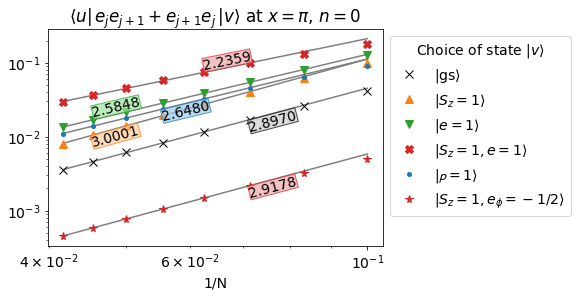}\includegraphics[height=5cm, trim={0 0 6.7cm 0},clip]{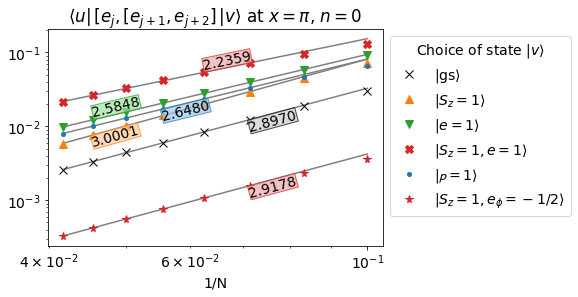}
\includegraphics[height=5cm]{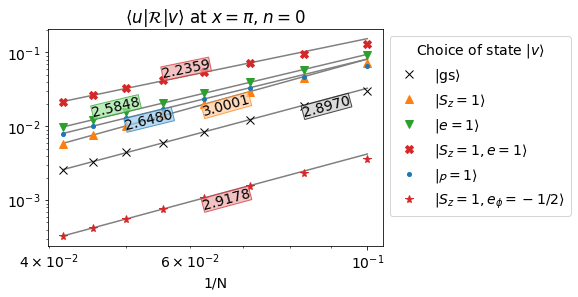}
\caption{Absolute value of matrix elements, plotted using the same conventions as in Figure \ref{separate_terms} up to the choice of $|u\rangle$: For each state $|v\rangle$ the state $|u\rangle$ is taken to be the first excited energy state corresponding to the same choice of lattice parameters that were used for $|v\rangle$. (We recall that $|v\rangle$ is taken to be the lowest-energy state). In the lowest plot $\remainder$ is as defined in \eqref{rest}.
}\label{allthree_0b} 
\end{figure}

Within this Appendix we collect figures providing further numerical evidence for the results \eqref{ndiff0_conj} and \eqref{cstar_conjecture}  given in Section \ref{limits_section} regarding the limit of the commutator $[\KSgen_{p+n}+\bar{\KSgen}_{-p-n},\KSgen_{-p}-\bar{\KSgen}_{p}] $. Throughout this Appendix we consider $x=\pi$, so that $\q$ is not a root of unity. 
In Figures \ref{allthree_2ab} and \ref{remainder_2ab} we show results for $n=-2$, while in Figure \ref{allthree_0b}  we show results for $n=0$ in the case where we do not project back on the original state, but rather on the first excited state corresponding to the same choices of lattice parameters. The figures indicate that the slope $r$ obeys $r>1$ for matrix elements of $e_j$, $e_j e_{j+1} + e_{j+1}e_j$ and $[e_j, [ e_{j+1}, e_{j+2}] ]$, while it obeys $r>2$  for the remainder $\mathcal{R}$.  
We note that in some figures there are large finite size effects making it harder to discern the tendency of the curves, in particular those involving $[e_j, [ e_{j+1}, e_{j+2}] ]$ in which we have chosen to exclude the first data points entirely.  These effects are larger for the states $|S_z=1,e=1\rangle$ and $| \mathcal{p} = 1 \rangle$, which have momenta $\mathcal{p} \notin \{0,N/2\}$ making results involving these states more sensitive to finite size effects.

It is the case for all figures, including those in Section \ref{limits_section}, that the estimated slope will vary slightly when data points for larger sizes are included.
We note that if one follows and extrapolates the estimation of the slope as a function of the sizes used for the estimate, the bounds $r>1$ and $r>2$ are still expected to hold in the continuum limit, indicating that the finite size effects have not influenced the overall conclusions regarding convergence. 



\FloatBarrier

\FloatBarrier

\section{The chiral-antichiral commutator}\label{Chiral-Antichiral}


In \cite{KooSaleur} the limit of $[\KSgen_{p+n}+\bar{\KSgen}_{-p-n},\KSgen_{-p}+\bar{\KSgen}_{p}]$ was checked, and was found to be correct. Meanwhile in Section \ref{Anomalies} and Appendix \ref{limits_appendix} we found strong evidence for $[\KSgen_{p+n}  +\bar{\KSgen}_{-p-n},\KSgen_{-p}-\bar{\KSgen}_{p}]$ having the correct limit up to the central term. In order to isolate the chiral-antichiral commutator $[\KSgen_{p+n},\bar{\KSgen}_{-p}]$
it remains to find the behaviour of $[\KSgen_{p+n}-\bar{\KSgen}_{-p-n},\KSgen_{-p}-\bar{\KSgen}_{p}]$.
We shall proceed in the same manner as in Section \ref{Anomalies} and Appendix \ref{limits_appendix}. However, while we there showed figures for two different cases at $n=0$, once case at $n=-1$ and two at $n=-2$, we shall here only reproduce the figures corresponding to one of the cases at $n=0$ and the case at $n=-1$. This is partly to save space, and partly (for the $n=-2$ cases) due to increasingly disruptive finite size effects---as was already noted in Appendix \ref{limits_appendix}, such effects become generally more pronounced for products of several Temperley-Lieb generators.

\begin{figure}[h!]
\center
\includegraphics[height=5cm, trim={0 0 6.8cm 0},clip ]{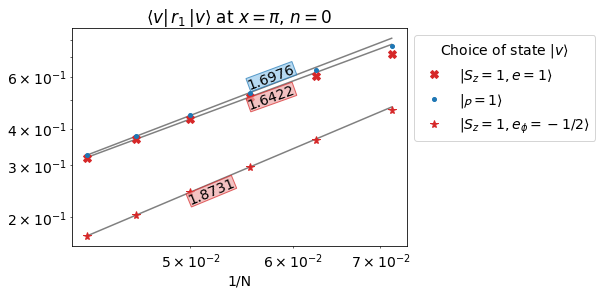}\includegraphics[height=5cm, trim={0 0 6.8cm 0},clip]{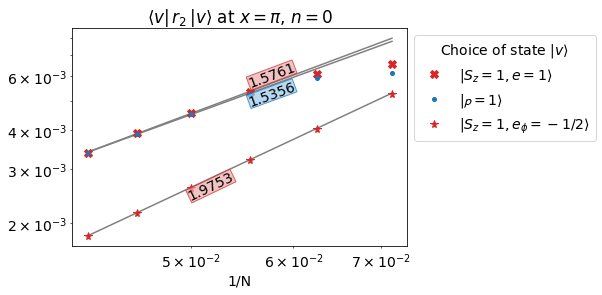}
\includegraphics[height=5cm]{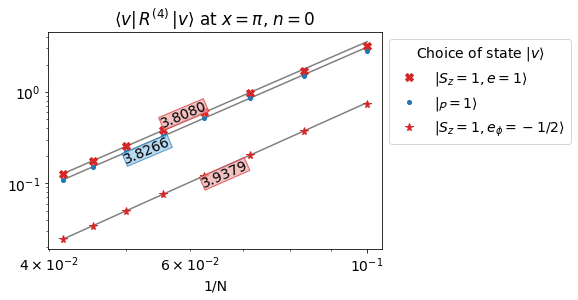}
\caption{Absolute value of matrix elements, plotted using the same conventions as in Figure \ref{separate_terms} but with the choice of $|u\rangle = |v \rangle$. 
}\label{h4_allthree_0} 
\end{figure}
\begin{figure}[h!]
\center
\includegraphics[height=5cm, trim={0 0 6.8cm 0},clip ]{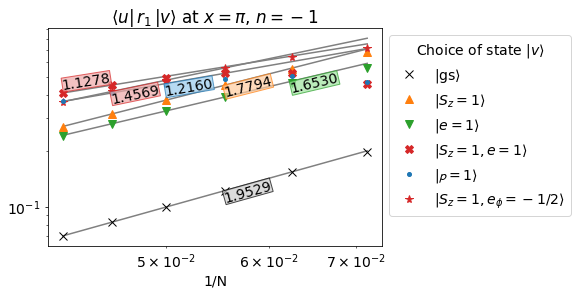}\includegraphics[height=5cm, trim={0 0 6.8cm 0},clip]{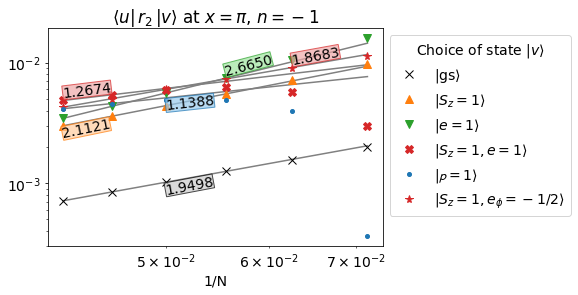}
\includegraphics[height=5cm]{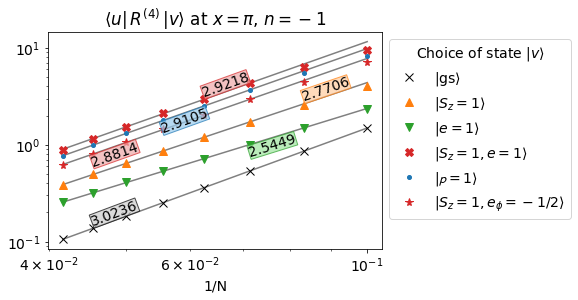}
\caption{Absolute value of matrix elements, plotted using the same conventions as in Figure \ref{separate_terms}. 
In case of choices of $|v\rangle$ showing particularly strong finite size effects ($|S_z=1,e=1\rangle$ and $|\mathcal{p}=1\rangle$ in upper two plots) the linear fit is performed using only the two leftmost points.
}\label{h4_allthree_1} 
\end{figure}

Similarly to \eqref{com_nonzero} we first expand the commutator under investigation. We obtain the expression 
\begin{equation}\label{com_of_P}
\begin{split}
&[\KSgen_{p+n}-\bar{\KSgen}_{-p-n},\KSgen_{-p}-\bar{\KSgen}_{p}] \\
& \quad = -2i\, \left({N\over2\pi}\right)^2 
\left( \frac{\gamma}{\pi \sin \gamma} \right)^4 \Bigg\{  \sin \left( \frac{4\pi p + 2\pi n}{N} \right) \sum_{j=1}^{N} e^{2i\pi n (j+3/2)/N}
 [[e_j,e_{j+1}],[e_{j+2},e_{j+3}]] \\
& \quad+ e^{- i \pi n /N}  \sin \left( \frac{2\pi p + \pi n}{N} \right) \sum_{j=1}^{N} e^{2i\pi n (j+3/2)/N}
\left( -[e_j,e_{j+1}] - [e_{j+1},e_{j+2}] + \sqrt{Q}(e_je_{j+1}e_{j+2} - e_{j+2}e_{j+1}e_j)  \right)
\Bigg\}.
\end{split}
\end{equation}
Provided that the matrix elements of the combinations of Temperley-Lieb operators appearing above are of order $\mathcal{O}(1/N^r)$ with $r>0$ for the first row and $r>1$ for the second row, 
we may restrict our attention to leading order terms the trigonometric functions as well as the exponential in the second row. To be concise, let us denote
\begin{equation}
\begin{split}
r_1& = [[e_j,e_{j+1}],[e_{j+2},e_{j+3}]] \\
r_2 &= -[e_j,e_{j+1}] - [e_{j+1},e_{j+2}] + \sqrt{Q}(e_je_{j+1}e_{j+2} - e_{j+2}e_{j+1}e_j) .
\end{split}
\end{equation} 
We plot the matrix elements of $r_1$ and $r_2$ for $n=0,-1$ in Figures \ref{h4_allthree_0},\ref{h4_allthree_1}.
 We find that $r>1$ in both cases. With this in mind we keep only leading terms in the expansion: 
\begin{equation}\label{com_of_P_leading}
\begin{split}
&[\KSgen_{p+n}-\bar{\KSgen}_{-p-n},\KSgen_{-p}-\bar{\KSgen}_{p}] 
 =  (2p+ n) (-i)   {N\over2\pi} 
\left( \frac{\gamma}{\pi \sin \gamma} \right)^4  \sum_{j=1}^{N} e^{2i\pi n (j+3/2)/N} ( 
 2 r_1  + r_2) +\mathcal{O}\left({1\over N^{r-1}}\right)\,.
\end{split}
\end{equation}

We now compare \eqref{com_of_P_leading} to what is denoted $\hat{h}^{(4)}$ in \cite{KooSaleur}, using this time the result KS (2.59)\footnote{Generalized to $n\neq 0$ in the same way as in KS (3.33).} 
\begin{equation}\label{h4_operator}
\begin{split}
-i\, {N\over2\pi}\left({\gamma\over\pi\sin\gamma}\right)^4 &\sum_{j=1}^{N}e^{2i\pi n(j+3/2)/N}  6\,[[e_j,e_{j+1}],[e_{j+2},e_{j+3}]] \\
 &\quad+ 6\sqrt{Q}[e_j(e_{j+1}e_{j+2} + e_{j+2}e_{j+1})] + (Q+2)[e_j,e_{j+1}] 
\mapsto  \left( L_n - \bar{L}_{-n}   \right) .
\end{split}
\end{equation}
We note that as is the case for $\Platt$ in \eqref{P_phi}, also denoted by $\hat{h}^{(2)}$ in \cite{KooSaleur}, the ground state expectation value of $\hat{h}^{(4)}$ is zero. For this reason we do not need to introduce normal ordering in \eqref{h4_operator}. It turns out that the ground state expectation values of $r_1, r_2$ are also zero, as are some other matrix elements that are for this reason excluded from the figures below.


As before we define a remainder, here given by the difference between the summand in \eqref{com_of_P_leading} and \eqref{h4_operator}:
\begin{equation}\label{remainder_h4}
\begin{split}
\mathcal{R}^{(4)} &= -4\, [[e_j,e_{j+1}],[e_{j+2},e_{j+3}]]  -6\sqrt{Q}[e_j(e_{j+1}e_{j+2} + e_{j+2}e_{j+1})]
-(Q+2)[e_j,e_{j+1}] \\
&\hspace*{5cm}-[e_j,e_{j+1}] - [e_{j+1},e_{j+2}] + \sqrt{Q}(e_je_{j+1}e_{j+2} - e_{j+2}e_{j+1}e_j) .
\end{split}
\end{equation}
For the limit of \eqref{com_of_P} to be correct we need the remainder to decay as $\mathcal{O}(1/N^r)$ with $r>2$. We see in Figures \ref{h4_allthree_0},\ref{h4_allthree_1} that the numerical results support this. Together with the previous results for $[\KSgen_{p+n}~+~\bar{\KSgen}_{-p-n},\KSgen_{-p}+\bar{\KSgen}_{p}]$  and  $[\KSgen_{p+n}  +\bar{\KSgen}_{-p-n},\KSgen_{-p}-\bar{\KSgen}_{p}]$ we can conclude that the numerical results strongly indicate that $[\KSgen_{p+n},\bar{\KSgen}_{-p}]~\mapsto~0$. Again we show the figures only for $n=0,-1$ in the interest of saving space, but we note that at $n=-2$ the finite size effects are significantly smaller for $\mathcal{R}^{(4)}$ than they are for $r_1$ and $r_2$, and the numerical results for $\mathcal{R}^{(4)}$ show a clear $r>2$ slope for $n=-2$ already at the sizes we can access.

\FloatBarrier

\section{The ordinary XXZ case}\label{UnitaryXXZ}


 


Going back to the bosonization in section \ref{Bosonization}, we can note that if we defined more general local interactions, replacing the factor $i{\sin\gamma\over 2}$ by $i\lambda$ in \eqref{TLspin}, we would obtain the twist
\begin{equation}\label{T_XXZ}
T^\lambda=-{1\over 4}:(\partial\varphi)^2:+i{2\lambda\over\sin\gamma}\alpha_0\partial^2\varphi
\end{equation}
and the central charge
\begin{equation}\label{c_XXZ}
c^\lambda=1-96\alpha_0^2 {\lambda^2\over \sin\gamma^2} \,.
\end{equation}
%
%
%
We can use this additional freedom to explore the ``ordinary'' XXZ case, which we shall define shortly, in terms of a unitary $c=1$ CFT, and to interpolate between this CFT and the $c=1-\frac{6}{x(x+1)}$ CFT described above. 

We now describe what we mean by the ``ordinary'' XXZ case. In this setting we forget for a moment the Temperley-Lieb algebra, and remove the telescoping terms in the definition of the Hamiltonian density. We consider instead
%
%
the XXZ Hamiltonian as a sum of different local, Hermitian  interactions. 
Based on \eqref{Ham-3.2} we define
\begin{equation}
{\mathcal h}_j^{XXZ}=-\frac{\gamma}{\pi\sin\gamma}f_j\,,
\end{equation}
where we recall that 
\begin{equation}
\eXXZ_j=-\sigma_j^{-}\sigma_{j+1}^{+}-\sigma_j^{+}\sigma_{j+1}^{-}
-\frac{\cos\gamma}{2}\sigma_j^{z}\sigma_{j+1}^{z} +\frac{\cos\gamma}{2} \,.
\end{equation}
%
%
%
We can once more consider twisted boundary conditions parametrized by $\phi$, ``smeared out'' over all $N$ sites for numerical convenience. Written out explicitly we find
\begin{equation}\label{fj_matrix}
f_j = \hdots\otimes\mathbf{1}\otimes
\begin{pmatrix}
0 & 0 & 0 & 0 \\
0 & \frac{\q+\q^{-1}}{2}  & -e^{i\phi/N} & 0\\
0 & -e^{-i\phi/N} & \frac{\q+\q^{-1}}{2} & 0\\
0 & 0 & 0 & 0 
\end{pmatrix} \otimes\mathbf{1}\otimes \hdots.
\end{equation}

By the same reasoning as in Section \ref{Discrete_Vir_Section} we next introduce a momentum operator
\begin{equation}\label{P_phi_XXZ}
\PN^{XXZ} = -i \left( \frac{\gamma}{ \pi \sin \gamma} \right)^2  \sum^{N}_{j=1} [f_j, f_{j+1} ] 
\end{equation}
with a lattice momentum density ${\mathcal p}_j^{XXZ} = i[{\mathcal h}_j^{XXZ},\mathcal{h}_{j-1}^{XXZ}]=$ \phantom{ }
 $ -i \big(\frac{\gamma}{\pi\sin\gamma} \big)^2 [\eXXZ_{j-1},\eXXZ_j]$. This leads to the following alternative discretization of the Virasoro generators:
\begin{subequations}\label{generators_hi}
\begin{eqnarray}
\KSgen_n^{\text{XXZ}} &=& \frac{L}{4\pi} \left[ -\frac{\gamma}{\pi \sin \gamma}  \sum^{L}_{j=1} e^{inj2\pi/L} \left( f_j - e_\infty + \frac{i\gamma}{\pi \sin \gamma} [f_j, f_{j+1} ] \right)\right] + \frac{1}{24} \delta_{n0} \,, \\
\bar{\KSgen}_n^{\text{XXZ}} &=& \frac{L}{4\pi} \left[ -\frac{\gamma}{\pi \sin \gamma}  \sum^{L}_{j=1} e^{-inj2\pi/L} \left( f_j - e_\infty - \frac{i\gamma}{\pi \sin \gamma} [f_j, f_{j+1} ] \right)\right] + \frac{1}{24} \delta_{n0} \,.
\end{eqnarray}
\end{subequations}
Notice that in these expressions we have used the value $c=1$ for the central charge, irrespective of $x$.

\subsection{Algebraic description of the ordinary XXZ case}


To find the algebra relevant to the new Hamiltonian densities $f_j$ we compare to the $R$-matrix defined in \cite{Candu}\footnote{Section 4.2. $O(2)$ spin model. Note that the sign convention for the off-diagonal elements in $E$ and $Q$ is here the opposite to that of \cite{Candu}, to be consistent with the rest of this paper.} as $R(u)=\mathbf{1}+f(u)Q+g(u)E$, where
\begin{subequations}
\begin{eqnarray}
f(u) &=& \frac{\sin\gamma - \sin u}{2\sin(\gamma-u)} - \frac{1}{2} \,, \\
g(u) &=&  \frac{\sin\gamma + \sin u}{2\sin(\gamma-u)} - \frac{1}{2}
\end{eqnarray}
\end{subequations}
contain the dependence on $u,\gamma$ and the operators $Q_j=\cdots\otimes\mathbf{1}\otimes Q\otimes\mathbf{1}\otimes\cdots$ and $E_j=\cdots\otimes\mathbf{1}\otimes E\otimes\mathbf{1}\otimes\cdots$, with
\begin{subequations}
\begin{eqnarray}
Q &=& \begin{pmatrix}
0 & 0 & 0 & 0 \\
0 & 1 & 1 & 0\\
0 & 1 & 1 & 0\\
0 & 0 & 0 & 0 
\end{pmatrix} \,, \label{Qmatrix} \\ 
E &=& \begin{pmatrix}
0 & 0 & 0 & 0 \\
0 & 1 & -1 & 0\\
0 & -1 & 1 & 0\\
0 & 0 & 0 & 0 
\end{pmatrix} \,,    \label{Ematrix}
\end{eqnarray}
\end{subequations}
both separately fulfil the Temperley-Lieb relations for $\mathfrak{m}=2$ corresponding to $\q=1$. Together with relations between operators $Q_j$ and $E_k$ they generate an algebra $\oldmathcal{V}_N$ that is further discussed in \cite{Candu}. 

As before the Hamiltonian is found from the derivative of the transfer matrix, and we obtain from the above R-matrix the desired Hamiltonian density 
\begin{multline}
{\mathcal h}_j^{XXZ} = -\frac{\gamma}{\pi} ( f'(0)Q + g'(0)E) \\
= -\frac{\gamma}{\pi} \frac{1}{2\sin\gamma} \left[ \left(\frac{\q+\q^{-1}}{2}-1\right)Q + \left(\frac{\q+\q^{-1}}{2}+1\right)E \right] = 
 - \frac{\gamma}{\pi \sin\gamma} 
\begin{pmatrix}
0 & 0 & 0 & 0 \\
0 & \frac{\q+\q^{-1}}{2}  & -1 & 0\\
0 & -1 & \frac{\q+\q^{-1}}{2} & 0\\
0 & 0 & 0 & 0 
\end{pmatrix}.
\end{multline} 
Twisted boundary conditions correspond to modifying the off-diagonal elements in $Q$ and $E$ as in \eqref{TLmatrix_spin} and \eqref{fj_matrix}.  

Small-size computations ($N=4,6$) indicate that the state space of the XXZ Hamiltonian at $S_z=0,\phi=0$ decomposes into a direct sum of at least two submodules under the action of the algebra $\oldmathcal{V}_N$. We shall further see in Table \ref{hi_gs} that at all sizes, the matrix elements $\langle a_{-1} \mathbf{1} | \KSgen^{XXZ}_{-1}|\mathbf{1}\rangle$ and $\langle \mathbf{1} | \KSgen^{XXZ}_{1}|a_{-1} \mathbf{1}\rangle$ are \emph{both} zero, in accordance with what would be expected from a unitary theory. Thus the structure of the scaling limit Virasoro modules seems to be found to at least some degree at the level of the lattice for the case of $c=1$ too, and not only for $c=1-\frac{6}{x(x+1)}$.



\subsection{Central charge and conformal weights for the ordinary XXZ case}

We can consider the same kind of trace as in \eqref{trace_eq} 
 in the twisted XXZ chain, this time considered as a $c=1$ theory. 
Calling ${\cal H}_{S_z}$ the subspace with third component of the spin fixed to $S_z$ we have
\begin{equation}
\text{Tr}_{{\mathcal H}_{S_z}}\,e^{-\beta_R H^{XXZ}_\phi}e^{-i\beta_i {\mathcal P}^{XXZ}_\phi} \xrightarrow{\, N\to\infty\,}\; G_{S_z,\phi}  
 \end{equation}
with 
 \begin{equation}\label{G-func}
 G_{S_z,\phi} = \frac{q^{-1/24}\bar{q}^{-1/24}}{P(q)P(\bar{q})} \sum_{e \in \mathbb{Z}} q^{h^{XXZ}_{(e-e_\phi),-S_z}}\bar{q}^{h^{XXZ}_{(e-e_\phi),S_z}},
\end{equation}
and  we have introduced the notation 
\begin{equation}
h^{XXZ}_{r,s}={[(x+1)r-xs]^2\over 4x(x+1)}={1\over 4} \left[\sqrt{x+1\over x}r-\sqrt{x\over x+1}s\right]^2 \,.
\end{equation}
Recalling $F$ from \eqref{F-func}, of course we have algebraically that $F=G$. The objects are however thought of quite differently. In particular, we expect now that, ``generically''
\begin{equation}
G(S_z,\phi)\mapsto \bigoplus_{e\in\mathbb{Z}} \Verma{e-e_\phi,-S_z}\otimes\Vermab{e-e_\phi,S_z} \,,
\end{equation}
where the Verma modules are now modules at central charge $c=1$. What ``generically'' means in this case is not so clear unless one delves into representation theory of the relevant lattice algebra: for the time being, we will consider $\q$ and $\phi$ generic to be defined through what is considered generic for Temperley-Lieb models. 

It is well expected that the fields associated with the conformal weights $h^{XXZ}$ in the sum (\ref{G-func}) are vertex operators with charges
$\alpha,\bar{\alpha}$ given as in \eqref{alpha} with $\alpha_0$ replaced by zero:
\begin{subequations}\label{alphaXXZ}
\begin{eqnarray}
\alpha^{XXZ} & = & \frac{1}{2}(e -  e_\phi )  \alpha_{+}  - \frac{1}{2}S_z \alpha_{-}  \\
\bar{\alpha}^{XXZ} & = & \frac{1}{2}  (e - e_\phi ) \alpha_{+} + \frac{1}{2}S_z \alpha_{-} .
\end{eqnarray}
\end{subequations}
When the weights $h^{XXZ}$ are non-degenerate the corresponding Verma modules are irreducible. 
Since we are dealing with central charge $c=1$, the only degenerate values are $h^{XXZ}={n^2\over 4}$. Even for generic $\q$, these occur when $\phi=0$, $S_z=0$. We will discuss below what happens in both the non-degenerate and the degenerate case.

\FloatBarrier

\subsection{Numerical results for the ordinary XXZ case}

Building the Koo-Saleur generators out of $f_j$ rather than $e_j$ we expect to obtain the theory with the previous background charge $\alpha_0$ replaced by zero. As a first numerical example we consider the generic case $S_z,\phi=1/10$ explored in Table \ref{generic_example}. We denote by $(\LnMatXXZ{-1}[N])_{ab}$ the matrix elements of the generators $\KSgen_n^{\text{XXZ}}$ defined by \eqref{generators_hi}, and choose the scaling states $|u_1\rangle$ and $|v_1\rangle$ as described in \eqref{p0states} and \eqref{p1states}. The conjectured value is $\sqrt{2}\alpha$ as before, where the charge $\alpha=\alpha^{XXZ}$ is now however given by \eqref{alphaXXZ} rather than \eqref{alpha}. 
The result is shown in Table \ref{hi}.


\begin{table}[h]
\center
\begin{minipage}{0.2\textwidth} 
\center
\begin{tabular}{ll}
$N$      & $\langle v_1 | \KSgen_{-1}^{XXZ} | u_1 \rangle$  \\
\hline
10 & 0.55059214 \\
12 & 0.56499765 \\
14 & 0.57410237 \\
16 & 0.58022239 \\
18 & 0.58453691 \\
20 & 0.58769572 \\
22 & 0.59008035 \\
24 & 0.59192667 \\
26 & 0.59338687 \\
28 & 0.59456275 \\
\vphantom{$\int^0$}\smash[t]{\vdots}  & \vphantom{$\int^0$}\smash[t]{\vdots} \\ 
\end{tabular}
\end{minipage}
\begin{minipage}{0.2\textwidth} 
\center
 \begin{tabular}{lll}
\vphantom{$\int^0$}\smash[t]{\vdots}  & \vphantom{$\int^0$}\smash[t]{\vdots} \\ 
30 & 0.59552449 \\
32 & 0.59632179 \\
34 & 0.59699064 \\
36 & 0.59755764 \\
38 & 0.59804281 \\
40 & 0.59846143 \\
42 & 0.59882537 \\ 
44 & 0.59914391 \\
46 & 0.59942447 \\
48 & 0.59967297 \\
\vphantom{$\int^0$}\smash[t]{\vdots}  & \vphantom{$\int^0$}\smash[t]{\vdots} \\ 
\end{tabular}
\end{minipage}
\begin{minipage}{0.2\textwidth} 
\center
 \begin{tabular}{lll}
\vphantom{$\int^0$}\smash[t]{\vdots}  & \vphantom{$\int^0$}\smash[t]{\vdots} \\ 
50 & 0.59989421 \\
52 & 0.60009214 \\
54 & 0.60026998 \\
56 & 0.60043043 \\
58 & 0.60057574 \\
60 & 0.60070781 \\
62 & 0.60082822 \\
64 & 0.60093836 \\
66 & 0.60103938 \\
68 & 0.6011323 \\
\vphantom{$\int^0$}\smash[t]{\vdots}  & \vphantom{$\int^0$}\smash[t]{\vdots} \\ 
\end{tabular}
\end{minipage}
\begin{minipage}{0.2\textwidth} 
\center
 \begin{tabular}{lll}
\vphantom{$\int^0$}\smash[t]{\vdots}  & \vphantom{$\int^0$}\smash[t]{\vdots} \\ 
70 & 0.60121798 \\
72 & 0.60129718 \\
74 & 0.60137054 \\
76 & 0.60143866 \\
78 & 0.60150202 \\
80 & 0.60156108 \\
$p_{25}$ & 0.60294403 \\
$p_{30}$ & 0.60294453 \\
$p_{35}$ & 0.60294326 \\
\textbf{conj} & 0.60293032 \\
\hline\\
\end{tabular}
\vspace{2pt}
\end{minipage}
\caption{ 
Matrix element $(\LnMatXXZ{-1}[N])_{v_1u_1}$, where the scaling states $|u_1\rangle$ and $|v_1\rangle$ follow the patterns of Bethe integers shown in \eqref{p0states} and \eqref{p1states}. The numerical values are given for the case of $S_z =  1, e = 0, x=\pi, \phi=1/10$.
For comparison, the matrix elements  $(\LnMat{-1}[N])_{v_1u_1}$ were given in Table \ref{generic_example}. The same conventions regarding extrapolation are used in this table.
}\label{hi}
\end{table}

\begin{table}[h]
\center
\begin{tabular}{lll}
  \vspace*{5pt}        & $\langle a_{-1}u | \KSgen_{-1} | u \rangle$  & $\langle \bar{a}_{-1}u |\bar{\KSgen}_{-1}| u \rangle$    \\
  $N$    & $\langle u| \KSgen_{1} | a_{-1}u \rangle$  & $\langle u |\bar{\KSgen}_{1}| \bar{a}_{-1}u \rangle$    \\
\hline
8 & 0.18927529 & 0.18927529 \\
10 & 0.19199126 & 0.19199126 \\
12 & 0.19341304 & 0.19341304 \\
14 & 0.1942385 & 0.1942385 \\
16 & 0.19475441 & 0.19475441 \\
18 & 0.19509508 & 0.19509508 \\
20 & 0.19532981 & 0.19532981 \\
22 & 0.1954971 & 0.1954971 \\
$p_7$ & 0.19601147 & 0.19601147 \\
 \textbf{conj} & 0.19603173 & 0.19603173 \\
\hline
\end{tabular}
\hspace*{1cm}
\begin{tabular}{lll}
   \vspace*{5pt}       & $\langle a_{-1}u | \KSgen_{-1}^{XXZ} | u \rangle$  & $\langle \bar{a}_{-1}u |\bar{\KSgen}_{-1}^{XXZ}| u \rangle$    \\
  $N$    & $\langle u| \KSgen_{1}^{XXZ} | a_{-1}u \rangle$  & $\langle u |\bar{\KSgen}_{1}^{XXZ}| \bar{a}_{-1}u \rangle$    \\
 \hline
 \\
  \\
   \\
    \\
8-22 &$ \mathcal{O}(10^{-15})$&$ \mathcal{O}(10^{-15})$\\
\\
 \\
  \\
   \\
 \textbf{conj} & 0 & 0 \\
\hline
\end{tabular}

\vspace*{0.5cm}

\begin{tabular}{lll}
     \vspace*{5pt}     & $\langle a_{-1}u_{\q^2} | \KSgen_{-1}^{XXZ} | u_{\q^{\pm 2}} \rangle$  & $\langle \bar{a}_{-1}u_{\q^2} |\bar{\KSgen}_{-1}^{XXZ}| u_{\q^{\pm 2}} \rangle$    \\
  $N$    & $\langle u_{\q^{\pm 2}}| \KSgen_{1}^{XXZ} | a_{-1}u_{\q^2} \rangle$  & $\langle u_{\q^{\pm 2}} |\bar{\KSgen}_{1}^{XXZ}| \bar{a}_{-1}u_{\q^2} \rangle$    \\
\hline
8 & 0.18692177 & 0.18692177 \\
10 & 0.19029496 & 0.19029496 \\
12 & 0.1921169 & 0.1921169 \\
14 & 0.19320597 & 0.19320597 \\
16 & 0.193906 & 0.193906 \\
18 & 0.19438114 & 0.19438114 \\
20 & 0.1947176 & 0.1947176 \\
22 & 0.19496405 & 0.19496405 \\
$p_7$ & 0.19602359 & 0.19602359 \\
 \textbf{conj} & 0.19603173 & 0.19603173 \\
\hline
\end{tabular}

\caption{Matrix elements  of $\KSgen_{\pm 1}$, $\bar{\KSgen}_{\pm 1}$, $\KSgen_{\pm 1}^{XXZ}$ and $\bar{\KSgen}_{\pm 1}^{XXZ}$ with the same conventions as in Table \ref{ED_comparisons}, in the sector of $S_z=0$ at $x=\pi$. We call $|u\rangle$ the primary state at $e=0$ and periodic boundary conditions $e_\phi=0$ corresponding to the module $\AStTL{0}{1}$ (top), and $|u_{\q^{\pm 2}}\rangle$ the primary state at $|e|=1$ and twisted boundary conditions $|e_\phi|=-\alpha_{-}/\alpha_{+}$ corresponding to the modules $\AStTL{0}{\q^{2}}$ and $\AStTL{0}{\q^{-2}}$(bottom) for which the results are the same. In the latter case, the results for $\KSgen_{\pm 1}$ and $\bar{\KSgen}_{\pm 1}$ are given in Table \ref{level1twisted}. 
%
}\label{hi_gs}
\end{table}

Next we make the same comparison in the sector of $S_z=0$, considering both a primary state $|u\rangle$ corresponding to $e=e_\phi=0$ and a primary state $|u_{\q^{\pm2}}\rangle$ corresponding to $|e|=1, |e_\phi|=-\alpha_-/\alpha_+$. For the twisted boson we expect to obtain a non-zero value when lowering $|u\rangle$, since this state is not the identity state but rather has $\alpha=\alpha_0$. For the $c=1$ case we expect to obtain zero. The opposite scenario holds for $|u_{\q^{\pm2}}\rangle$ up to having two possibilities within the twisted boson theory, depending on the sign of $e_\phi$. The results are shown in Table \ref{hi_gs} and in the earlier Table \ref{level1twisted}, where we in both tables focus on smaller sizes to show how we obtain exact results at finite size.
At $c=1$ the matrix elements at $e=e_\phi=0$ are not only zero for $\KSgen_{\pm1}$, but for each $f_j$ individually. This can be compared to how the matrix elements were zero for individual $e_j$ within indecomposable modules of the affine Temperley-Lieb algebra. The difference is here that both $\langle a| f_j|b\rangle$ and $\langle b| f_j|a\rangle$ are zero, while for $e_j$ only one of these matrix elements is zero. Thus the results at $c=1$ would indicate a reducible module replacing the indecomposable module at $c=\frac{6}{x(x+1)}$ in this sector.


\subsection{Interpolation between $c=1$ and $c=1-\frac{6}{x(x+1)}$}\label{interpolation}

We have seen that the ``ordinary'' XXZ chain corresponds to $\lambda=0$ in \eqref{T_XXZ}--\eqref{c_XXZ}. Here we instead consider $\lambda= t \frac{\sin\gamma}{2}$, varying $t$ between 0 and 1 to obtain values of the central charge between $c=1$ and $c=1-\frac{6}{x(x+1)}$.\footnote{
With $\alpha_0 \neq 0$ it is necessary for conformal charges to be quantized in terms of the screening charges $\alpha_\pm$ in order to obtain a nonzero four-point function of the corresponding field \cite{DotsenkoFateev}. For this reason the interpolation may not be physical, since it breaks the relation between $\alpha_0$ and $\alpha_\pm$.
} On the lattice we define
\begin{equation}\label{interpolated}
M_j(t) = -\sigma_j^{-}\sigma_{j+1}^{+}-\sigma_j^{+}\sigma_{j+1}^{-}
-\frac{\cos\gamma}{2}\sigma_j^{z}\sigma_{j+1}^{z} -t\frac{i\sin\gamma}{2}(\sigma_j^{z}-\sigma_{j+1}^{z})+\frac{\cos\gamma}{2} \,,
\end{equation}
such that 
\begin{equation}\label{Ln_interpolating}
M_j(t) = 
\begin{pmatrix}
0 & 0 & 0 & 0 \\
0 & \q^{-1} + (1-t)\frac{\q-\q^{-1}}{2}  & -1 & 0\\
0 & -1 & \q- (1-t)\frac{\q-\q^{-1}}{2} & 0\\
0 & 0 & 0 & 0 
\end{pmatrix} 
=
\begin{cases}
f_j &\text{if } t=0 \,, \\
e_j &\text{if } t=1 \,.
\end{cases}
\end{equation}
Again, twisted boundary conditions correspond to modifying the off-diagonal elements of $M_j(t)$ as in \eqref{TLmatrix_spin} and \eqref{fj_matrix}. 
We call the corresponding Koo-Saleur generators $\KSgen^t_n$ and $\bar{\KSgen}^t_n$.

\begin{figure}[h]
\center
\includegraphics[height=6cm]{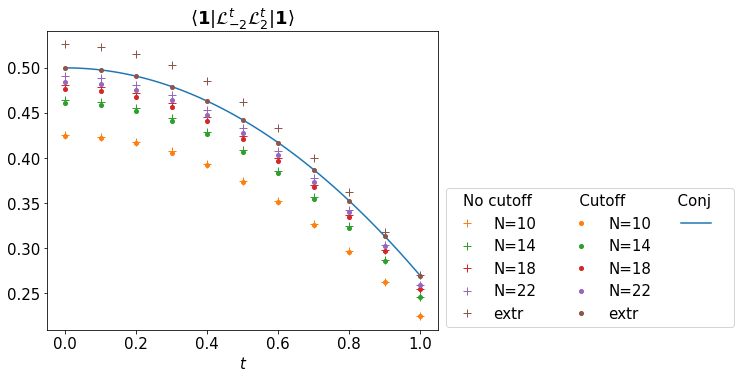}
\caption{Finite-size estimate for the central charge $c$ as a function of interpolation parameter $t$ at $x=\pi$, using the same conventions as in Figure \ref{c_plot}. The generators $\KSgen_n^t$ are defined below \eqref{Ln_interpolating}.}\label{c_vs_t}
\end{figure}

As seen from \ref{c_XXZ}, we expect the central charge to be quadratic as a function of $t$, 
\begin{equation}
c(t)=1- 24 (t\alpha_0)^2.
\end{equation}
To explore this numerically we consider matrix elements 
\begin{equation}
\langle \mathbf{1}|\KSgen^t_{2} \KSgen^t_{-2}|\mathbf{1}\rangle.
\end{equation}
We thus need to interpolate the identity state $\mathbf{1}$ continuously as we vary $t$. Both $S_z$ and $e$ taking integer values only, we are left with varying the twist $\phi$. We may start the interpolation either from the known identity state $S_z=0,\,e=0,\,e_\phi=0$ at $t=0,\, c=1$ or from the known identity state $S_z=0,\,|e|=1,\,e_\phi=\alpha_-/\alpha_+$ at $t=1,\,c=1-\frac{6}{x(x+1)}$. In these cases the result of applying $\KSgen_{-1}^{XXZ}$ or $\KSgen_{-1}$, respectively, yields identically zero on the lattice thanks to the structure of the modules of the respective algebras. 
It turns out that these two approaches yield identical results: the ``unsmeared'' twist $\phi=2\pi e_\phi$ being defined modulo $2\pi$, combined with the conjecture \eqref{alpha} depending only on the combination $e-e_\phi$, means that shifting $e$ from 0 to 1 is equivalent to choosing a different representative in the equivalence class for $\phi$.\footnote{We note that while all representatives are equivalent on the level of the ``unsmeared'' twist, one does need to be careful with the choice of representative when numerically ``smearing out'' the twist.} 
We choose here to start from the identity state at $t=0$, with $e=e_\phi=S_z=0$. Within the conjecture \eqref{alpha} the background charge $\alpha_0$ is replaced by $t\alpha_0$, meaning that as $t$ increases we must increase the twist as $e_\phi = 2t\alpha_0/\alpha_+$ to keep $\alpha=0$. Following this procedure we obtain the values of the central charge at a function of $t$ shown in Figure \ref{c_vs_t}. We note that the effects of ``parasitic couplings'' seem to be more pronounced for the case of the ``ordinary'' XXZ chain at $t=0$. That is, we find once more that these effects are the strongest for the largest value of $c$. 

\FloatBarrier

\end{document}